\documentclass[a4paper]{jpconf}
\pdfoutput=1

\usepackage{graphicx}

\begin{document}
\title{The A4 project: physics data processing using the Google protocol buffer library}

\author{Johannes Ebke$^1$ and Peter Waller$^2$}

\address{$^1$ LMU M\"unchen, Am Coulombwall 1, 85748 Garching, Germany}
\address{$^2$ University of Liverpool, Liverpool L69~7ZE, UK}

\ead{ebke@cern.ch, peter.waller@cern.ch}

\begin{abstract}
In this paper, we present the High Energy Physics data format, processing toolset and analysis library {\scshape a4}, providing fast I/O of structured data using the {\scshape Google} protocol buffer library. The overall goal of {\scshape a4} is to provide physicists with tools to work efficiently with billions of events, providing not only high speeds, but also automatic metadata handling, a set of {\scshape UNIX}-like tools to operate on {\scshape a4} files, and powerful and fast histogramming capabilities. At present, {\scshape a4} is an experimental project, but it has already been used by the authors in preparing physics publications. We give an overview of the individual modules of {\scshape a4}, provide examples of use, and supply a set of basic benchmarks. We compare {\scshape a4} read performance with the common practice of storing unstructured data in {\scshape ROOT} trees. For the common case of storing a variable number of floating-point numbers per event, speedups in read speed of up to a factor of six are observed.
\end{abstract}

\section{Introduction}

One common problem in High Energy Particle Physics computing is getting a reasonable balance between rapid and easy code development, usually the domain of scripting languages such as Python, and raw processing speed. We believe that new developments in computing outside High Energy Physics, in particular the adoption of the new C++11 standard and in general the availability of high-quality open source libraries, make it possible to improve the usability and readability of physics analysis codes without sacrificing processing speed.

The {\scshape a4} project was started with the goal of processing and analyzing the data taken with the ATLAS detector at the LHC in 2011. Including Monte Carlo simulations, the processing of approximately one billion ($10^9$) events was necessary. From each event $\approx 6$ kB of data were required for analysis, resulting in an expected dataset size of 6 TB. Criteria were a fast turn-around time for analyses, easy definition and generation of large numbers of diverse histograms and the possibility to quickly adapt the code to as yet unknown analysis requirements. 

To achieve this, a file format with a standalone I/O library was designed (Sections~\ref{sec:protobuf} and~\ref{sec:a4io}). Additional libraries enable fast processing (Section~\ref{sec:a4process}) and easy output handling (Sections~\ref{sec:a4store} and~\ref{sec:syst}). Conversion of results to the ROOT system is also provided (Section \ref{sec:a4root}). In the last section, some comparative numbers from basic benchmarks are presented.

At the time of writing, the {\scshape a4} project is still in the experimental phase. While it has been used in published analyses of the ATLAS experiment, it is still under heavy development, and the details presented in the following sections are still subject to change\footnote{Testing and collaboration are welcome: {\scshape a4} is available at \texttt{liba4.net}.}.

\section{The Protocol Buffer Library}\label{sec:protobuf}

\begin{figure}
    \begin{verbatim}
    message Lepton {
        optional double pt = 1;
        optional double eta = 2;
        optional double phi = 3;
        optional int32 charge = 4;
    }
    message PhysicsEvent {
        optional int32 run_number = 1;
        optional int32 event_number = 2;
        repeated Lepton electrons = 5;
        repeated Lepton muons = 6;
    }
    \end{verbatim}
    \caption{Example of a protobuf message definition for a Physics event. The numbers are the field identifiers in the binary format. The fact that they are specified explicitly allows renaming the fields without changing the data on disk. }
    \label{proto_example}
\end{figure}

Motivated by reports \cite{samvel} of higher performance and ease-of-use with respect to {\scshape ROOT} \cite{root} trees, the {\scshape Google} protocol buffer (protobuf) library (used heavily at {\scshape Google} \cite{protobuf}) was chosen as a serialisation format. The protobuf library defines a fast binary format for \textit{messages}. Message structures (so called `descriptors') are defined in a simple C-like language in \verb".proto" files. An example is given in Figure~\ref{proto_example}. From these, interfaces for different programming languages can be generated using the provided extensible compiler, \verb"protoc". A given interface contains the code describing a set of classes providing the (de)serialisation functionality tailored for a given {\verb".proto"} file and platform. The descriptors themselves can be serialised as protobuf messages, facilitating the inspection of arbitrary serialised messages without needing the descriptors at compile time.

In summary, the relevant features of the protobuf library are the following:

\begin{itemize}
    \item Fast serialisation and de-serialisation of structured data in the form of \textit{messages}
    \item Separation of data structure definition and code
    \item Extensible code generators for different programming languages\footnote{Java, C++ and Python are officially supported, others including C, Lisp, D, Go, Javascript, Matlab, Perl, and R, are available via add-ons}
    \item Messages can be nested
    \item Repeated fields can store data in a manner similar to dynamic arrays
    \item Content can be omitted from optional fields, and does not take up space in this case 
    \item Thread safety
    
\end{itemize}

One problem with multiple analysts working on the same dataset is ensuring that the files remain compatible when variables are added, removed or renamed. The descriptors of message fields also have some useful properties that assist in reusing files:

\begin{itemize}
    \item Extendable with new fields
    \item Fields can be renamed or removed without breaking binary compatibility
    \item Descriptors are available at run-time if necessary, and can be stored as protobuf messages themselves
    \item They support metadata on field definitions, which allows for example describing conversions from {\scshape ROOT} trees or how metadata should be merged
\end{itemize}

The auto-generated protobuf code in C++ is well designed and exposes the contents of messages in a way that encourages efficient code without using error-prone pointers for data access. However, protobuf is primarily an inter-process message format, and cannot directly be used in files without extra work, described in the following section.

\section{A4 I/O: offline storage of protocol buffer messages}\label{sec:a4io}

Individual protobuf messages are not suitable as a standalone offline data format. To make sense of them, they require external knowledge (usually compiled into the binary in the case of C++), they do not describe their own length, nor do they have ending delimiters.

\begin{figure}\begin{center}
\includegraphics[scale=0.95]{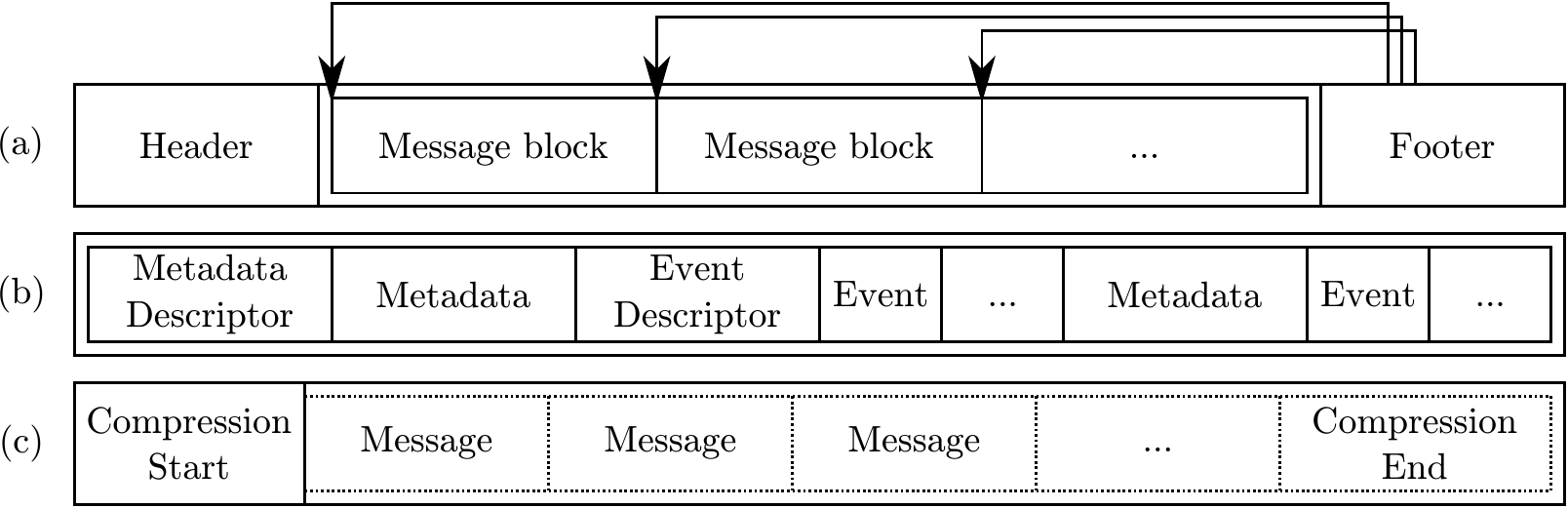}
\caption{
The {\scshape a4} data structure primitives. Each segment represents a protobuf message. An ellipsis is used to indicate repetition. (a) Overall logical file structure. The header contains the file type and version. It is followed by a number of message blocks. The footer contains information about the written messages, including byte offsets back to the header, to class descriptors and to metadata, making it possible to seek directly to the metadata. (b) An example message block, which may or may not be compressed (see (c)). The descriptors are only written once per protobuf class per output stream. (c) Example of a compressed series of messages. Message compression begins with an uncompressed message indicating the compression type. The compression is halted with an end-compression message. Compression is halted and immediately resumed when offsets are requested, e.g. to put descriptor offsets in the footer. Message compression is handled transparently in the reading library.
}
\label{fileformat}
\end{center}\end{figure}

Since no container format for protobuf messages was available, the {\scshape a4} message file and stream format was defined. It is illustrated in Figure~\ref{fileformat}. 
The following bullet points outline the primary design considerations:
\begin{itemize}
    \item Store protobuf messages of arbitrary types
    \item Store the descriptors for messages, making the format self-describing and enabling the use of format-independent tools
    \item Store metadata for blocks of messages
    \item Transparent compression using different algorithms\footnote{Currently implemented are {\scshape zlib}, {\scshape gzip} and {\scshape snappy}.}
    \item Binary concatenation of {\scshape a4} files yields a valid {\scshape a4} file with all metadata quickly available, for trivial merging of large numbers of small files and efficient network transport
    \item Splitting of {\scshape a4} files by a metadata field key (for example by data taking period) is possible with a command line tool
    \item Support a linear no-seeking mode of operation, suitable for network streaming
\end{itemize}

An experimental converter for ATLAS events to an ``event message'' was written in Python in the ATLAS Athena analysis framework\footnote{Python proved useful for rapidly prototyping a working converter at the expense of runtime speed. Faster conversion is now available with \texttt{root2a4} if the data is already in the form of {\scshape ROOT} trees.} and run on ATLAS data using Ganga \cite{ganga} to submit jobs to the LHC computing grid. The resulting {\scshape a4} files are stored on a dCache \cite{dcache} system. These data were processed by distributing compiled executables using the {\scshape a4} I/O and processing libraries via the local batch system. Both reduced or derived data-sets and complex sets of histograms have been produced, and enabled contributions to official ATLAS results.

Both C++ and Python interfaces are provided for the {\scshape a4} format. The C++ interface can use the remote access libraries \verb"rfio" and \verb"dcap" to access grid storage elements. Experimental support for the Apache Hadoop distributed filesystem is also implemented. It also provides input and output classes that distribute files to any number of threads and combine output automatically.

Since the message structure is stored in the file, it does not need to be known at compile time if speed is not critical. Included in the {\scshape a4} I/O module are the command-line tools \verb"a4dump" and \verb"a4info", which print messages and metadata stored in {\scshape a4} files in a human-readable format. 

To summarise, the standalone {\scshape a4} I/O module allows arbitrary protobuf messages to be written and read at high speed from multiple threads. A large amount of experimental data has been stored in {\scshape a4} files, and was utilised to produce physics results. Building on this foundation, the {\scshape a4} processing module described in the next section provides infrastructure elements to facilitate common tasks.

\section{A4 processing and automatic book-keeping}\label{sec:a4process}

In High Energy Physics, data are typically analysed in an event loop: each event is loaded in turn and processed by an analysis function. The {\scshape a4} processing module attempts to make it as easy as possible to write powerful, configurable programs that analyse files containing events using multiple threads which run simultaneously. Metadata is also stored in a protobuf message class, which can be annotated to define how the metadata may be merged as illustrated in Figure~\ref{eventmetadata}. The merging of the metadata itself is illustrated in Figure~\ref{metadatamerge}. This allows automatic propagation of arbitrary information such as run numbers, Monte Carlo IDs and initial event counts through to the final histograms. This automatic metadata handling drastically simplifies the bookkeeping necessary to produce physics results from data.

\begin{figure}
\begin{verbatim}
import "a4/io/A4.proto";

message EventMetaData {
    optional bool simulation = 1 [(a4.io.merge)=MERGE_BLOCK_IF_DIFFERENT];
    repeated int32 mc_channel = 11 [(a4.io.merge)=MERGE_UNION];
    repeated string period = 3 [(a4.io.merge)=MERGE_UNION];
    
    optional int32 event_count = 6 [(a4.io.merge)=MERGE_ADD];
    optional double sum_mc_weights = 7 [(a4.io.merge)=MERGE_ADD];
    optional double reweight_lumi = 8 [(a4.io.merge)=MERGE_BLOCK_IF_DIFFERENT];
}
\end{verbatim}
\caption{
An excerpt of the metadata descriptor used for the authors analysis. The {\scshape a4} extensions to the protobuf field descriptors, \texttt{a4.io.merge}, describe how two metadata messages can be combined into one. The presence of \texttt{MERGE\symbol{95}BLOCK\symbol{95}IF\symbol{95}DIFFERENT} prevents histograms from data and simulation being summed. The \texttt{event\symbol{95}count} contains the sum of events which were processed in the original files, before any events were removed from the file during skimming.
}
\label{eventmetadata}
\end{figure}

\begin{figure}\begin{center}

\includegraphics[scale=0.95]{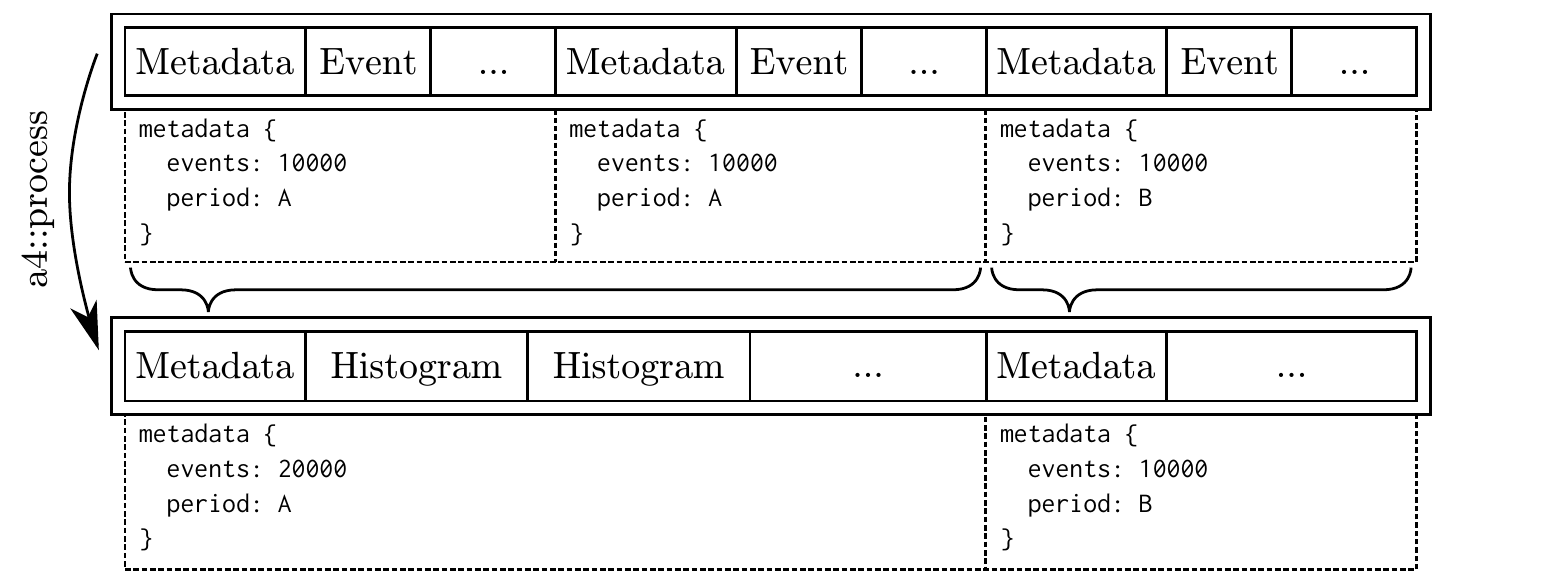}
\caption{
A simplified example of metadata propagation. Analysis code using \texttt{a4::process} generates histograms or processed events from the input events using the \texttt{--per=period} command-line switch. The metadata is automatically combined according to the definitions on the metadata descriptor (see Figure~\ref{eventmetadata} for examples). If the metadata key had instead been \texttt{--per=simulation} (or another field which was uniform across files), the resulting histograms would contain entries for all events described by a single metadata: \texttt{\{ events: 30000, period: A, period: B \}}.
}
\label{metadatamerge}
\end{center}\end{figure}

\begin{figure}
\begin{verbatim}
#include <a4/application.h>
#include "Event.pb.h"
#include "smear.h"
class SkimSlimThinSmear : public a4::process::ProcessorOf<Event, MetaData> {
  public: 
    void process(const Event & event) {
      // Cut on at least two muons (skim)
      if (event.size_muons() < 2) return;
      // Remove some fields (slim)
      Event new_event = event; new_event.clear_tracks();
      foreach(auto & muon, *new_event.mutable_muons()) {
        // Smear muons before writing them
        if (metadata().simulation()) smear(muon);
        muon.clear_id_hits(); // Remove hits from muons (thin)
      }
      write(new_event); // Write modified event to output file
    }
};

int main(int argc, const char * argv[]) {
  return a4::process::a4_main_process<SkimSlimThinSmear>(argc, argv);
}
\end{verbatim}
\caption{Listing of an example program utilising the {\scshape a4} processor class, skimming events with two muons, slimming these by removing tracks, removing the hits on the muons, and writing the events to an output file. The functions referring to the physics objects are generated automatically by \texttt{protoc}. The ``smear'' function is a hypothetical analysis function which modifies the contents of the written muon objects.}
\label{example_proc}
\end{figure}

The {\scshape a4} processing module provides a \verb"processor" base class. This class is available as a C++ template, which allows custom event and metadata protobuf message classes to be specified. The source code of an example processor is given in Figure~\ref{example_proc}. This code compiles into an executable program. Command-line arguments can be used to set input and output files, disable {\scshape a4} output entirely, set the number of threads, limit the number of events processed for testing purposes and control the metadata management. Additional program options can be added by the analysis writer. The popular C++ {\scshape Boost} library \cite{boost} is used to provide this feature, making it possible to provide \verb".ini" files to specify command line options.

Using this library, the common tasks of skimming (selecting only specific events to copy), slimming (selecting only specific physics objects to copy) and thinning (dropping unneeded variables from physics objects) can be performed simply by copying the event object or parts of it, modifying it, and calling \verb"write". In addition, new fields of the event can be filled, or even a different message type written out. One advantage of {\scshape a4} in this approach is that the event definition does not have to be changed in the case of slimming or thinning, since in protobuf messages fields that are not set do not use any space\footnote{It is possible to check if fields have been set.}. In all cases, the metadata will be preserved and passed on to the output files. In addition, the metadata applicable to the current event is always available in the processor.

In short, the processor provides a quick way to define common tasks or more complex processing stpdf. Automatic handling of metadata reduces manual bookkeeping. Programs are compiled into executables which accept customisable command-line options, listed by \verb"-h/--help", and can additionally be stored in an \verb".ini" file.

Histograms are the primary output of High Energy Physics analyses. Since defining, filling and storing histograms makes up a large part of typical analysis codes by line count, simplifying this process is a worthwhile task. In the following section, the {\scshape a4} store and {\scshape a4} histogramming modules are presented, and their use in the processor architecture described.

\section{Reusing histogram definitions with the A4 histogram store}\label{sec:a4store}

\begin{figure}
\begin{center}
\begin{math}
\underbrace{\vphantom{\texttt{"\symbol{95}\{T\}"}}\texttt{S("electrons/")}}_{\textbf{A}}
\underbrace{\vphantom{\texttt{"\symbol{95}\{T\}"}}\texttt{.T<H1>}}_{\textbf{B}} 
\underbrace{\vphantom{\texttt{"\symbol{95}\{T\}"}}\texttt{("pt")}}_{\textbf{C}}
\underbrace{\vphantom{\texttt{"\symbol{95}\{T\}"}}\texttt{(100,0,100,"p\symbol{95}\{T\} [GeV]")}}_{\textbf{D}}
\underbrace{\vphantom{\texttt{"\symbol{95}\{T\}"}}\texttt{.fill(electron.pt())}}_{\textbf{E}}
\end{math}

\vspace{\baselineskip}

\begin{math}
\underbrace{\vphantom{\texttt{"\symbol{95}\{T\}"}}\texttt{S("e/")}}_{\textbf{A}}
\underbrace{\vphantom{\texttt{"\symbol{95}\{T\}"}}\texttt{.T<H2>}}_{\textbf{B}}
\underbrace{\vphantom{\texttt{"\symbol{95}\{T\}"}}\texttt{("eta\_phi")}}_{\textbf{C}} 
\underbrace{\vphantom{\texttt{"\symbol{95}\{T\}"}}\texttt{(10,-5,5,"\#eta")(100,-PI,PI,"\#phi")}}_{\textbf{D}}
\underbrace{\vphantom{\texttt{"\symbol{95}\{T\}"}}\texttt{.fill(eta,phi)}}_{\textbf{E}}
\end{math}

\vspace{\baselineskip}

\begin{math}
\underbrace{\vphantom{\texttt{"\symbol{95}\{T\}"}}\texttt{S}}_{\textbf{A}}
\underbrace{\vphantom{\texttt{"\symbol{95}\{T\}"}}\texttt{.T<Cutflow>}}_{\textbf{B}} 
\underbrace{\vphantom{\texttt{"\symbol{95}\{T\}"}}\texttt{("ee","\_","channel",i)}}_{\textbf{C}}
\underbrace{\vphantom{\texttt{"\symbol{95}\{T\}"}}\texttt{.passed("cut\_",3)}}_{\textbf{E}}
\end{math}
\end{center}

\caption{Examples of the {\scshape a4} store invocation rules. \texttt{S} is an object of the type \texttt{ObjectStore}, representing a directory or prefix. \textbf{A} represents the location where the histogram should be stored, and its return type is the same as the original type of \texttt{S} which is cheap to copy. This may be efficiently passed through to a function accepting an \texttt{ObjectStore} as a parameter, allowing reuse of histogram definitions. \textbf{B} requests a particular object type, in this case a one- or two-dimensional histogram or a histogram indexed by label. \textbf{C} names the object in the store, and may contain a variable number of string or integer arguments. \textbf{D} specifies the axis range, and if available, may specify variable binning with C++11's new initializer lists. The axis label specified in this example can be omitted if desired. \textbf{D} is repeated once for each dimension of the target histogram. \textbf{E} fills the histogram at the desired quantity or label. The resulting code has a performance close to that of the \textbf{E} call after the histogram has been encountered for the first time.}
\label{a4store} 
\end{figure}

The {\scshape a4} store uses features of the new C++11 standard to provide a way to define, initialize, fill and store histograms and similar objects on one line inside the event loop, greatly simplifying the necessary bookkeeping. For this, a \textit{store} object, usually called \verb"S" is defined, representing a directory or a prefix to the name of an object intended to be saved as a result. In Figure~\ref{a4store}, the use of the {\scshape a4} store using the lightweight {\scshape a4} histogram classes \verb"H1", \verb"H2" and \verb"Cutflow" is illustrated\footnote{A tool to convert {\scshape a4} histograms to {\scshape ROOT} histograms is included.}.

These one-line definitions can not only be used to define single histograms, but can be prefixed with different subdirectories and reused in different functions. It is possible to conditionally add a prefix to all of the following histograms in the current event using, for example, the statement \verb'if (in_control_region) S = S("control_region/");'. All histograms from this point will be created and filled in the subdirectory ``control\_region''. It is also possible to pass a store object to a function in which common distributions are filled, enabling the reuse of common plots at different points in the analysis.

A common approach to simplify histogram management - used e.g. in the ATLAS Athena analysis framework - is to use a map keyed by the name of the histogram, making it necessary to do expensive string operations in the event loop. In {\scshape a4}, the store object uses a specially designed hash table instead, which uses the numerical value of the \verb"const char *" pointer to the given strings instead of comparing characters. On first insertion of any pointer, it is checked to see if it points into a memory region designated read-only by the operating system and rejected otherwise. This protects the user of a library against errors if a slower dynamic string is inadvertently used. Common operations usually requiring dynamic strings - concatenation and numbering - can be done instead using the store itself, e.g. \verb'S.T<H1>("hist_", "nr_", 4)' or \verb'S("subdir_",3,"/")' are valid calls. This fast string lookup is also used for the \verb"Cutflow" class to provide a histogram class where the bins are indexed by label: \verb'S.T<Cutflow>("cf").passed("my_cut")' indicates that this event has passed ``my\_cut''.

When using a processor object, the contents of the {\scshape a4} store are written as protobuf messages to an {\scshape a4} file specified on the command line. The results are written into blocks with the same metadata key as specified on the command line with the \verb'--per' switch. This again reduces the bookkeeping effort, since results can be obtained per data taking run even if files contain multiple runs, or runs are spread over multiple files. The command-line tool \verb"a4merge" is provided to merge stores (and the histograms in them). A file containing multiple keys can be split into multiple files with the \verb"--split-per" parameter. In the final files, each set of histograms is  associated with the metadata of all the events used to produce it. Histograms can then be re-weighted depending on metadata using \verb"a4reweight". This tool uses cross-section information as well as metadata to re-weight histogram entries to a desired luminosity.

\section{Handling of systematic variations and multiple channels}\label{sec:syst}

Using the {\scshape a4} store, the processor implements another key feature: handling of \textit{channels} and \textit{systematics}. The function call \verb'bool x = channel("electron");' later causes a re-run of the whole processing function, with the same initial event, but with an additional store prefix of \verb"channel/electron/". Only during this analysis pass, the return value of the function is \verb"true". This is useful to obtain histograms created for example in object selection just for events that pass selection criteria further on, without having to copy and paste these histograms. This function can also be used to study the effect of alternative cuts on the analysis.

The \verb'systematics("scale_up")' function has similar semantics, but does not trigger a re-run by itself. If a re-run is scheduled with a command line option, this function returns true during that rerun, and the \verb"systematic/scale_up/" prefix is added to the store. This enables conditional application of systematic uncertainty factors, and evaluation of these factors without recompilation.

To summarise, the {\scshape a4} histogram store enables fast definition and filling of histograms on one line. Store objects as prefixes make it possible to reuse histogram definition and fill code in functions, and fill them e.g. after each cut. The processor functions \verb'channel' and \verb'systematic' simplify optimisation and the evaluation of systematic uncertainties. The histograms are stored in {\scshape a4} files, which keep metadata information about the events the histograms were filled with. The {\scshape a4} store can also be used in {\scshape ROOT} analysis using an adapter for {\scshape ROOT} histograms, with a limited set of features.

\section{A4 ROOT}\label{sec:a4root}

The {\scshape a4} {\scshape ROOT} module contains programs to convert both {\scshape a4} event data and {\scshape a4} histograms to and from the popular {\scshape ROOT} file format. The command-line tool \verb"a4results2root" converts histogram stores, whereas \verb"a42root" auto-generates a {\scshape ROOT} tree structure to hold event data\footnote{It is planned to merge these two tools in the near future}. Conversion from {\scshape ROOT} trees to {\scshape a4} messages in a flat format would also be automatically possible, but in general, flat {\scshape ROOT} trees usually used in analysis do not contain sufficient information to reconstruct the object structure. Adding such information to a special \verb".proto" file is necessary, see Figure~\ref{root_proto}. For all ATLAS flat {\scshape ROOT} data formats (D3PDs), this file can be generated by the ATLAS Athena analysis framework with D3PDMakerA4 which is included in a4root. The \verb"root2a4" program then converts any {\scshape ROOT} tree using the given structure.

\begin{figure}
\begin{verbatim}
package a4.atlas.ntup.photon;
import "a4/root/ROOTExtension.proto";

message Photon {
    optional float E = 1;
    optional float px = 7; optional float py = 8; optional float pz = 9;
    extensions 100000 to max;
}

message Event {
    optional uint32 run_number = 1;
    optional uint32 event_number = 2 [(root_branch)="EventNumber"];
    repeated Photon photons = 100 [(root_prefix)="ph_"];
    extensions 100000 to max;
}
\end{verbatim}

\caption{An illustrative \texttt{.proto} file which can be used to convert an existing flat {\scshape ROOT} tree to an {\scshape a4} event file. The {\scshape a4} extensions to the protobuf field descriptors, \texttt{root\symbol{95}branch} and \texttt{root\symbol{95}prefix} are used to indicate the names of the leaves on the {\scshape ROOT} side. For example, this input {\scshape ROOT} file has a \texttt{std::vector<float>} branch called \texttt{ph\symbol{95}px[i]}, corresponding to \texttt{event.photons(i).px()} in the resulting protobuf class structure. The protobuf \texttt{extensions} keyword reserves numbers to be used in user extensions to the class, allowing the format to be extended at runtime.}
\label{root_proto}
\end{figure}

\section{Benchmarks}

Since one of the primary design goals of {\scshape a4} is high processing speed, we have performed a set of benchmarks using synthetic events. For each event, all fields are filled with a random number generated using the {\scshape glibc} \verb"random()" function. The events are ``processed'' by calculating the sum of all their fields. As a comparison, we perform the same procedure using {\scshape ROOT} version~5.32.

For all these benchmarks it must be considered that {\scshape ROOT} trees are also designed to quickly retrieve all instances of one or a few fields for each event, an advantage that is effectively disabled by the requirement to load all fields. In our case, we assumed that the relevant information was already selected and processed into a smaller file for local analysis.

The benchmark setup is similar to the flat ntuple format used in ATLAS analysis. The number of single, non-repeated fields of a certain type in the synthetic message is denoted by $n_{flat,float} $. The number of repeated fields - represented by \verb"std::vector<T>" objects in {\scshape ROOT} - is $n_{rep,float}$. The number of entries in a repeated field is $n_{nfill,float}$. All benchmarks are performed on an unloaded system on a RAM disk. The reported runtimes are the minimum value measured over three runs, and the error bars shown are the standard deviation. For the first part of the benchmarks, no compression was used. The {\scshape ROOT} basket size was left at the default of 32k, but was varied in a separate run from zero to 1MB, resulting in speed differences of $< 3\%$. No other attempts to tune the performance of {\scshape ROOT} have been made, representing common non-expert usage.

\begin{figure}
\begin{minipage}{\textwidth}
\begin{center}
\includegraphics[width=0.39\textwidth]{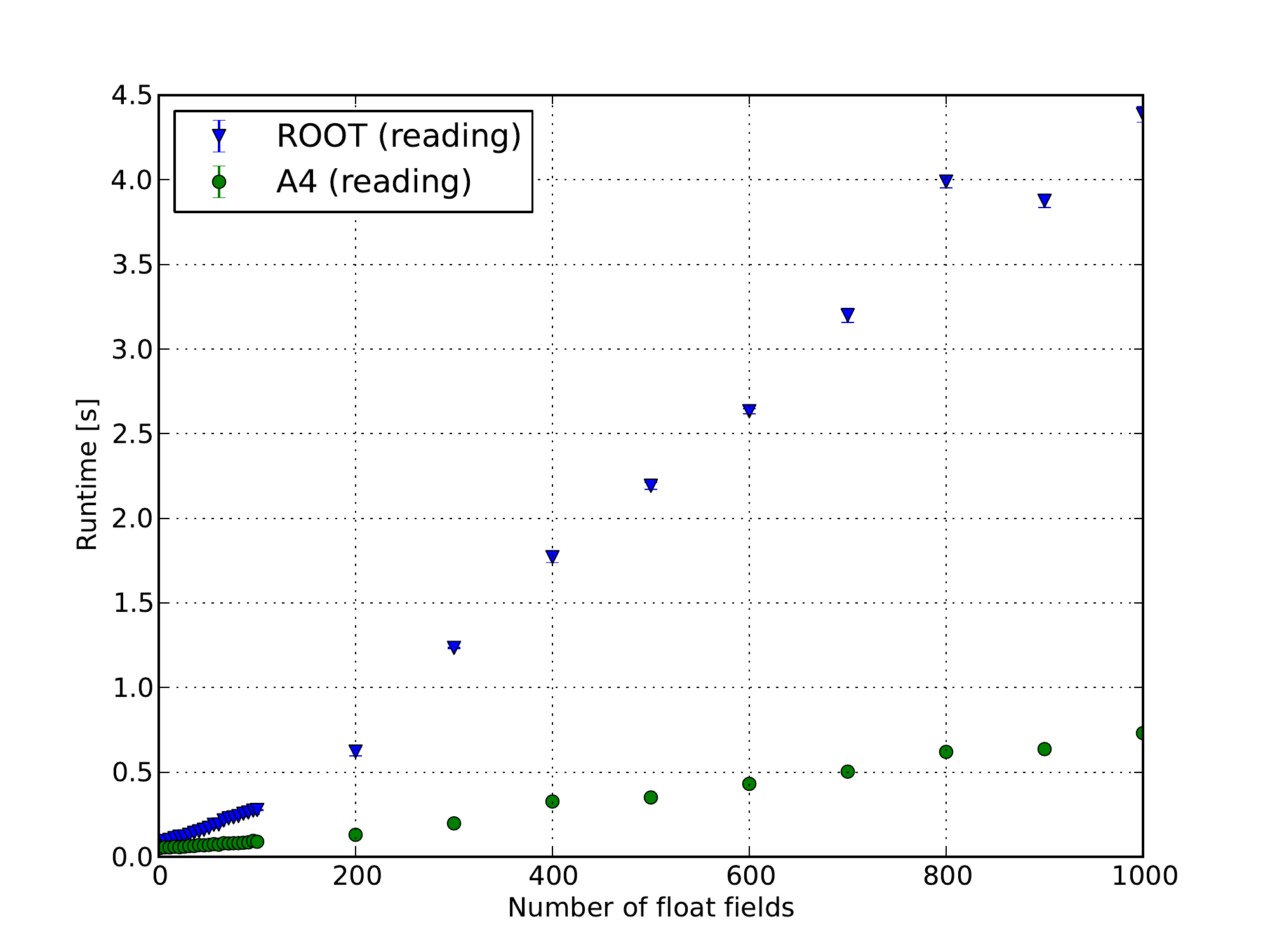}
\includegraphics[width=0.39\textwidth]{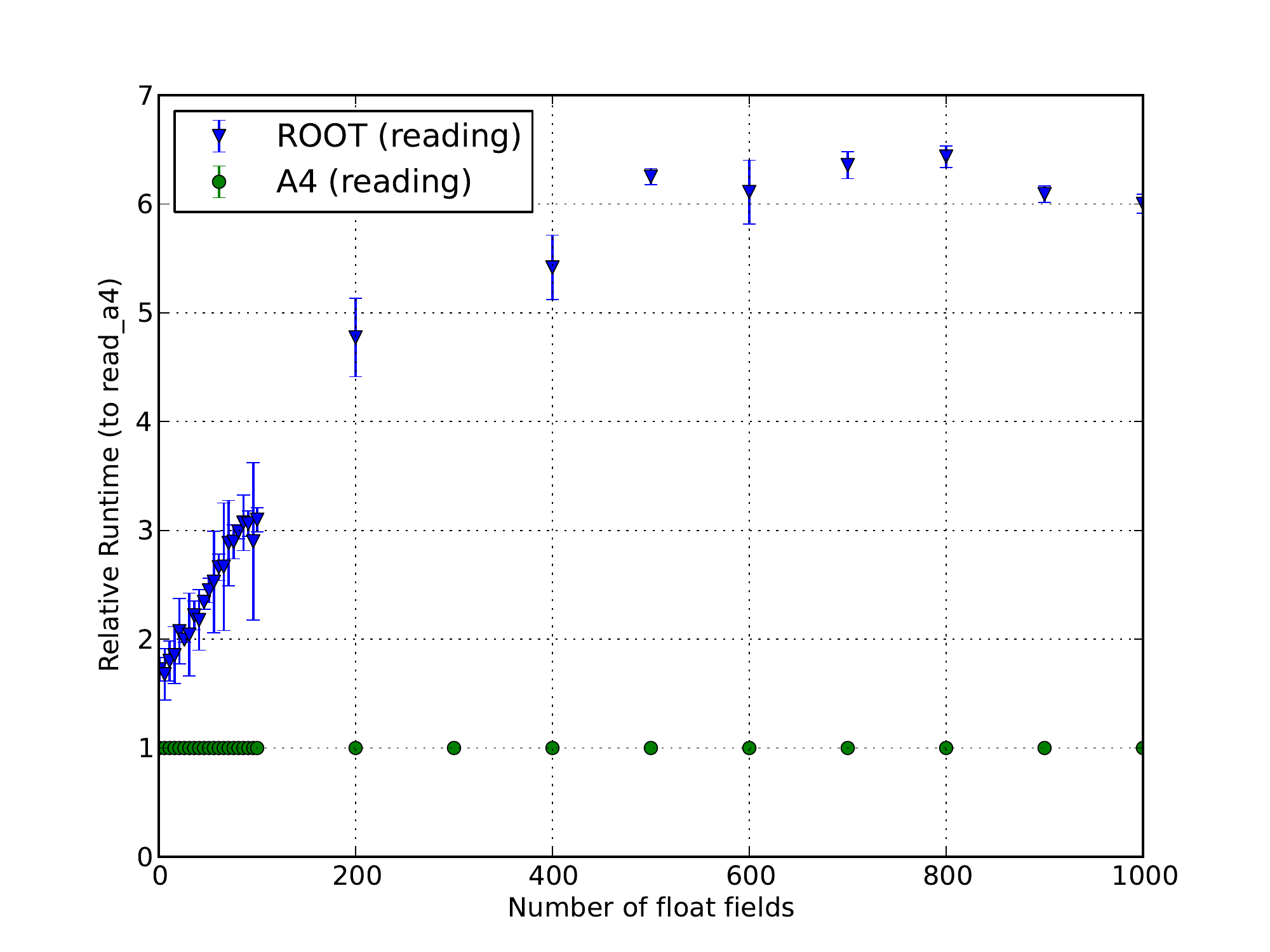}
\end{center}
\caption{\label{fig:float_flat}Processing time in seconds for $100000$ events versus $n_{flat,float}$. On the right-hand side, the time is normalized to the {\scshape a4} runtime.}
\end{minipage}\hspace{2pc}
\begin{minipage}{\textwidth}
\begin{center}
\includegraphics[width=0.39\textwidth]{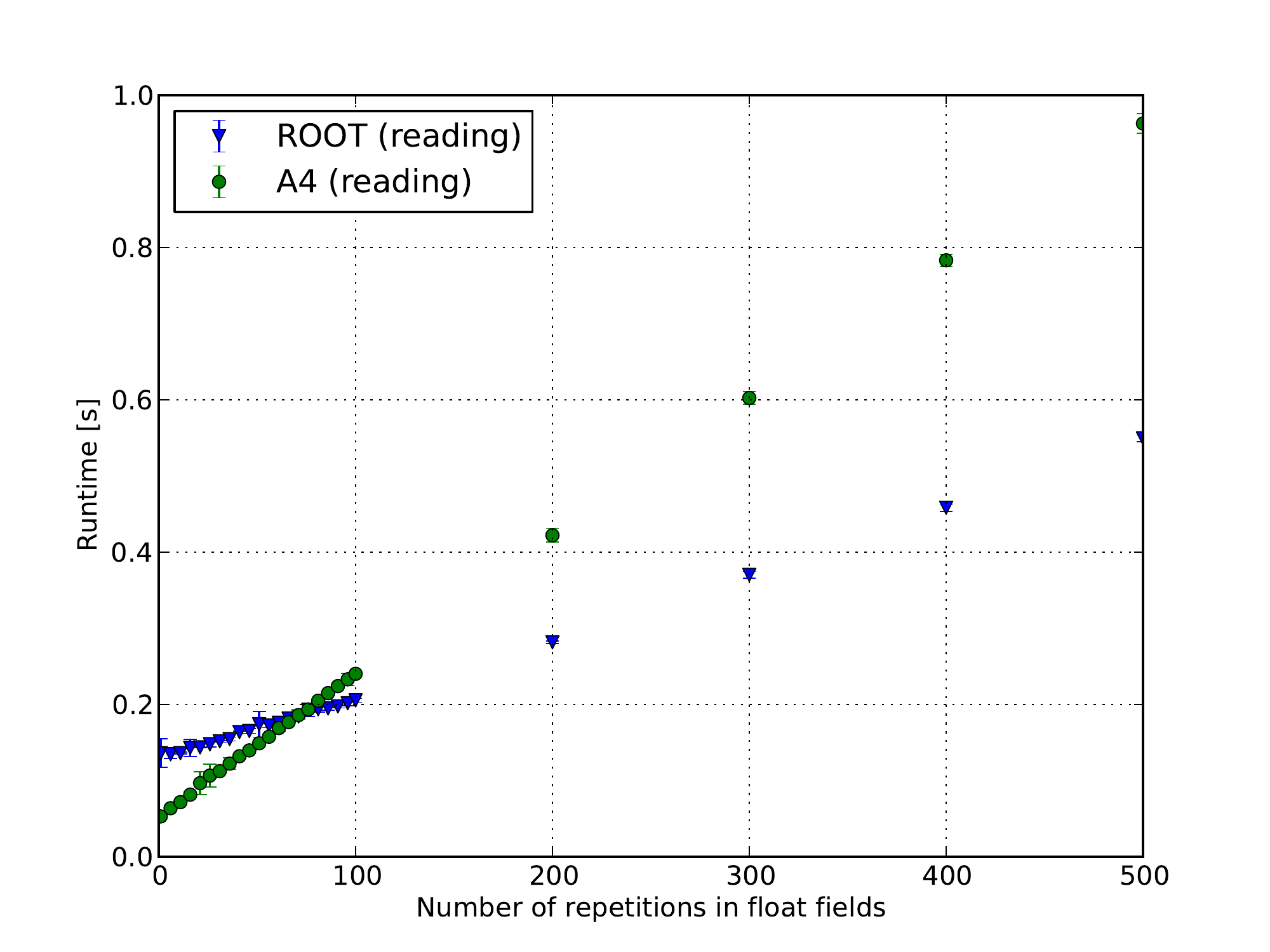}
\includegraphics[width=0.39\textwidth]{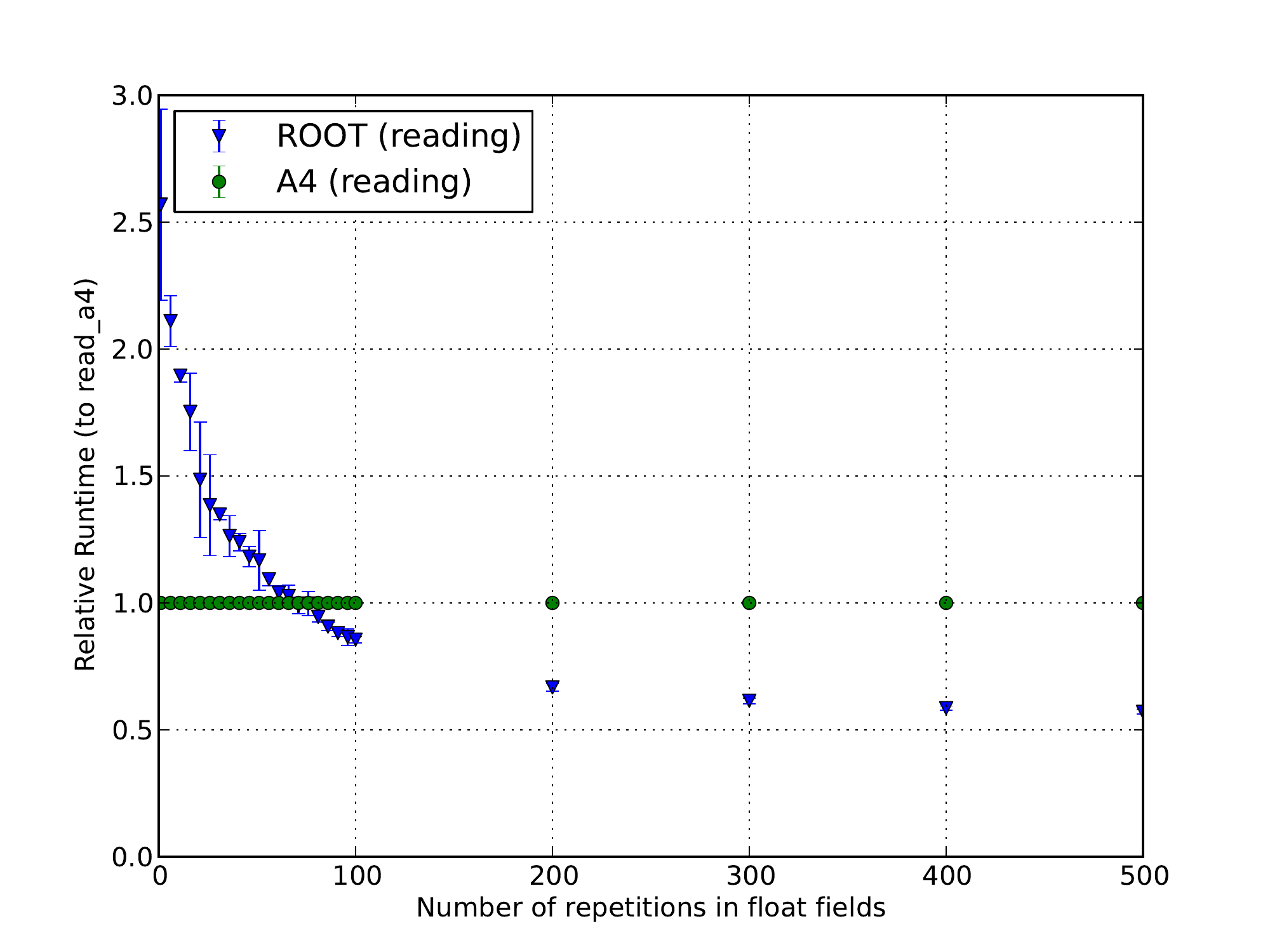}
\end{center}
\caption{\label{fig:float_varrep}Processing time in seconds for $100000$ events with $n_{rep,float} = 4$ versus $n_{nfill,float}$. On the right-hand side, the time is normalized to the {\scshape a4} runtime.}
\end{minipage}\hspace{2pc}
\end{figure}

There are two features of interest in the benchmark data: A speedup with respect to {\scshape ROOT} of a factor 3 up to a factor of 6 for events with $>100$ flat or array-like fields of any type (Figure~\ref{fig:float_flat}) and a slowdown with respect to {\scshape ROOT} for large array-like fields to 0.9 (double) 0.6 (float), and 0.5 (integers) (Figure~\ref{fig:float_varrep}). The observed behaviour is not yet fully understood, but indicates that gains in speed are possible, in situations where high bandwidth is available.

In the Appendix, the corresponding plots for all types are shown. One unexpected result is the relatively inefficient handling of integers, where further investigation into protobuf performance is necessary. In addition, plots with {\scshape zlib} level 1 compression enabled are shown. Since the same compression is used in {\scshape ROOT} and {\scshape a4}, the difference in speed decreases. The maximum speedup of {\scshape a4} with respect to {\scshape ROOT} is now $\approx 1.8$, except for boolean variables where it remains at $\approx 4$.

To check the behavior under more realistic conditions, we obtained a typical ATLAS ntuple file based on {\scshape ROOT} trees, converted it to an {\scshape a4} file\footnote{The {\scshape a4} file in this case did not have the structure described in the benchmark above, but had more complex structured definitions for physics objects.}, and wrote a minimal event loop for both {\scshape ROOT} and {\scshape a4}. The speedup of {\scshape a4} in this case was 2.6. After this, the branches of the {\scshape ROOT} tree were manually disabled, until the same processing speed was reached. At this point, 40\% of the branches were enabled, indicating approximately linear scaling of runtime with the number of active branches in {\scshape ROOT}.

We can conclude that in situations where a majority of data in an event are used in an analysis, {\scshape a4} can provide significant speedups. Even in other cases, the simple slimming and thinning can quickly lead to a situation where again the majority of data in an event is required. Large arrays are not yet handled efficiently, and need further attention.

\section{Summary and outlook}

In this paper, we presented the first overview of the {\scshape a4} library, a toolkit for data analysis in High Energy Physics. A fast I/O format is described, and we demonstrate that it is able to perform comparably to {\scshape ROOT} trees for flat event analysis. The bookkeeping requirements at any step in data analysis are minimised by automated treatment of metadata for events and histograms. Creation of histograms and common procedures such as evaluation of systematic uncertainties are simplified by the {\scshape a4} store in combination with the {\scshape a4} processor class. Interoperability with {\scshape ROOT} is achieved by a set of conversion tools.

Many features of the architecture\footnote{e.g. chaining processors, threading single files} have not yet been fully exploited. Consolidation and documentation of the existing codebase and extending the test suite is currently a focus of development. Finally, no attempt has yet been made to modify the protobuf library message parsing code for our specific problem, an approach that might improve performance or usability in some cases even further, and could then be submitted for inclusion in the protobuf library.

We believe that {\scshape a4} has already proven itself to be a useful tool in analysing High Energy Physics data despite its experimental status. Using collaborative development tools as \verb"git" and platforms such as \verb"github" enables anyone to modify {\scshape a4}, publish their changes, and submit requests to review and include these changes back. We believe that this model of development suits typical High Energy Physics organisational structures, and warmly invite further use and collaboration.

\clearpage

\ack

The authors would like to thank Samvel Khalatyan for the idea of using the protobuf library for physics analysis and also for other concepts from his initial implementation. Thanks also goes to {\scshape Google} Inc. for publishing the protobuf library under an open source license in the first place. This work was supported by the DFG and by the U.K. Science and Technology Facilities Council.

\newpage
\appendix

\section{Additional benchmark results, no compression}

\begin{figure}[ht]\begin{center}
\includegraphics[width=0.40\textwidth]{plot_float_flat_nc.pdf}
\includegraphics[width=0.40\textwidth]{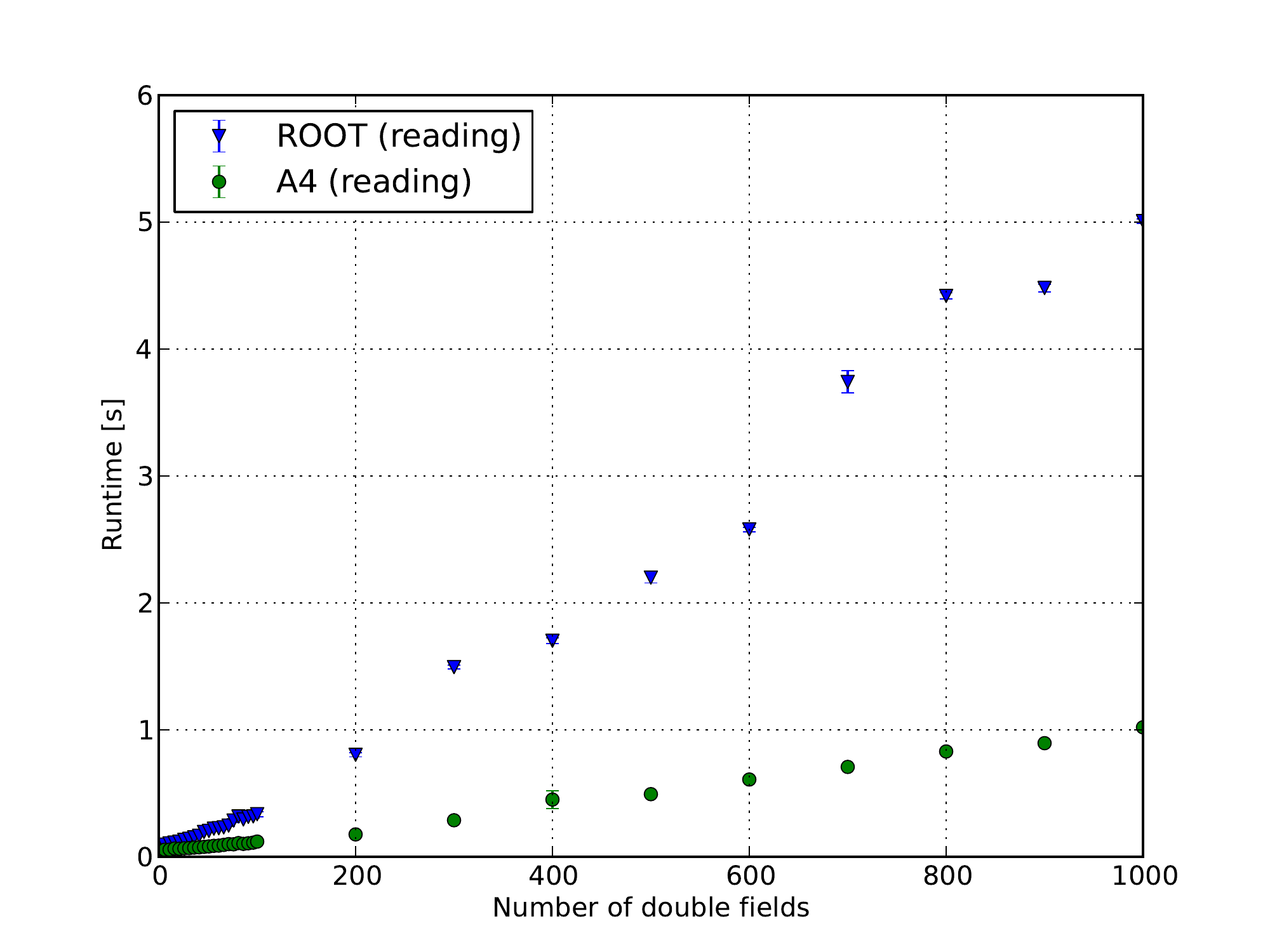}
\includegraphics[width=0.40\textwidth]{plot_wrt_read_a4_float_flat_nc.pdf}
\includegraphics[width=0.40\textwidth]{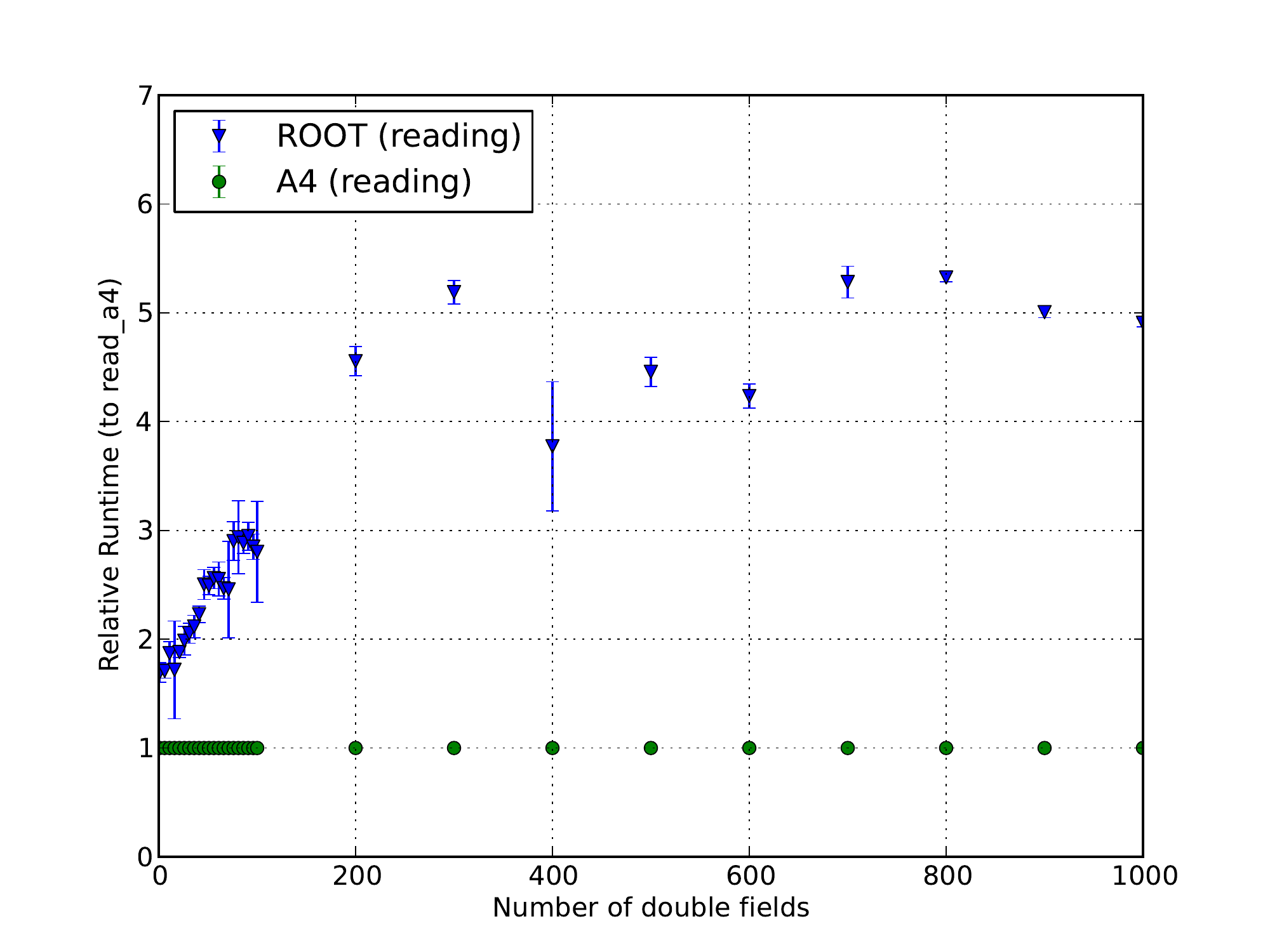}
\begin{tabular*}{\textwidth}{c}\hline\end{tabular*}
\includegraphics[width=0.40\textwidth]{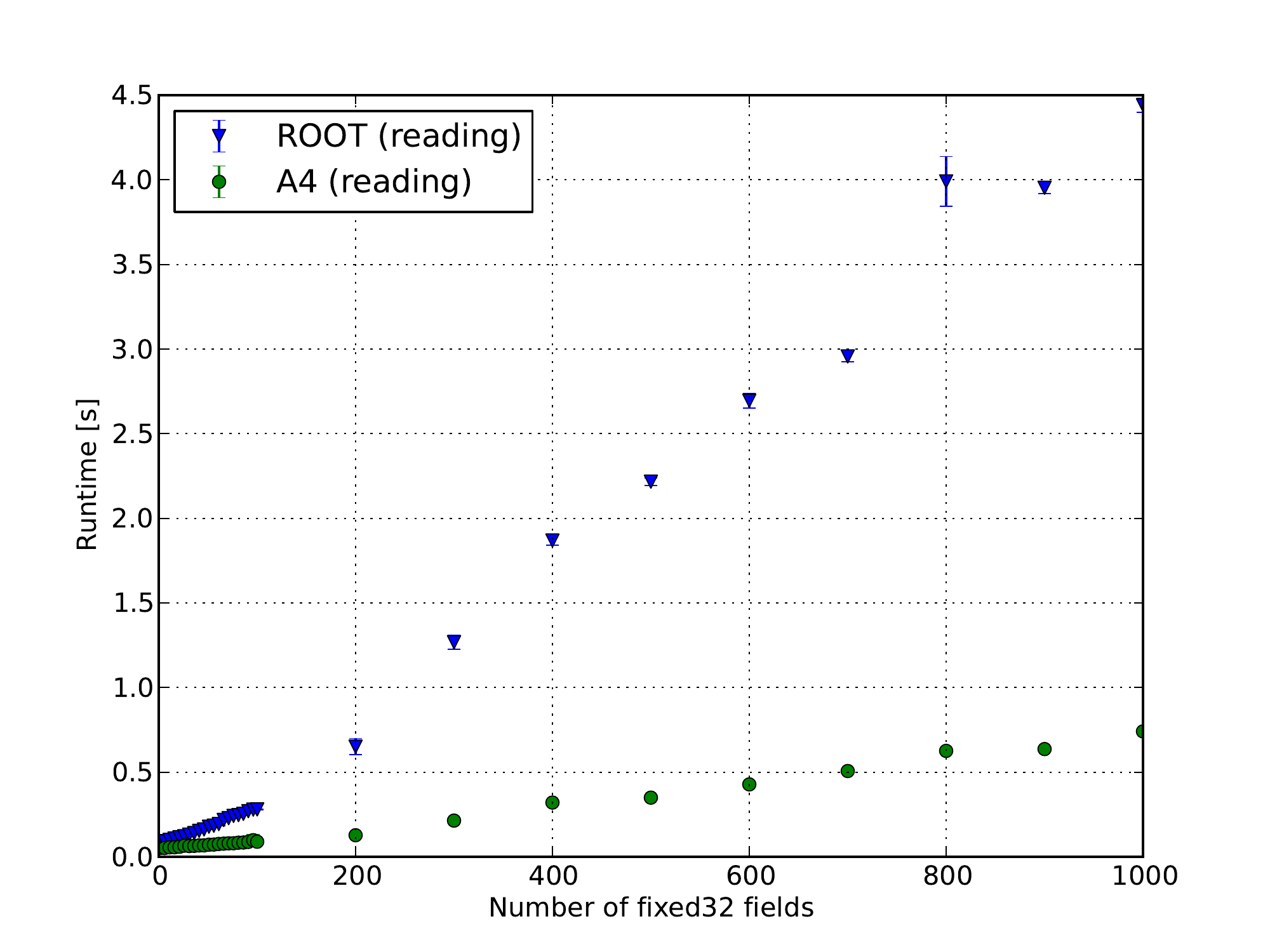}
\includegraphics[width=0.40\textwidth]{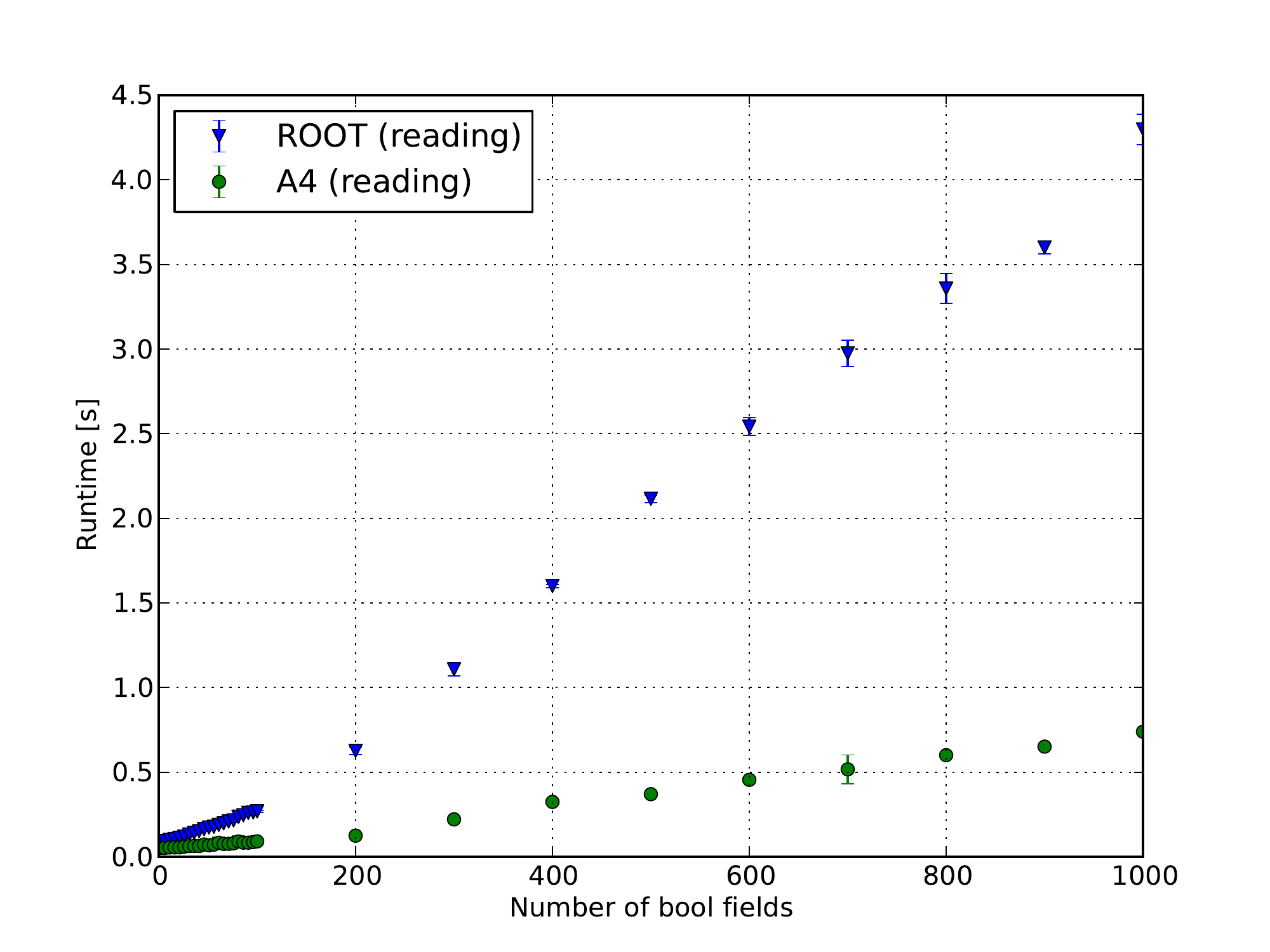}\\
\includegraphics[width=0.40\textwidth]{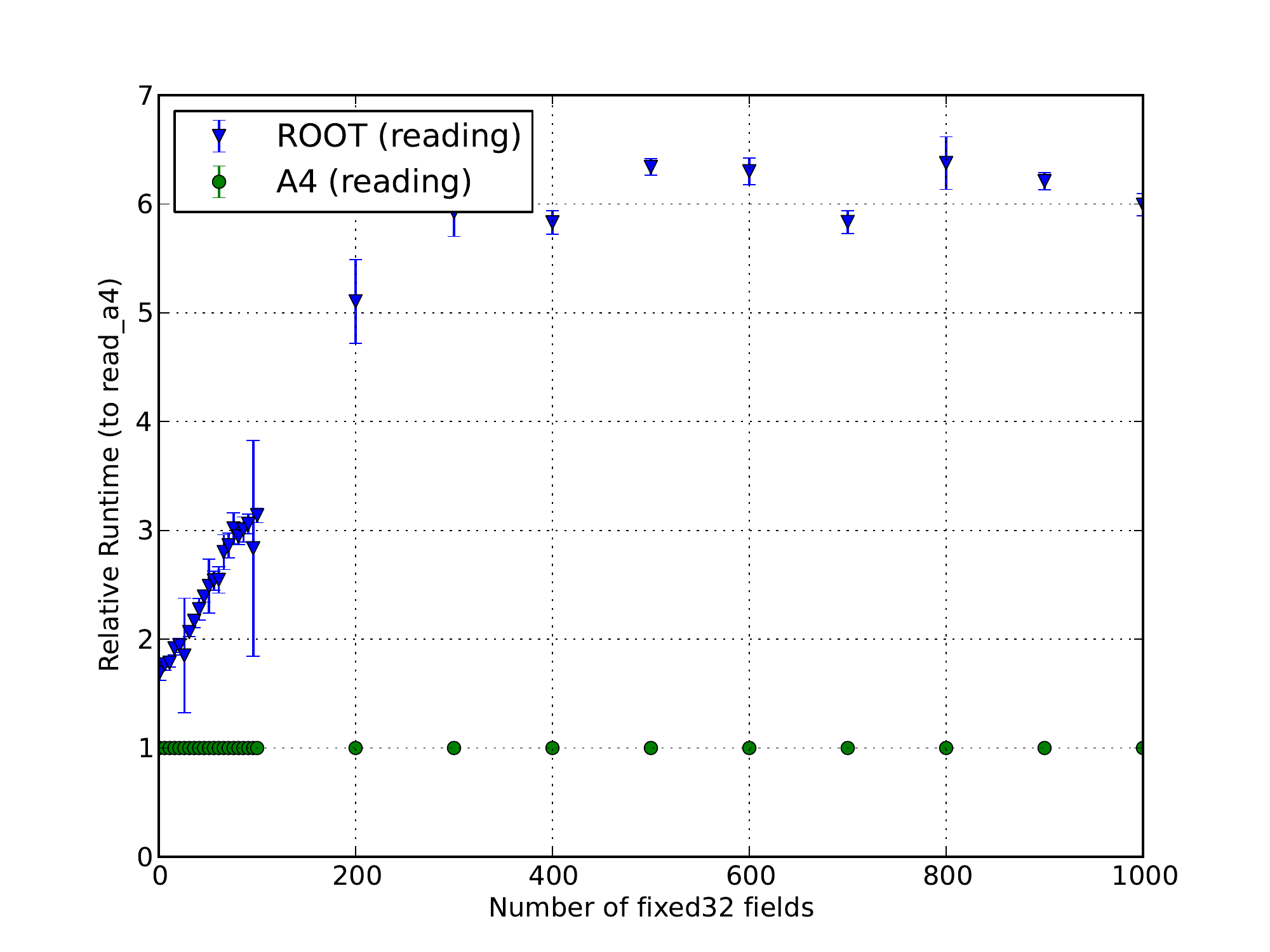}
\includegraphics[width=0.40\textwidth]{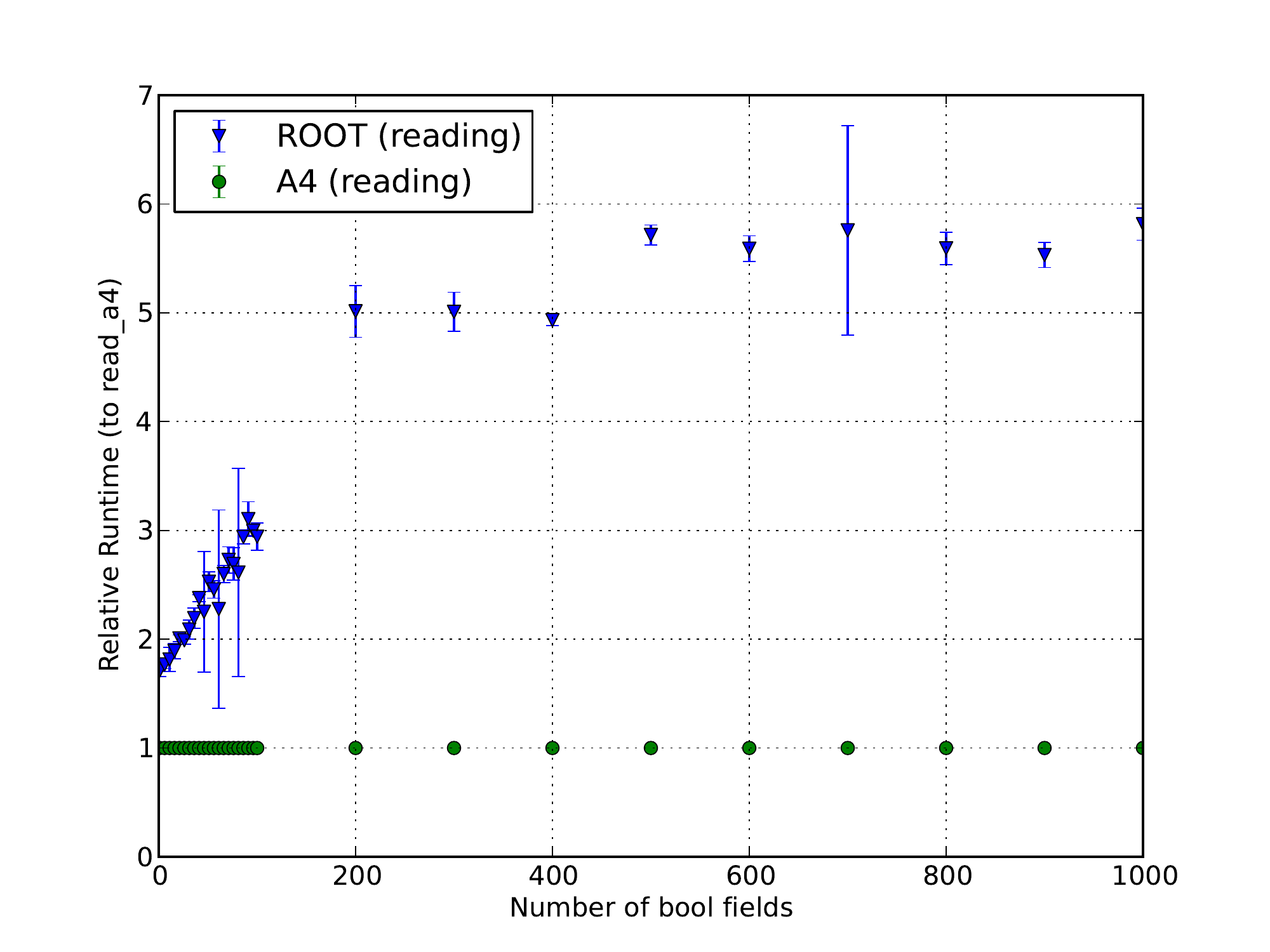}
\caption{Processing time in seconds for $100000$ events versus $n_{flat}$, for floats, doubles, integers and booleans from top left to bottom right. The top row shows absolute runtime, the lower row runtime relative to {\scshape a4}.}
\end{center}\end{figure}
\begin{figure}[ht]\begin{center}
\includegraphics[width=0.40\textwidth]{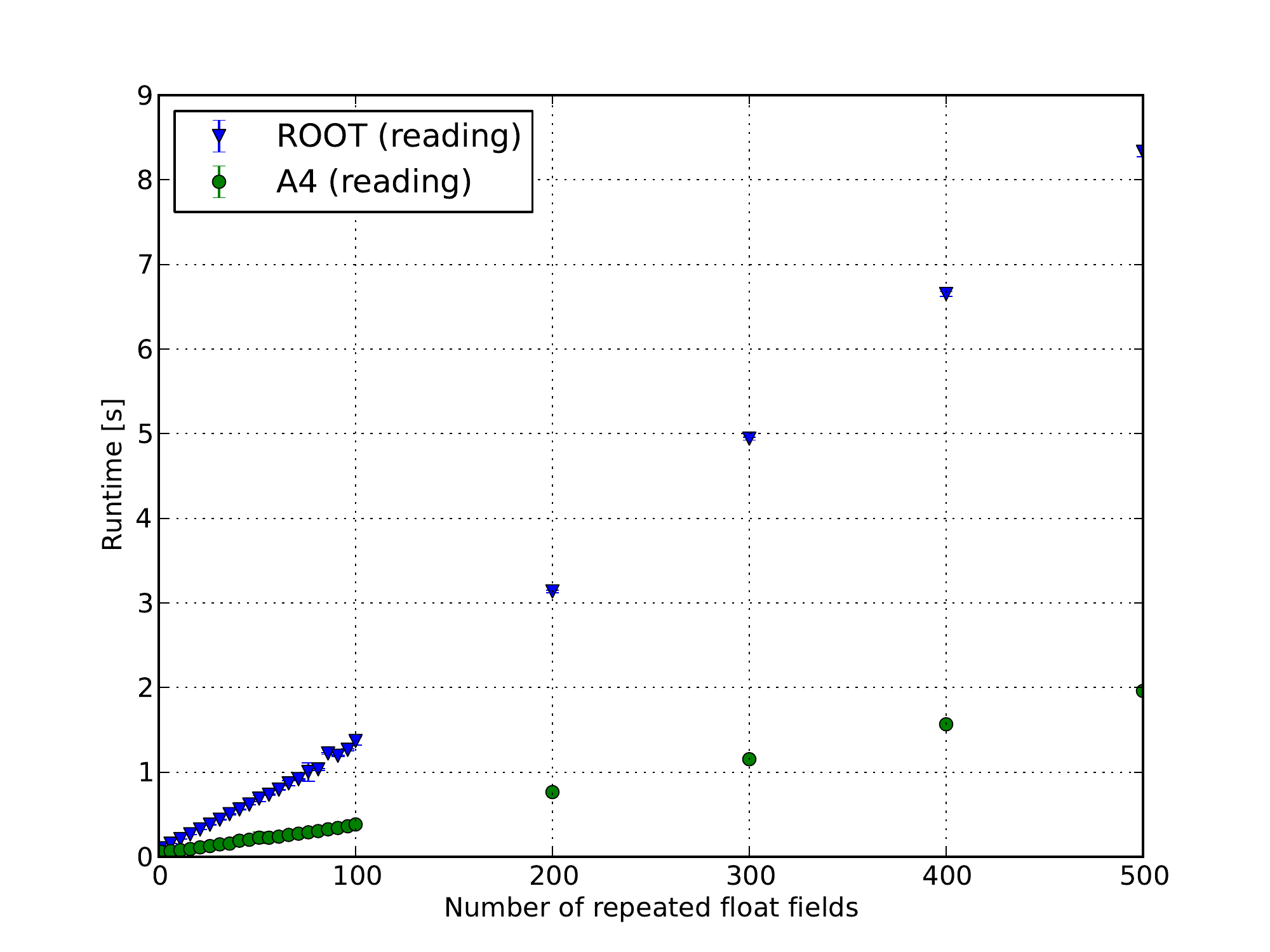}
\includegraphics[width=0.40\textwidth]{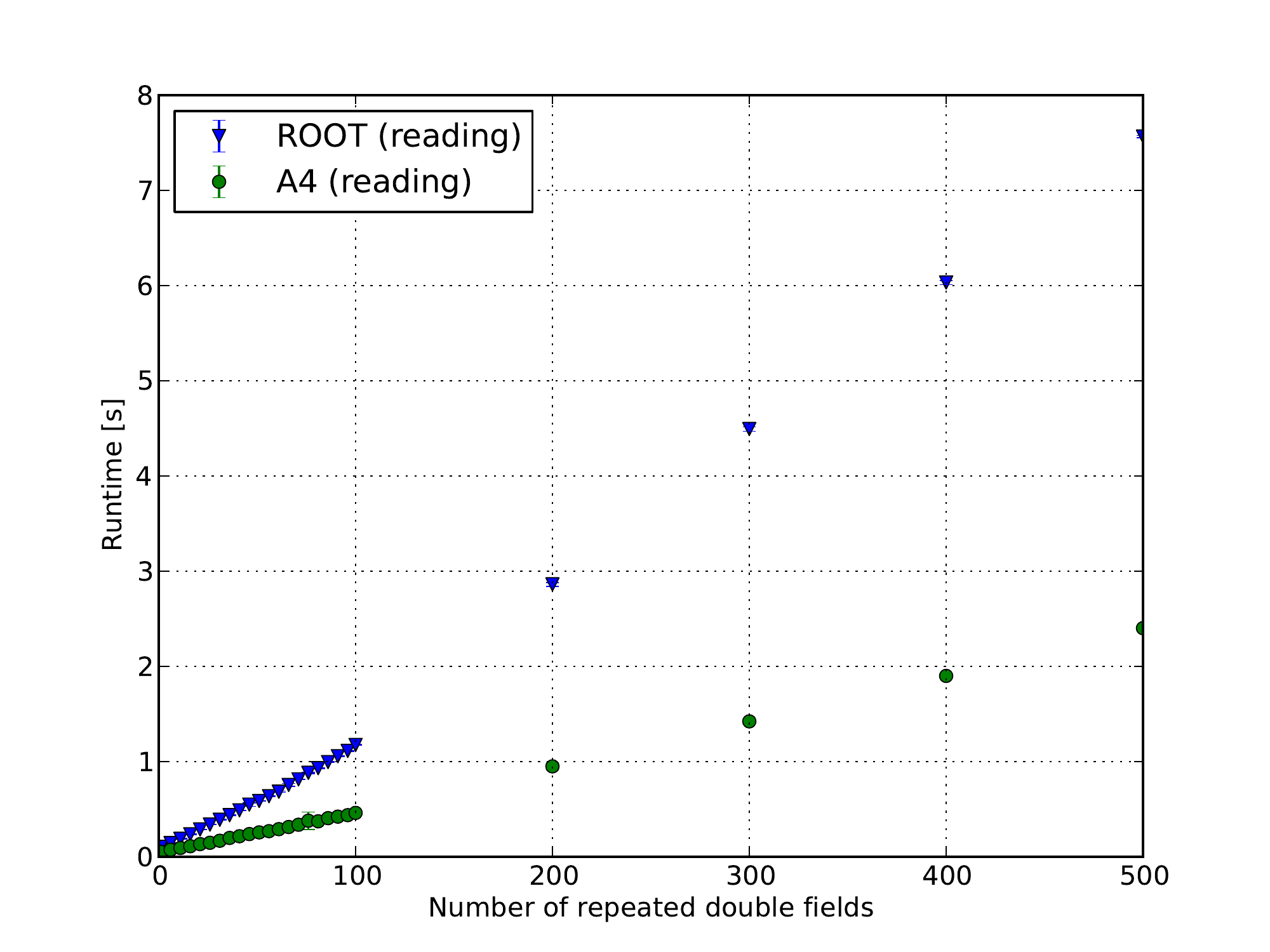}
\includegraphics[width=0.40\textwidth]{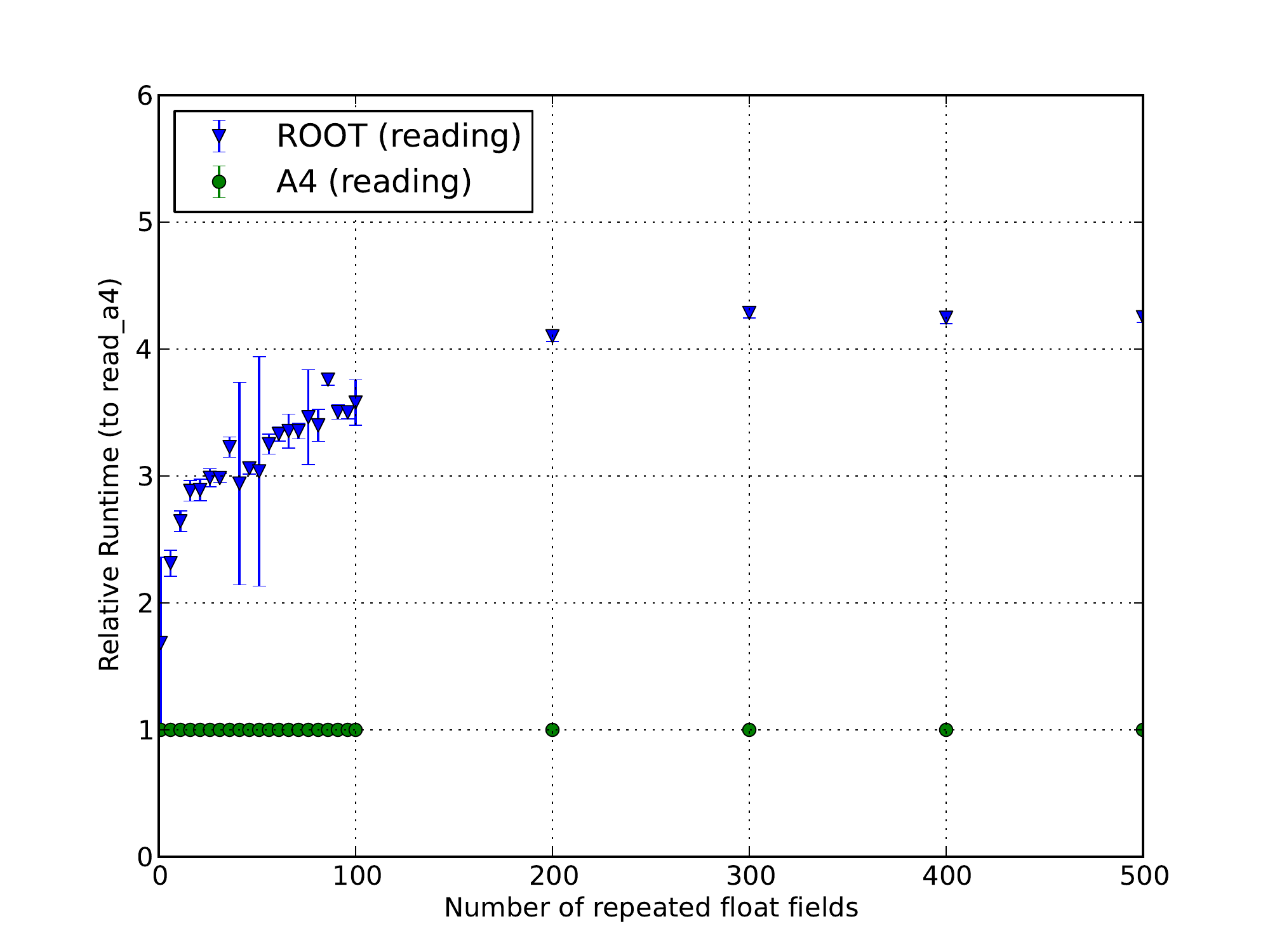}
\includegraphics[width=0.40\textwidth]{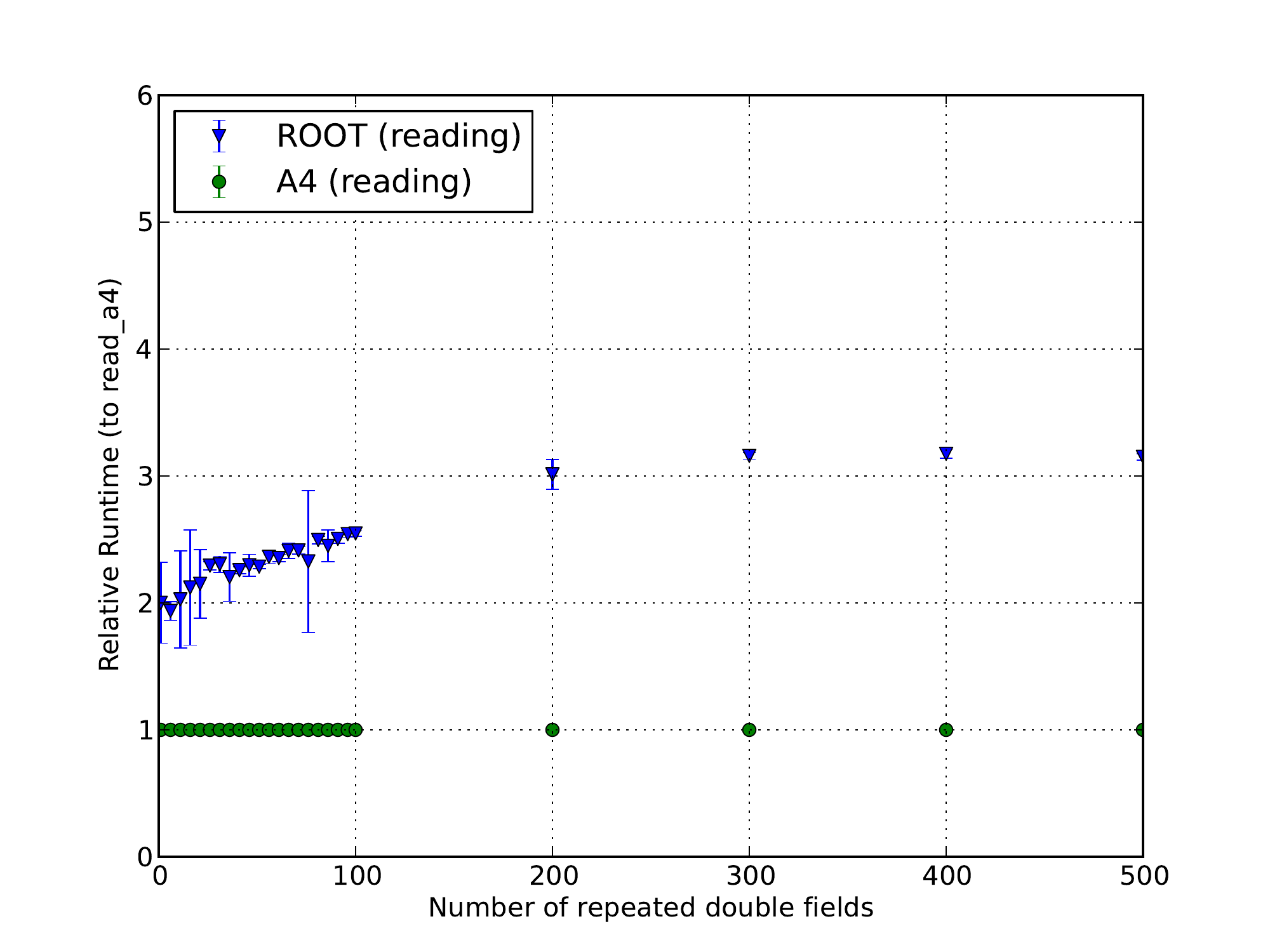}
\begin{tabular*}{\textwidth}{c}\hline\end{tabular*}
\includegraphics[width=0.40\textwidth]{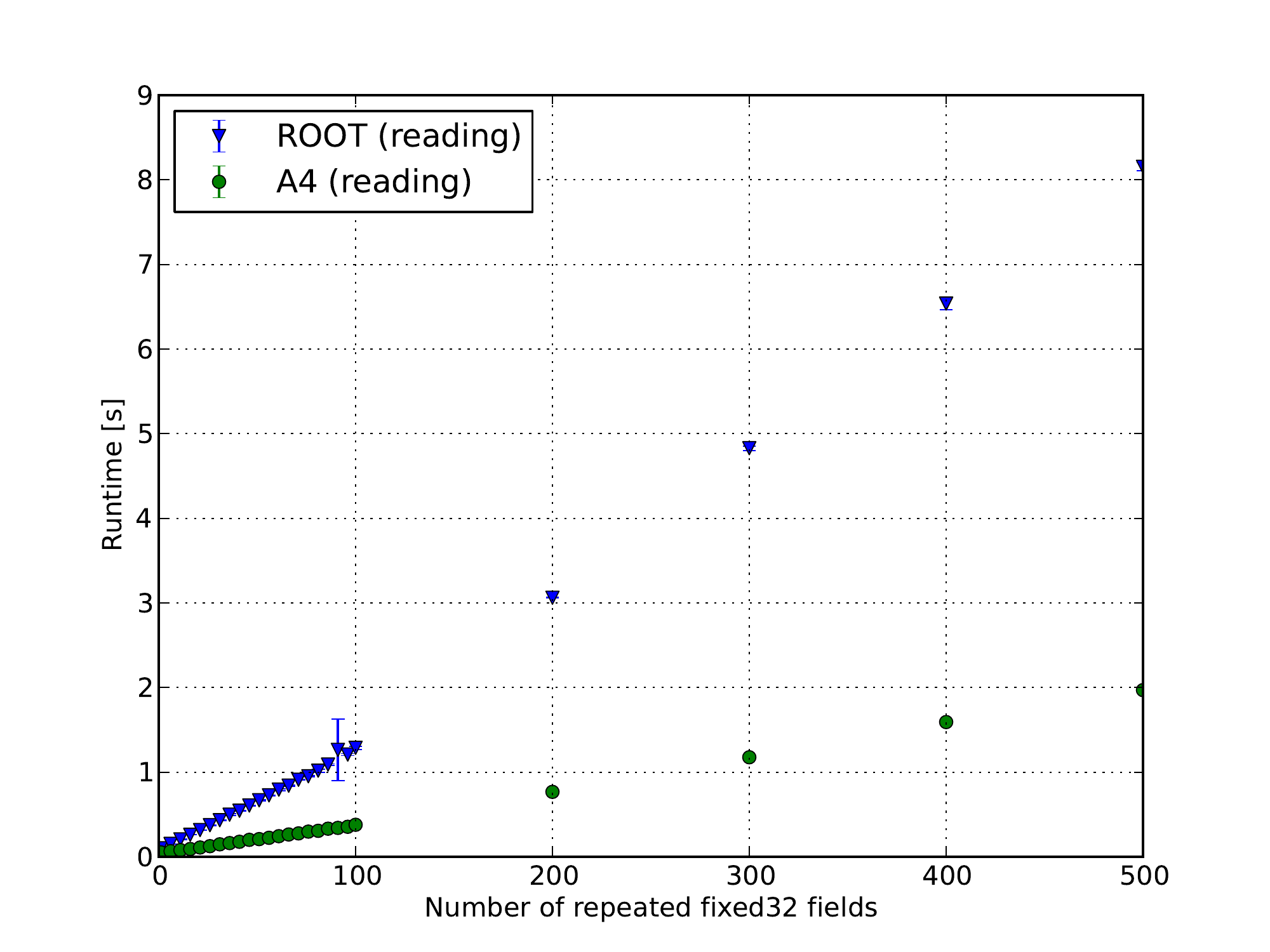}
\includegraphics[width=0.40\textwidth]{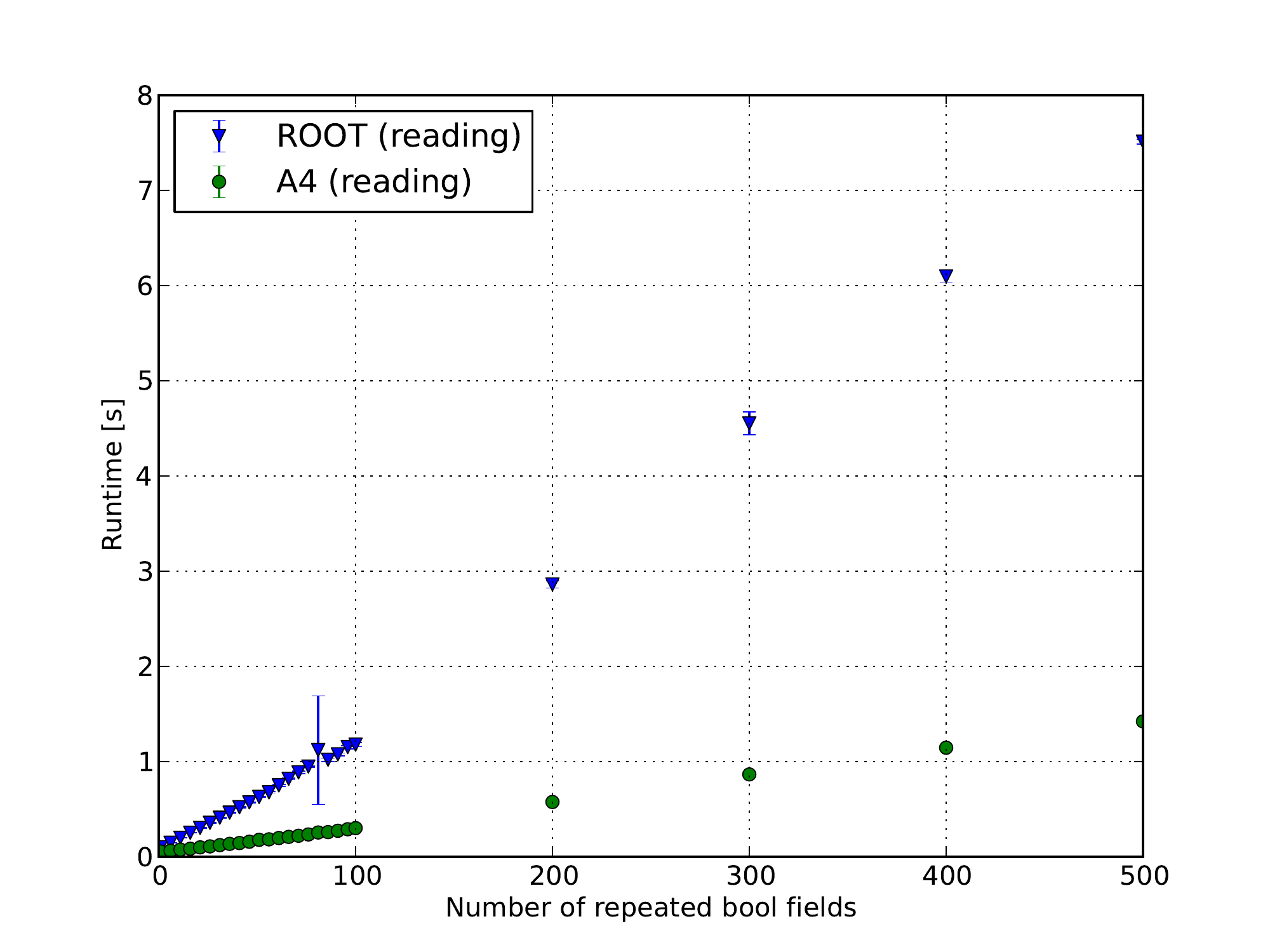}\\
\includegraphics[width=0.40\textwidth]{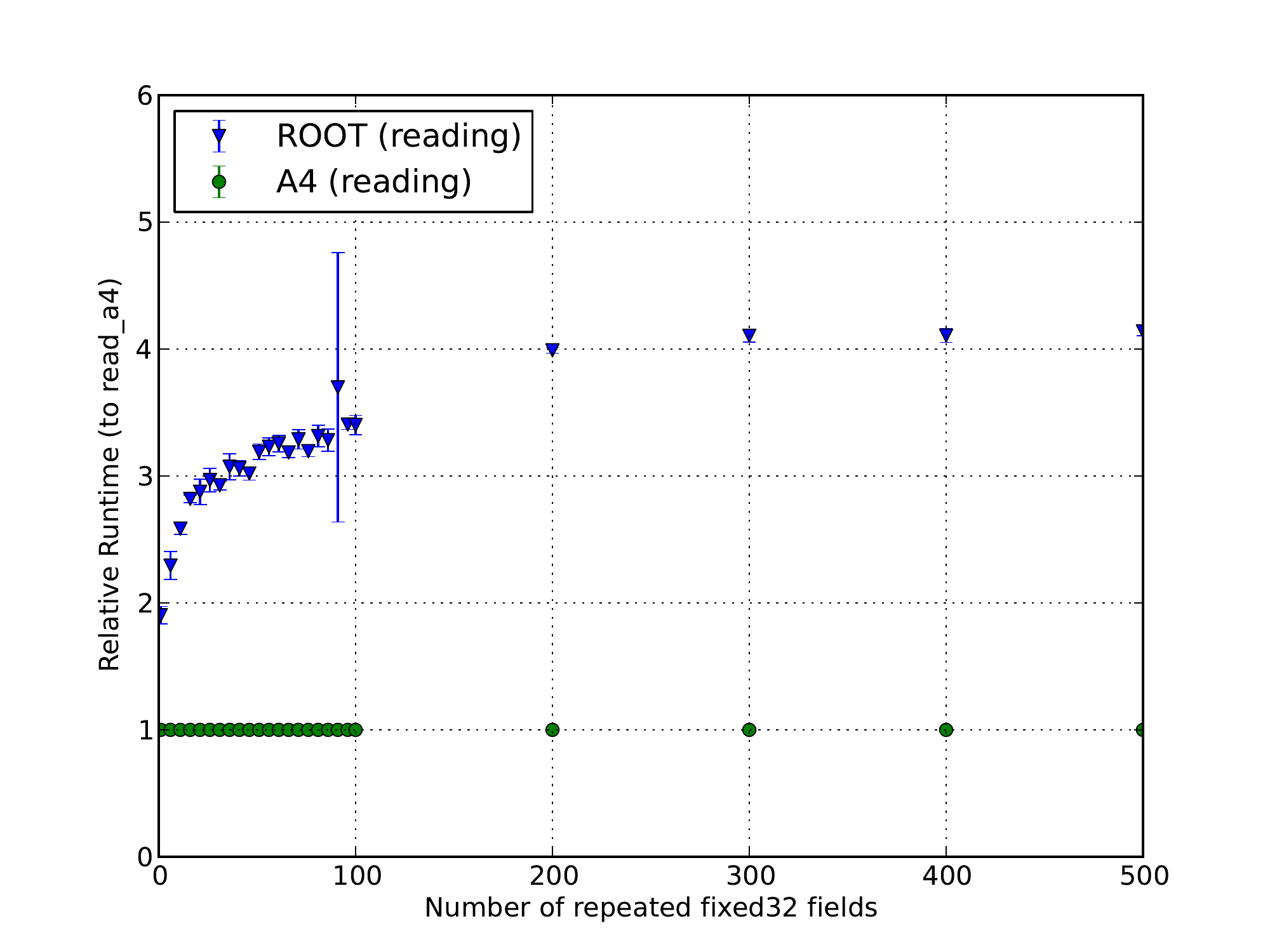}
\includegraphics[width=0.40\textwidth]{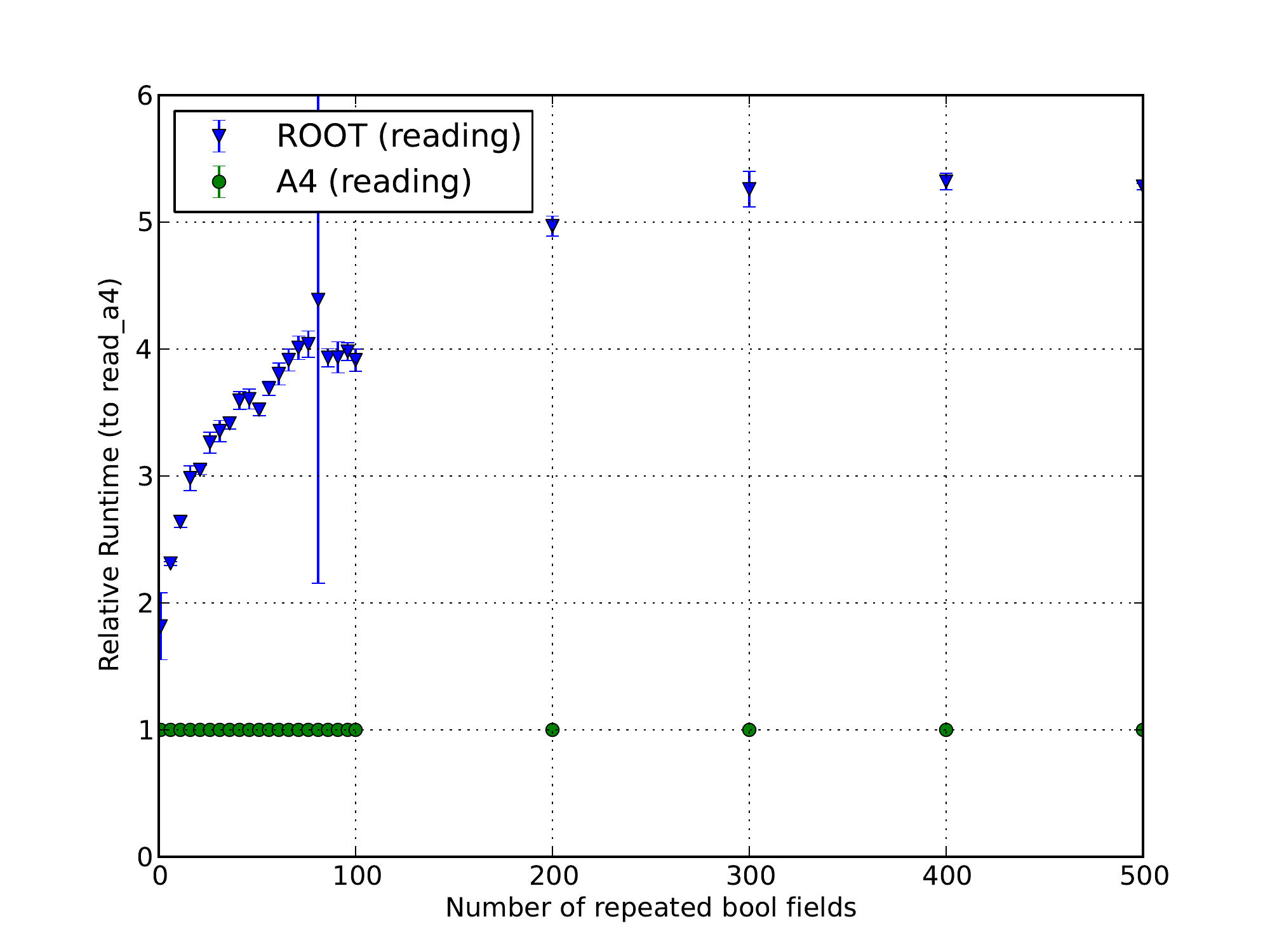}
\caption{Processing time in seconds for $100000$ events versus $n_{rep}$, for floats, doubles, integers and booleans from top left to bottom right. The top row shows absolute runtime, the lower row runtime relative to {\scshape a4}.}
\end{center}\end{figure}
\begin{figure}[ht]\begin{center}
\includegraphics[width=0.40\textwidth]{plot_float_var_rep_nc.pdf}
\includegraphics[width=0.40\textwidth]{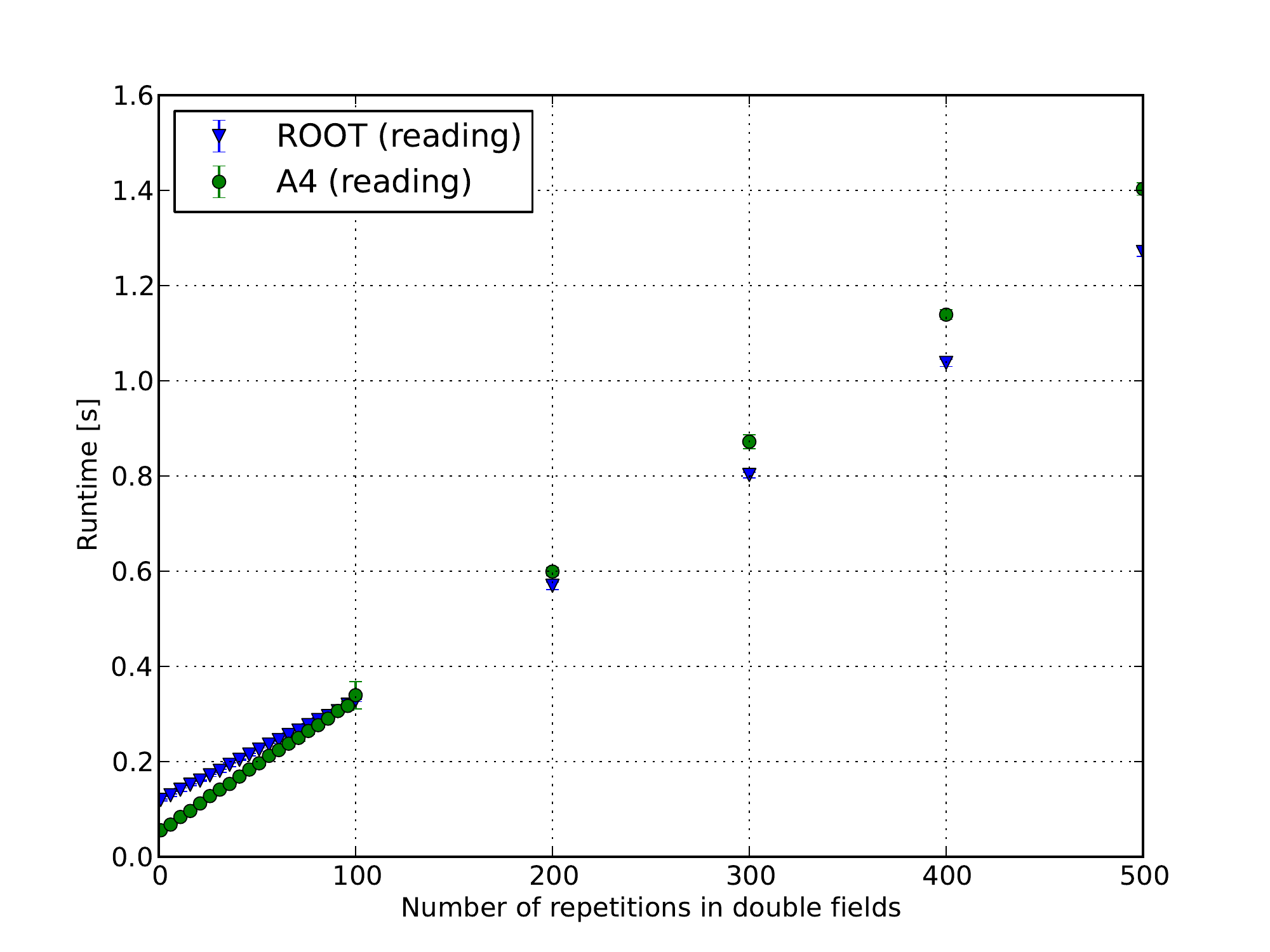}
\includegraphics[width=0.40\textwidth]{plot_wrt_read_a4_float_var_rep_nc.pdf}
\includegraphics[width=0.40\textwidth]{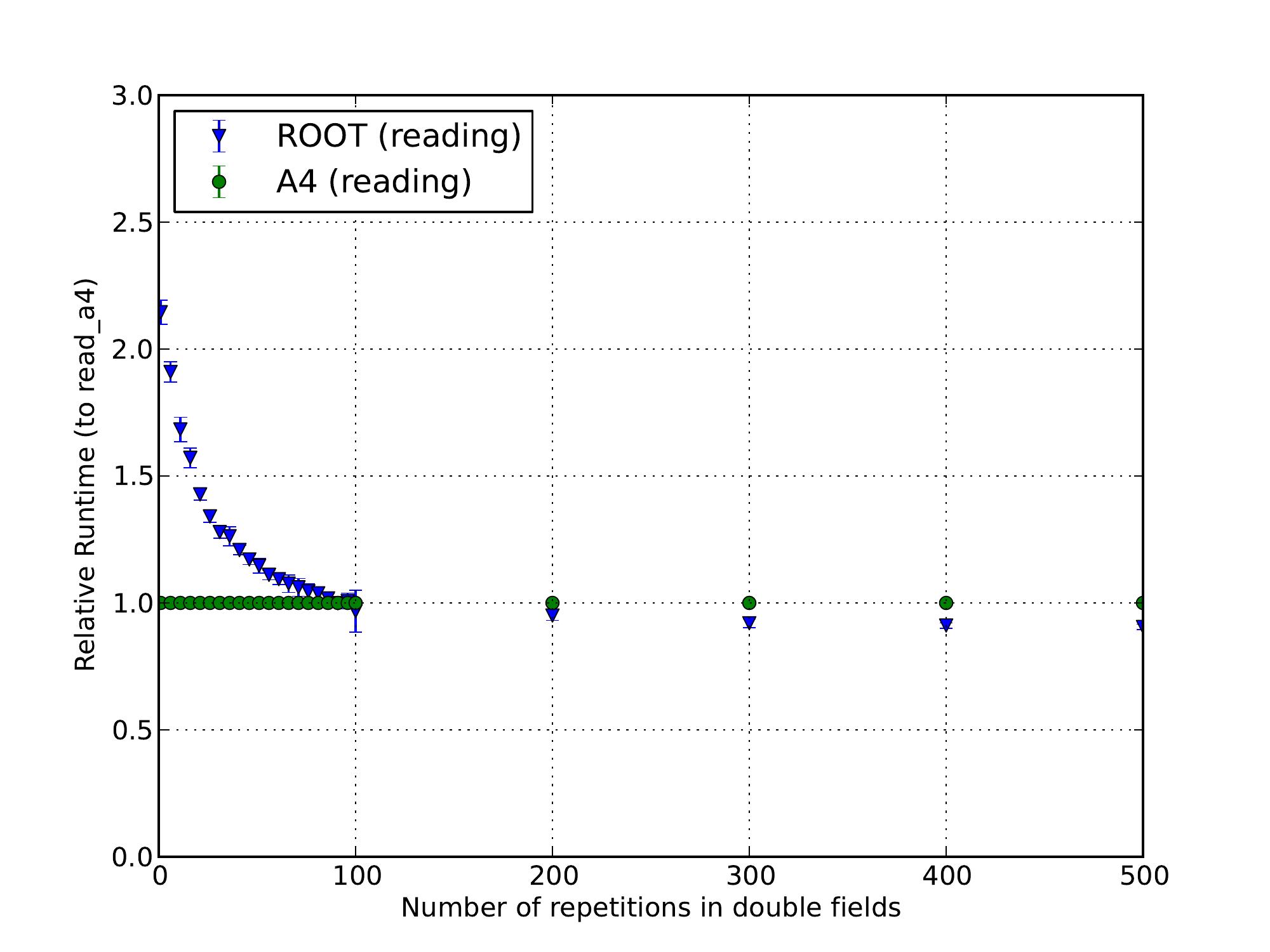}
\begin{tabular*}{\textwidth}{c}\hline\end{tabular*}
\includegraphics[width=0.40\textwidth]{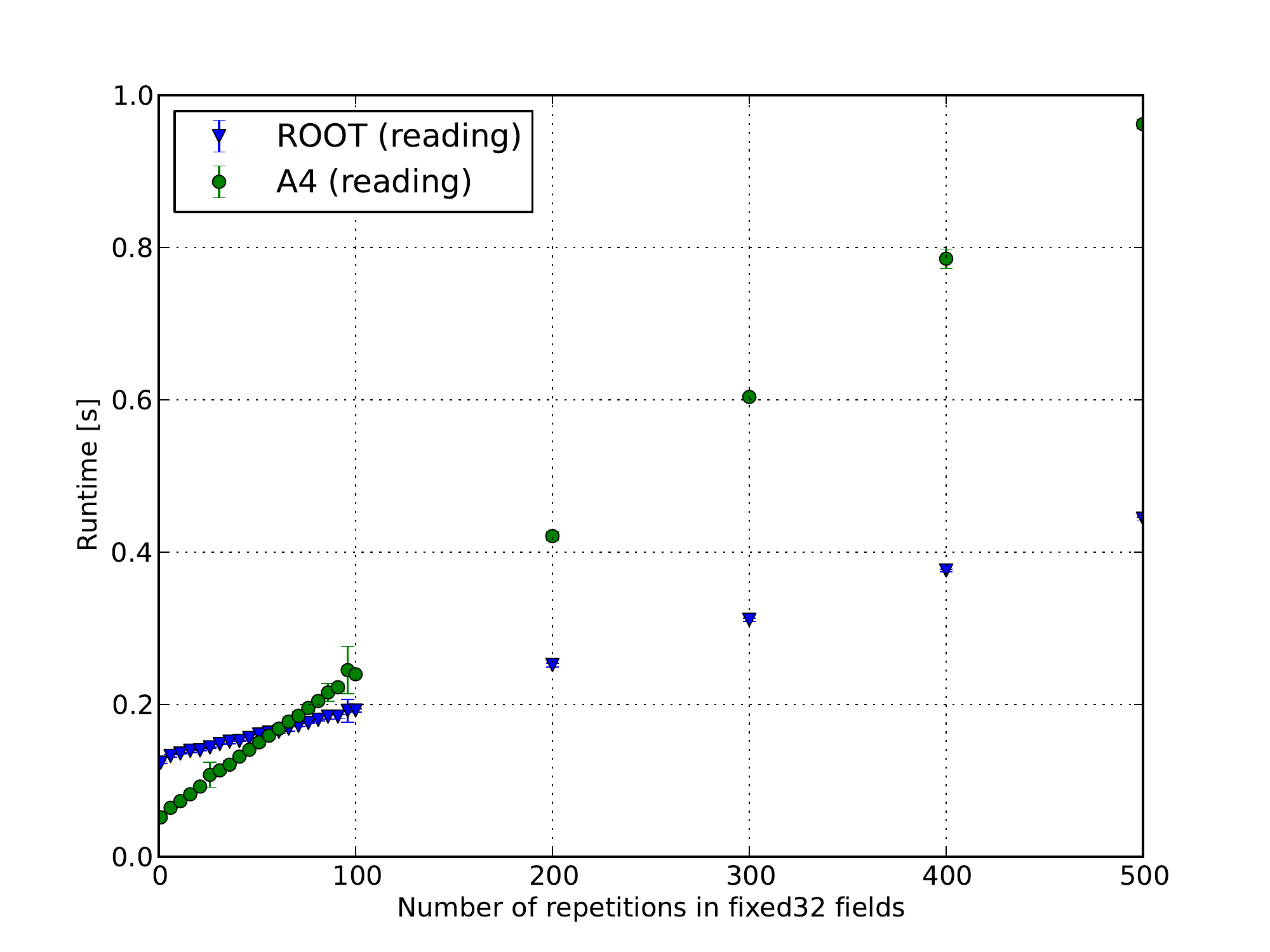}
\includegraphics[width=0.40\textwidth]{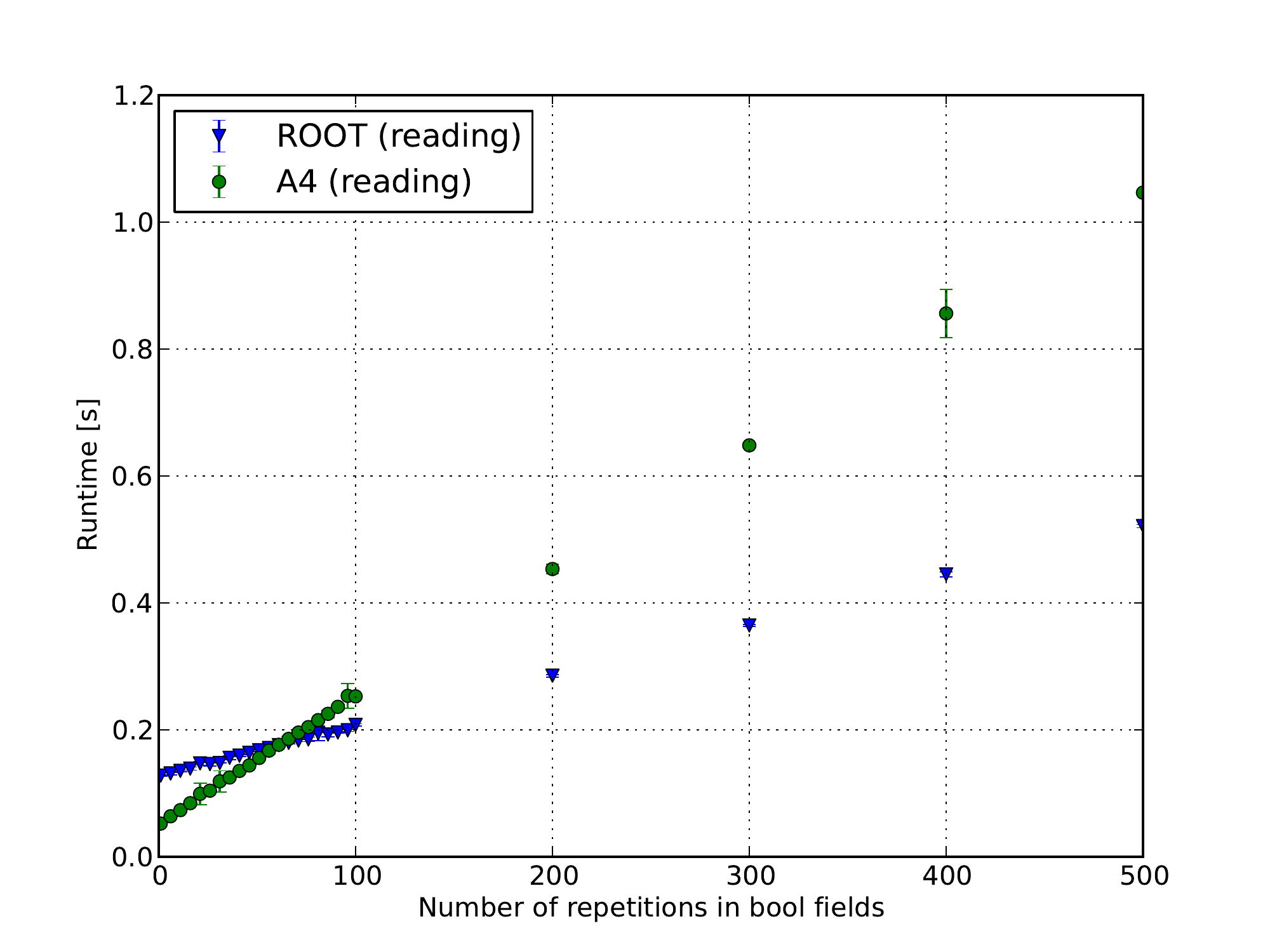}\\
\includegraphics[width=0.40\textwidth]{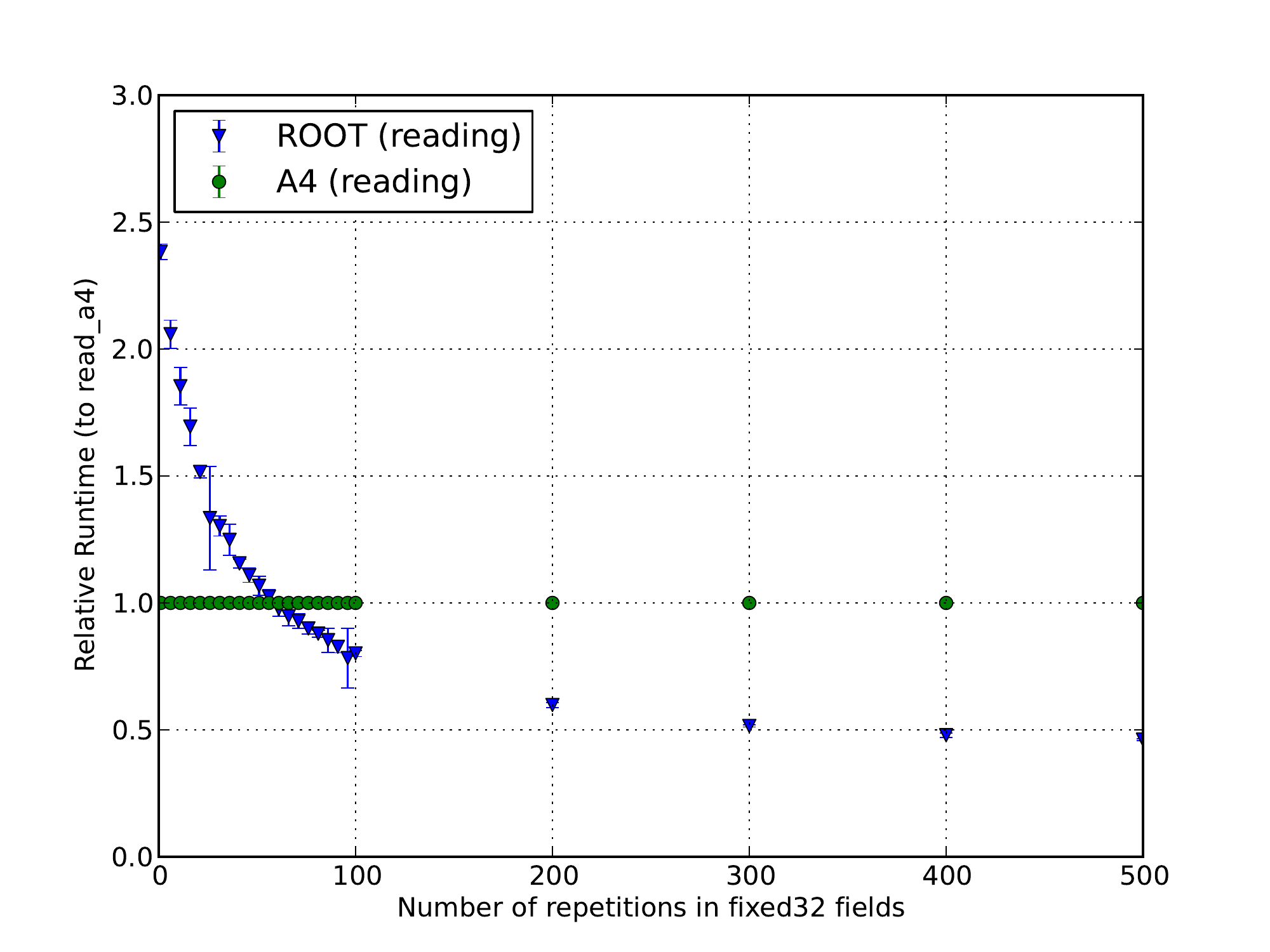}
\includegraphics[width=0.40\textwidth]{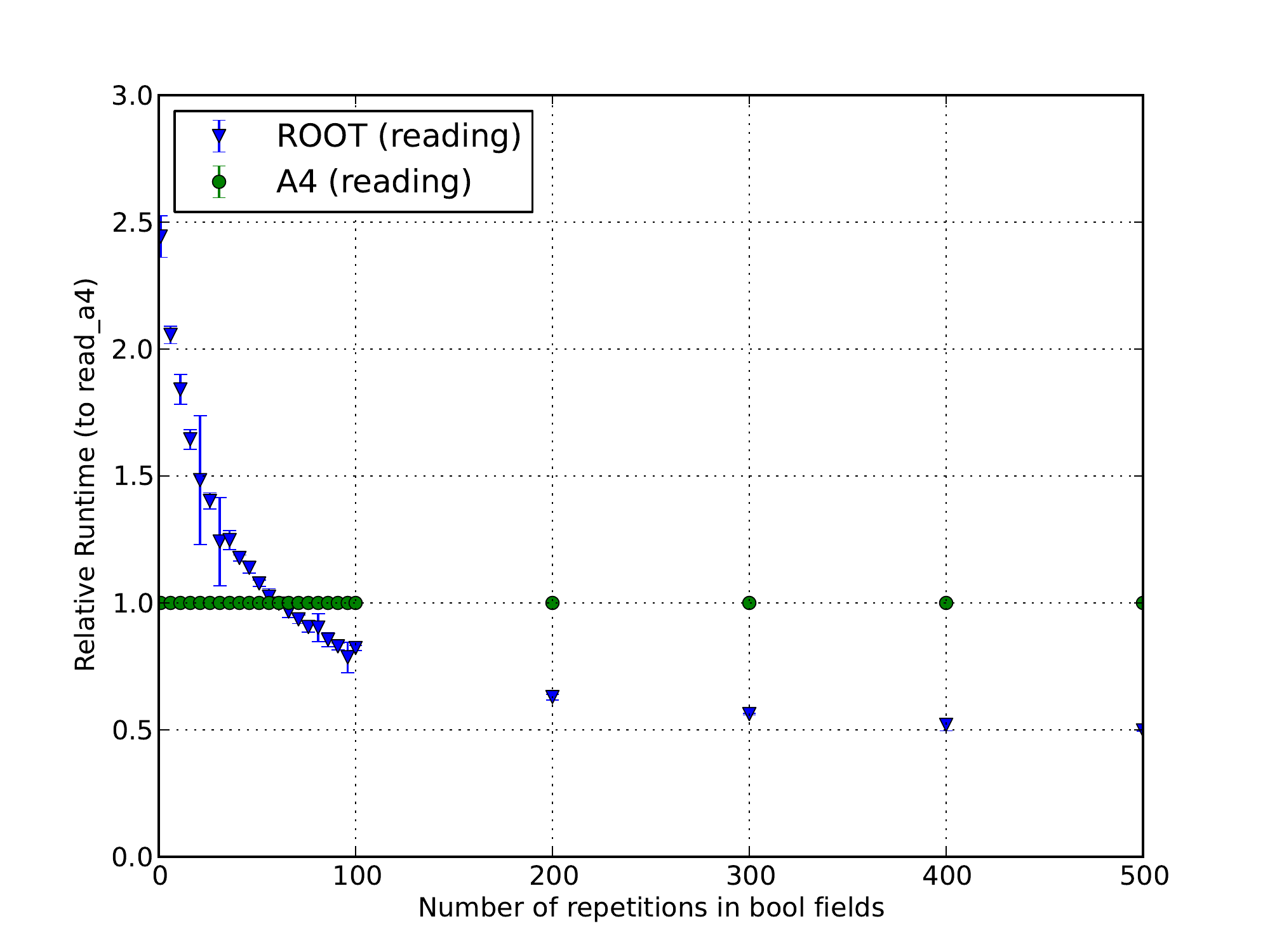}
\caption{Processing time in seconds for $100000$ events versus $n_{float}$ for $n_{rep} = 4$, for floats, doubles, integers and booleans from top left to bottom right. The top row shows absolute runtime, the lower row runtime relative to {\scshape a4}.}
\end{center}\end{figure}
\begin{figure}[ht]\begin{center}
\includegraphics[width=0.40\textwidth]{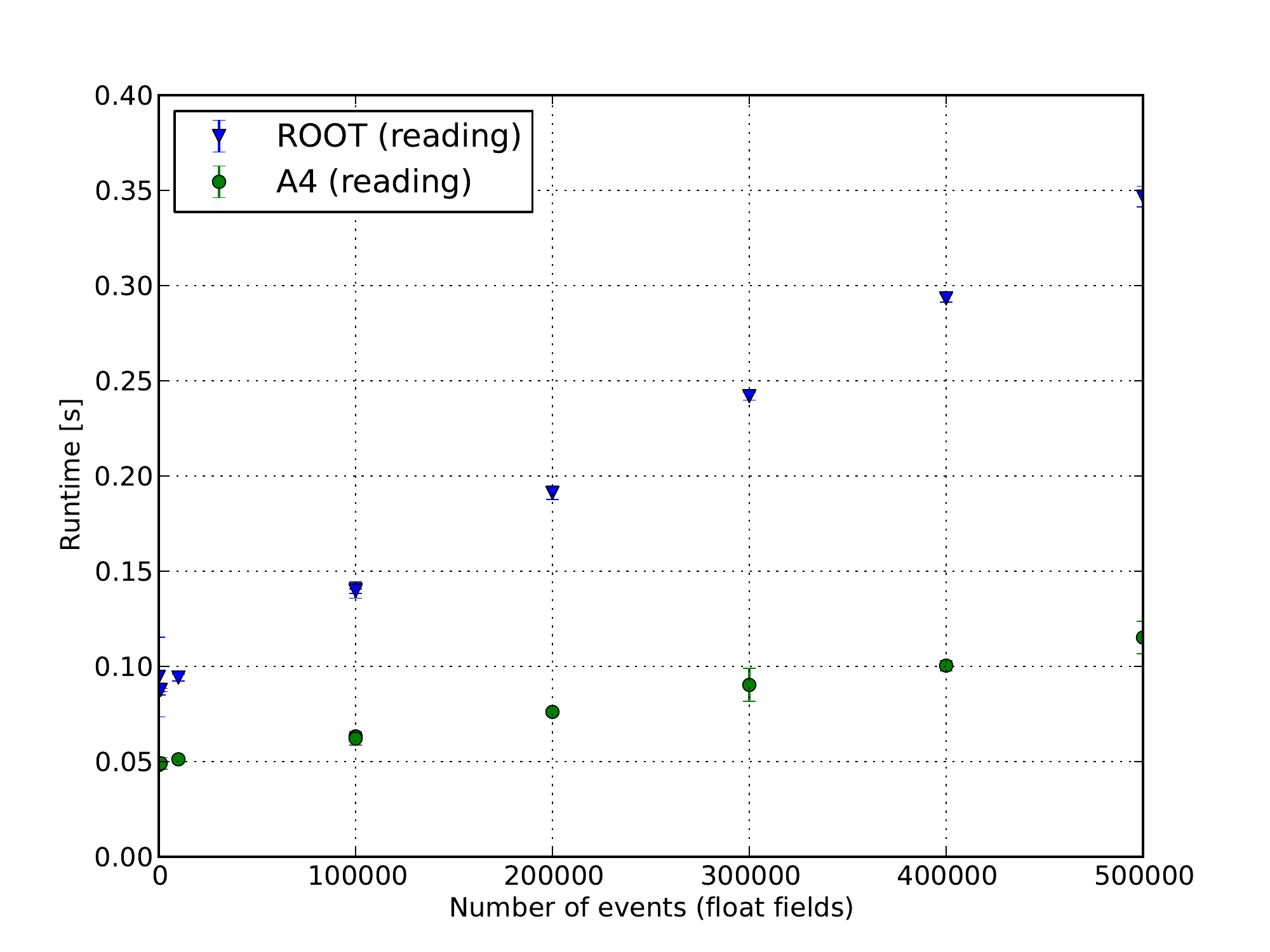}
\includegraphics[width=0.40\textwidth]{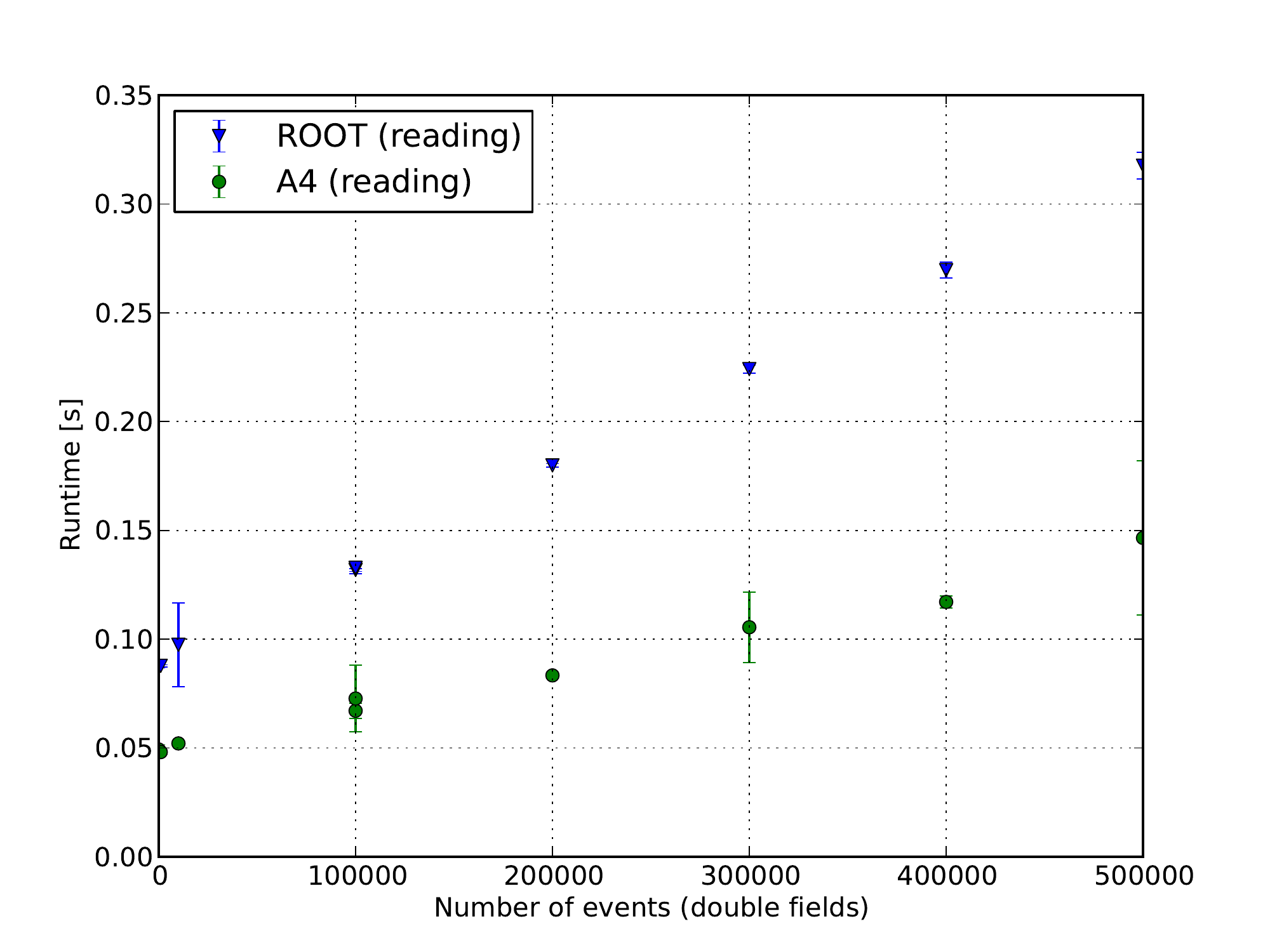}
\includegraphics[width=0.40\textwidth]{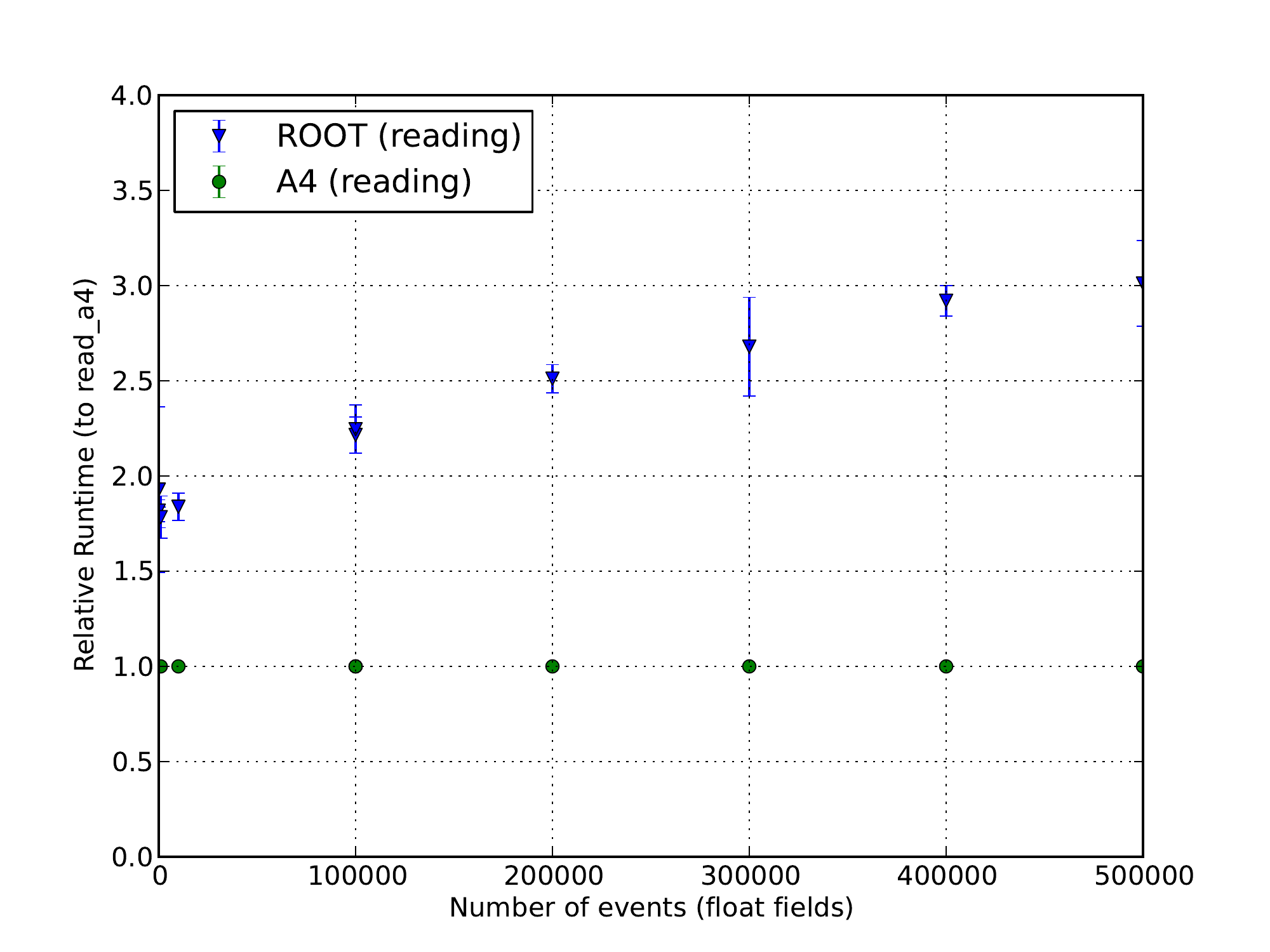}
\includegraphics[width=0.40\textwidth]{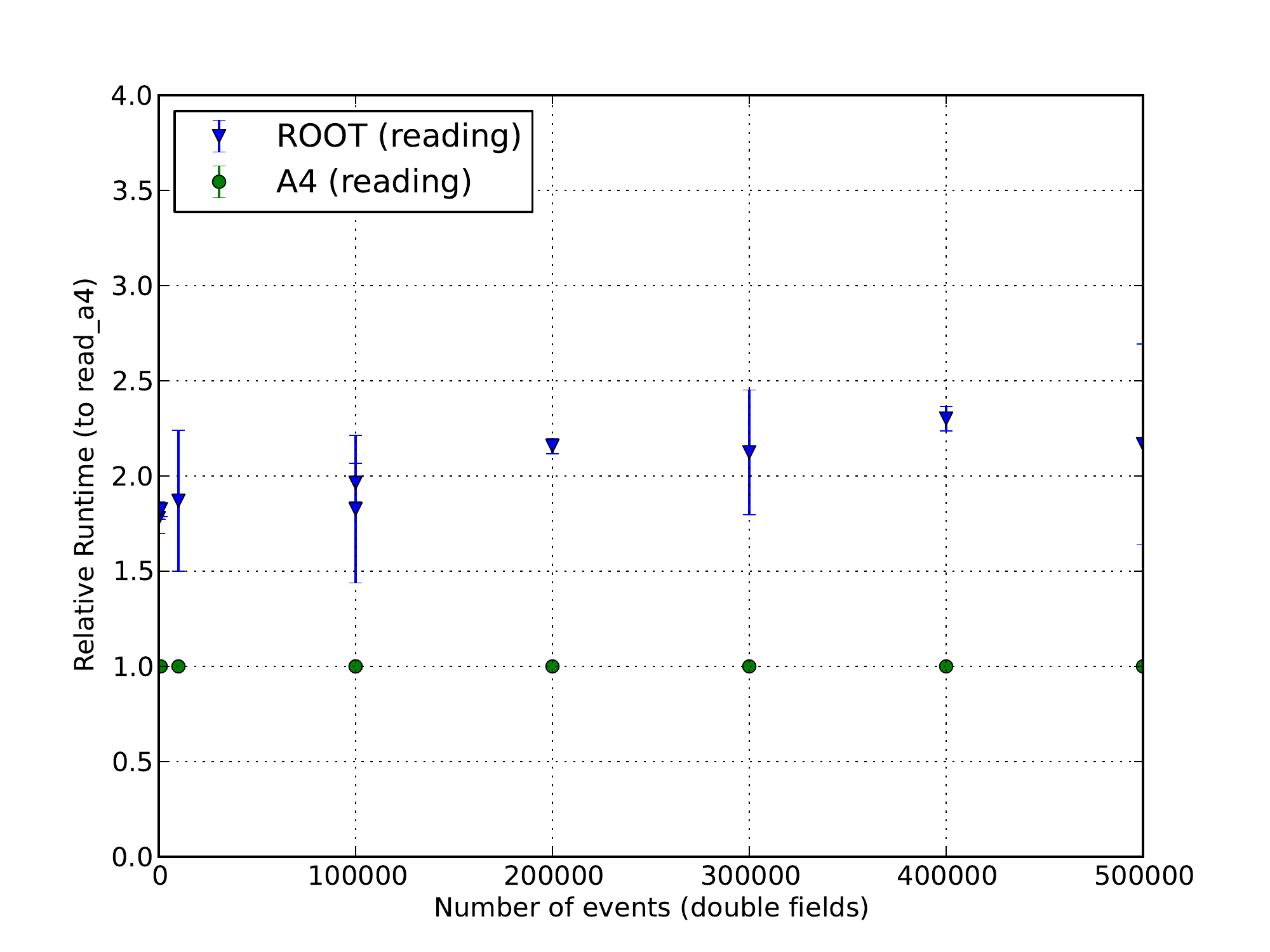}
\begin{tabular*}{\textwidth}{c}\hline\end{tabular*}
\includegraphics[width=0.40\textwidth]{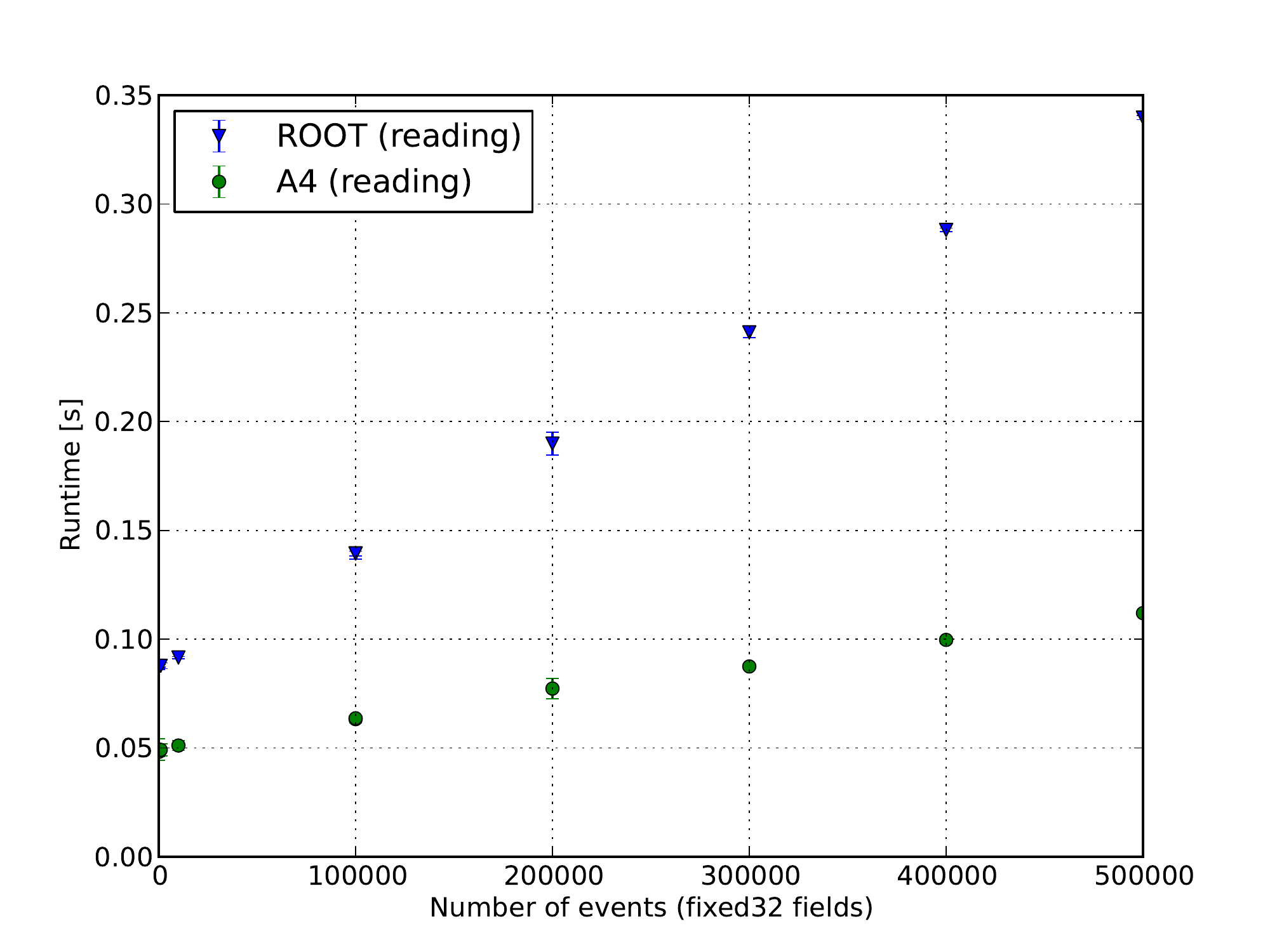}
\includegraphics[width=0.40\textwidth]{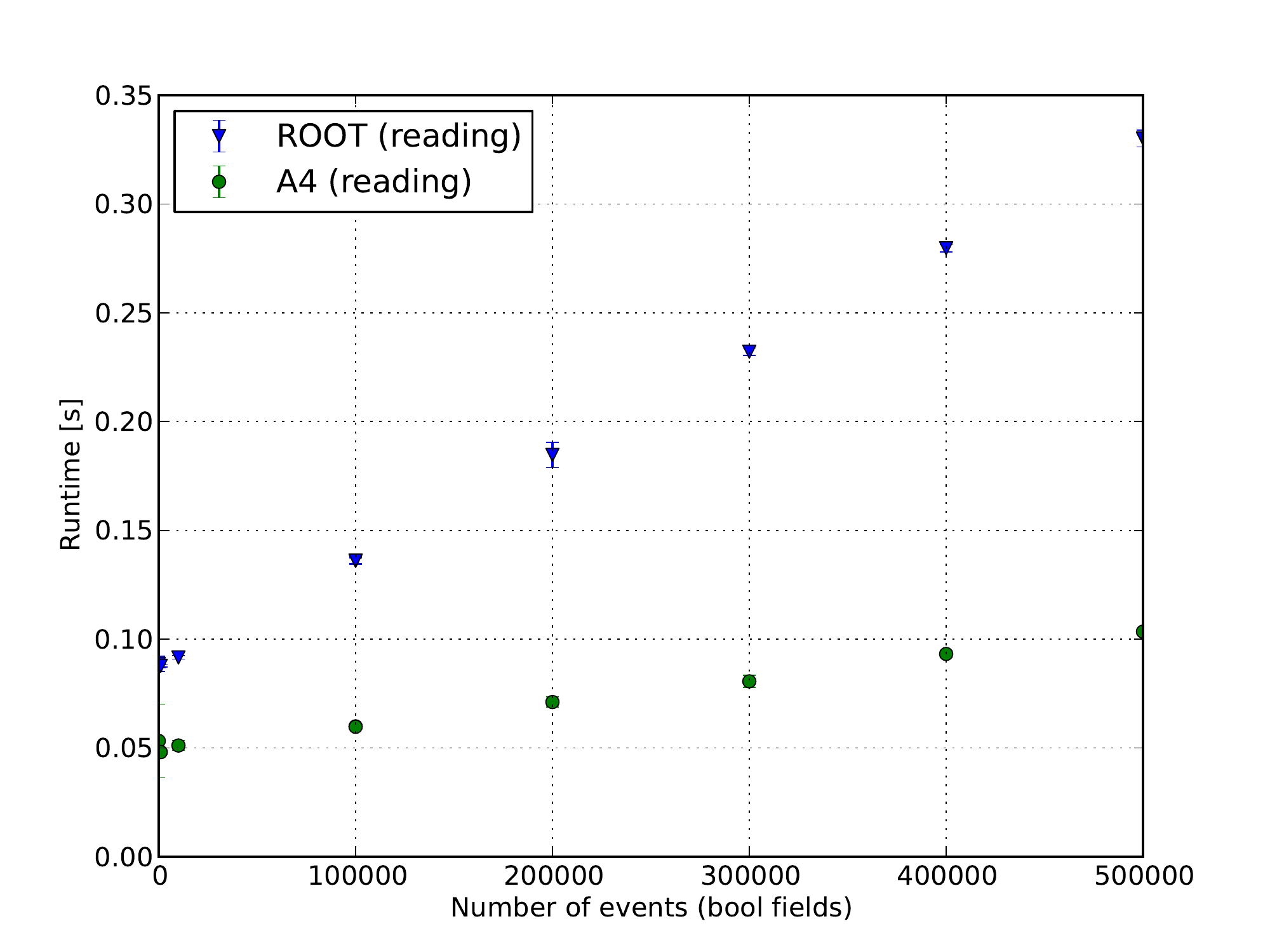}\\
\includegraphics[width=0.40\textwidth]{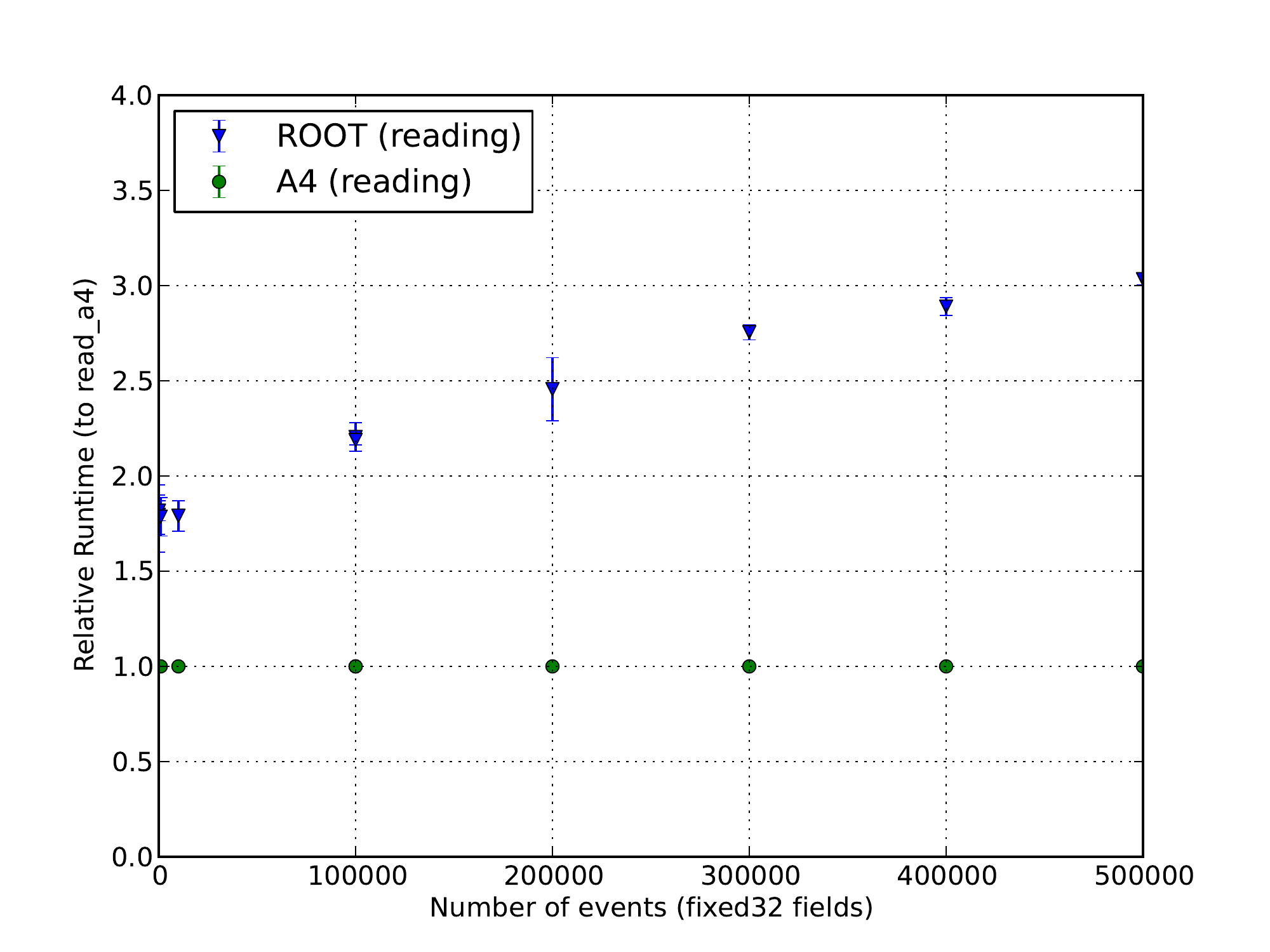}
\includegraphics[width=0.40\textwidth]{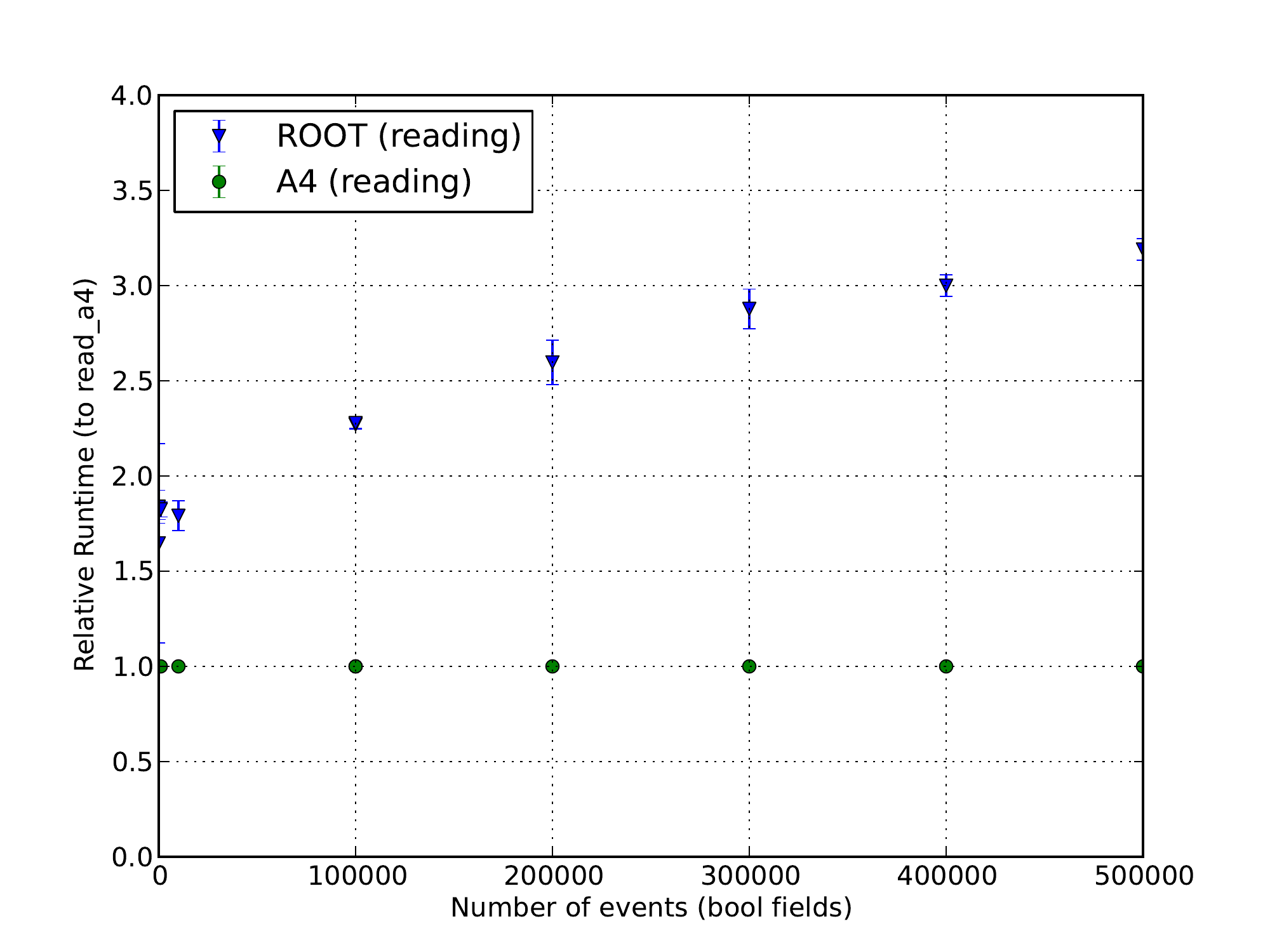}
\caption{Processing time in seconds vs number of events for $n_{flat} = 4, n_{rep} = 4$ and $n_{nfill} = 4$, for floats, doubles, integers and booleans from top left to bottom right. The top row shows absolute runtime, the lower row runtime relative to {\scshape a4}.}
\end{center}\end{figure}

\clearpage

\section{Additional benchmark results, zlib level 1 compression}
\begin{figure}[ht]\begin{center}
\includegraphics[width=0.40\textwidth]{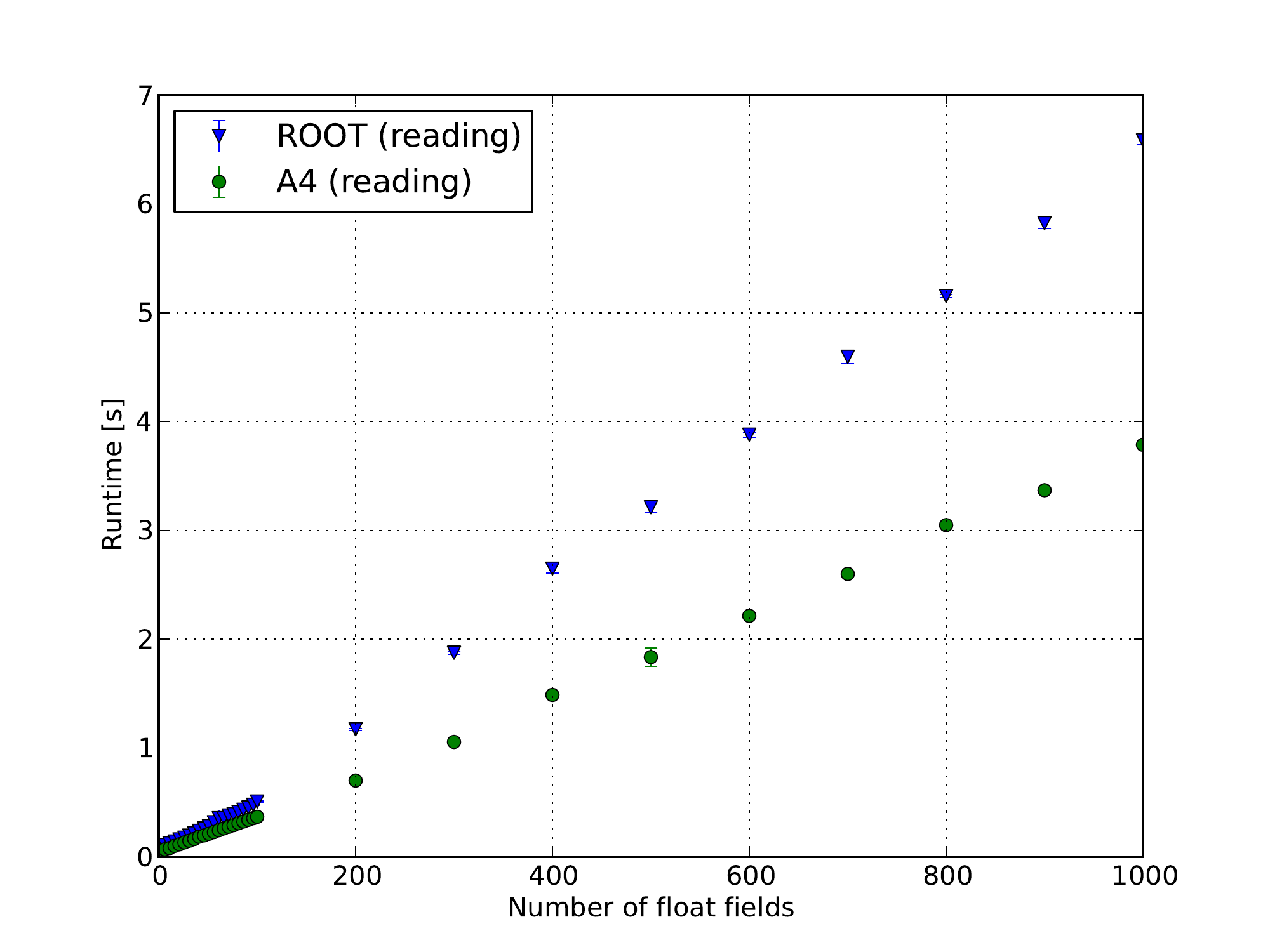}
\includegraphics[width=0.40\textwidth]{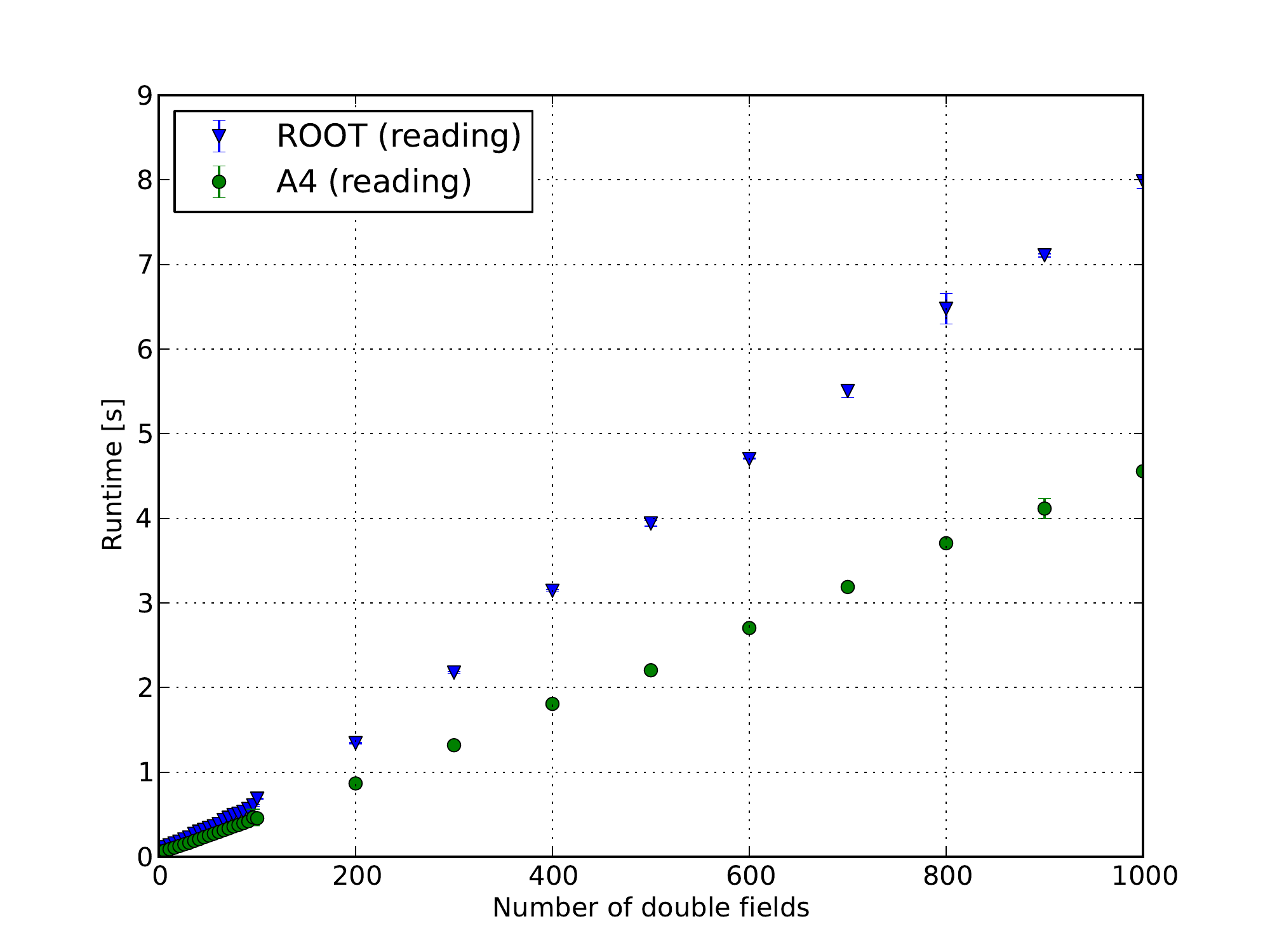}
\includegraphics[width=0.40\textwidth]{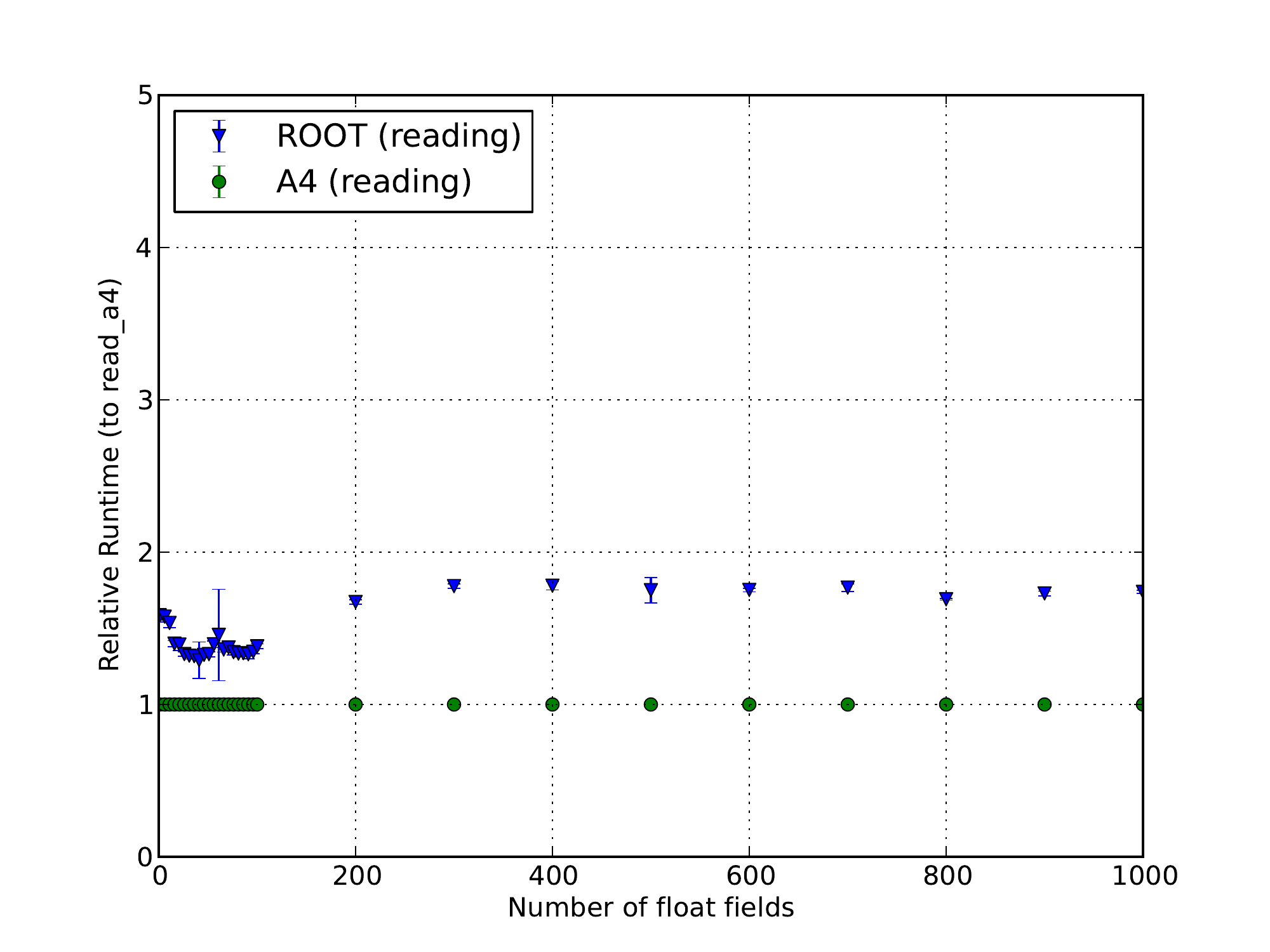}
\includegraphics[width=0.40\textwidth]{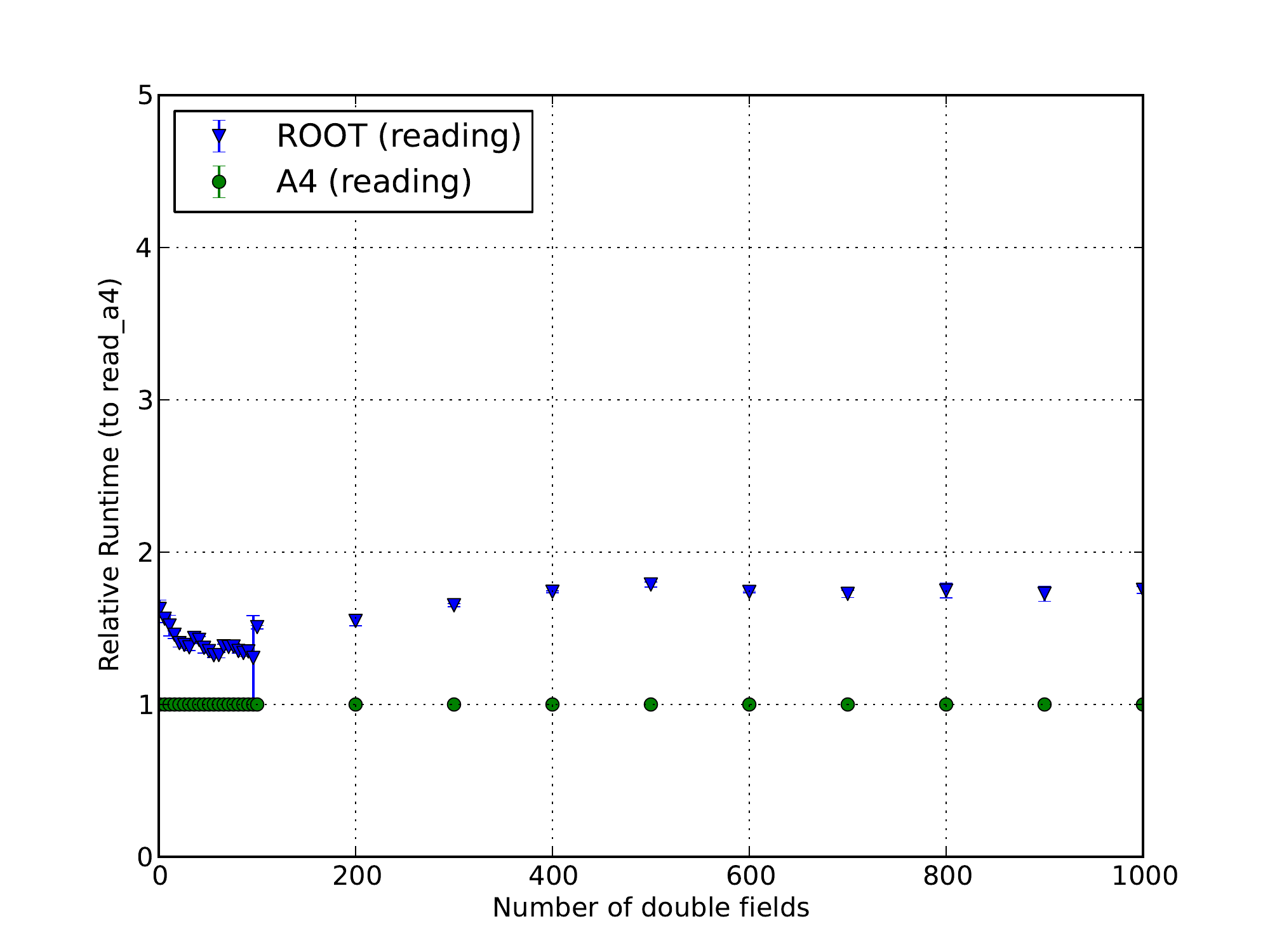}
\begin{tabular*}{\textwidth}{c}\hline\end{tabular*}
\includegraphics[width=0.40\textwidth]{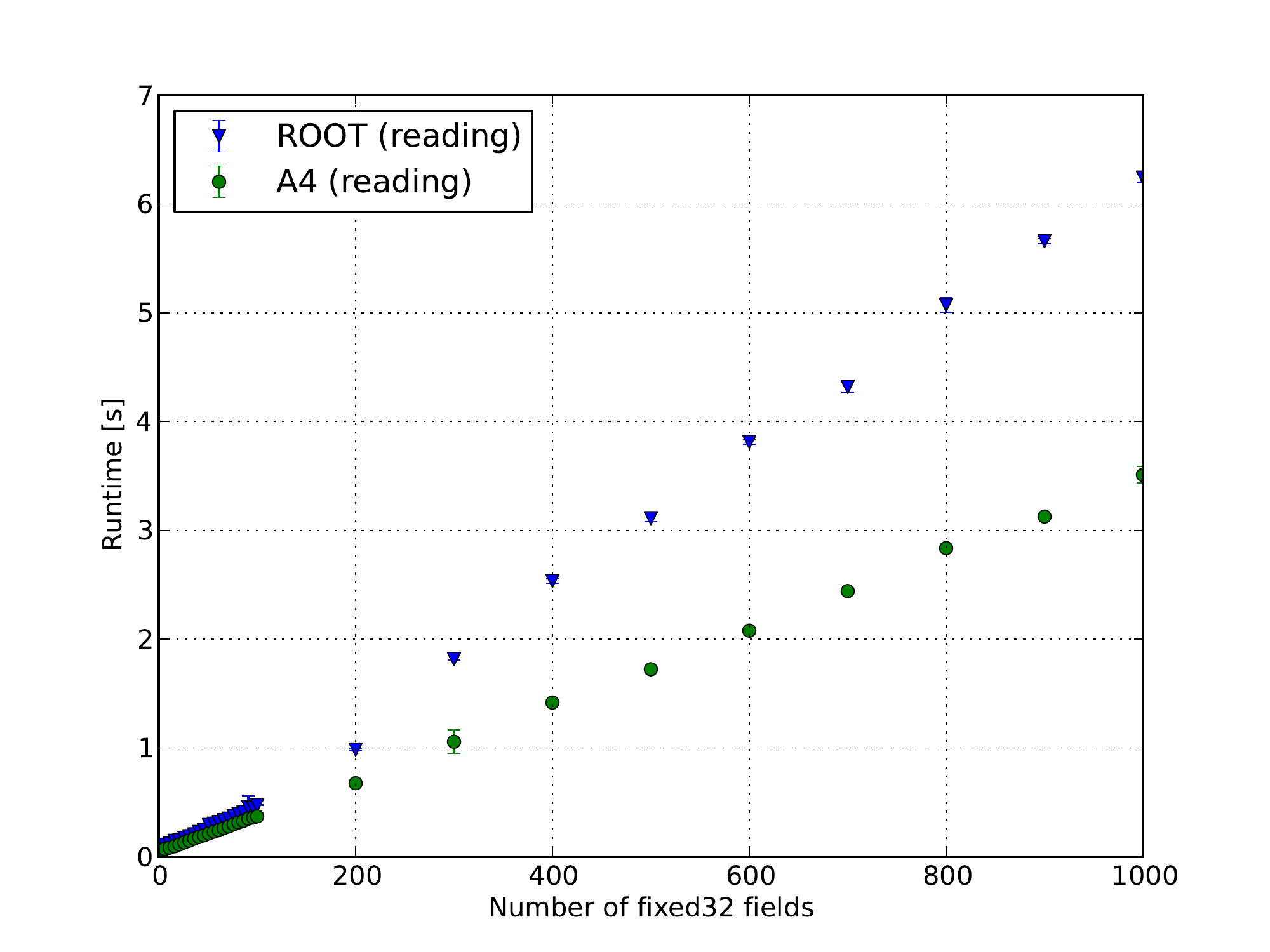}
\includegraphics[width=0.40\textwidth]{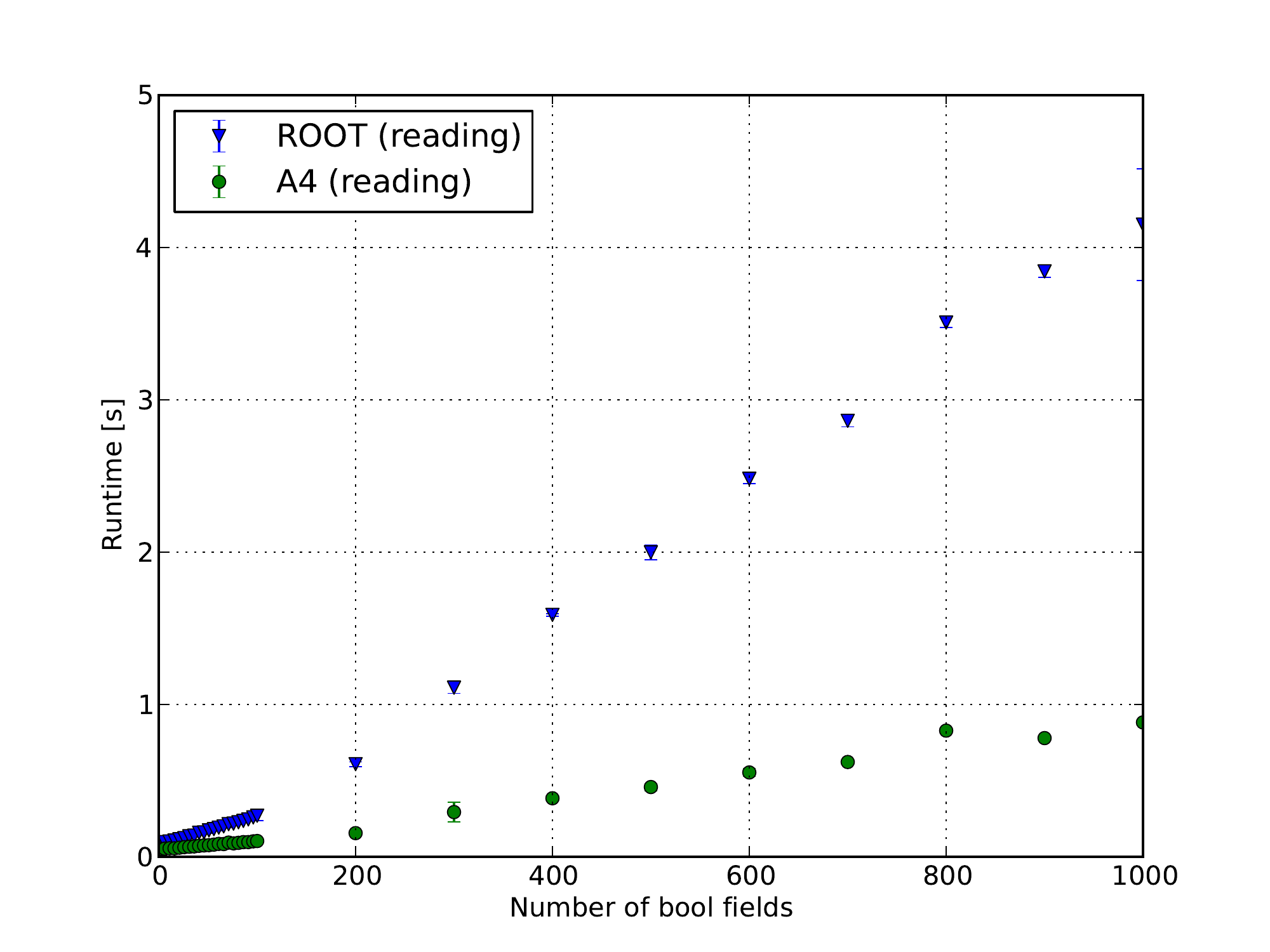}\\
\includegraphics[width=0.40\textwidth]{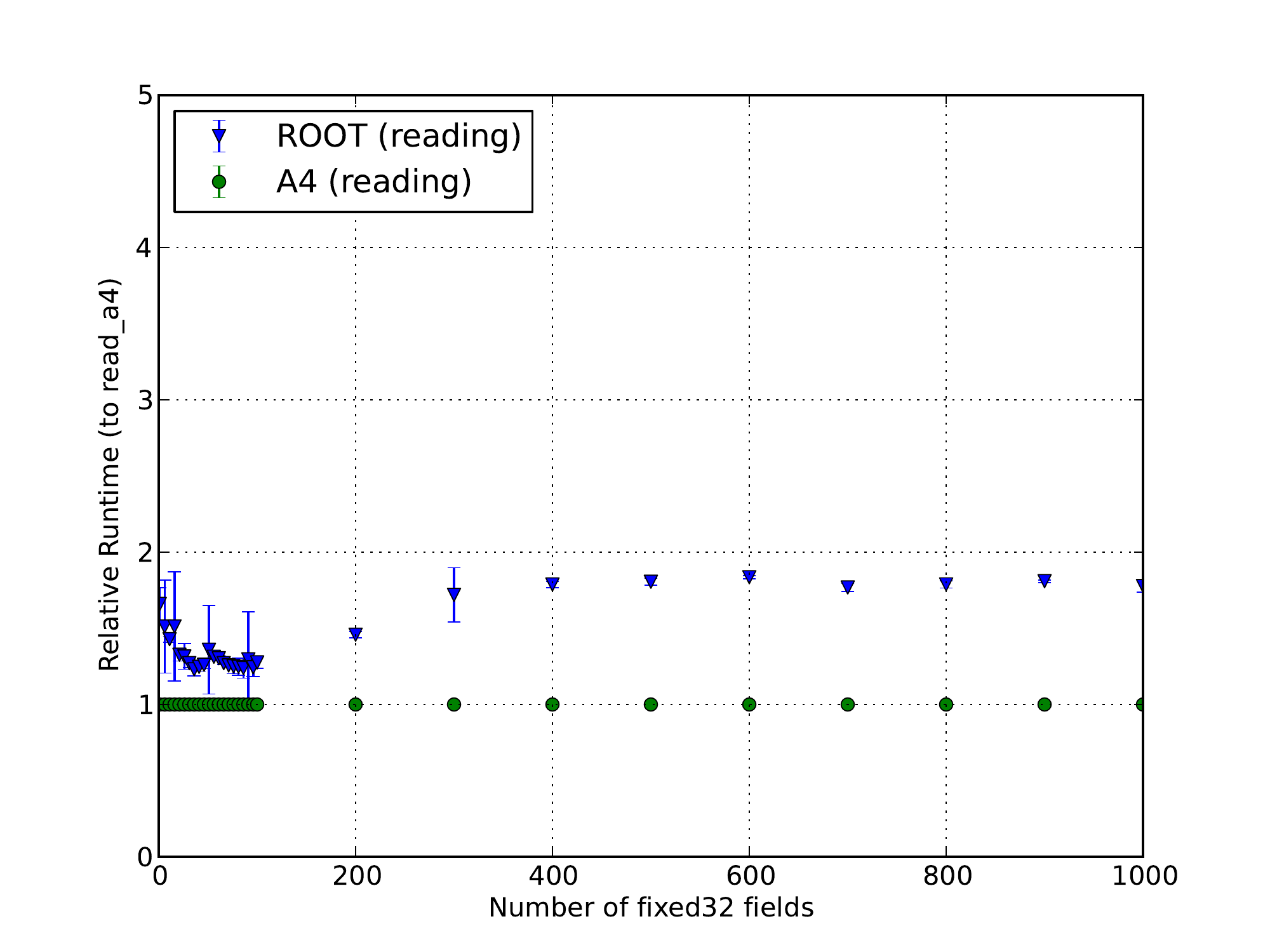}
\includegraphics[width=0.40\textwidth]{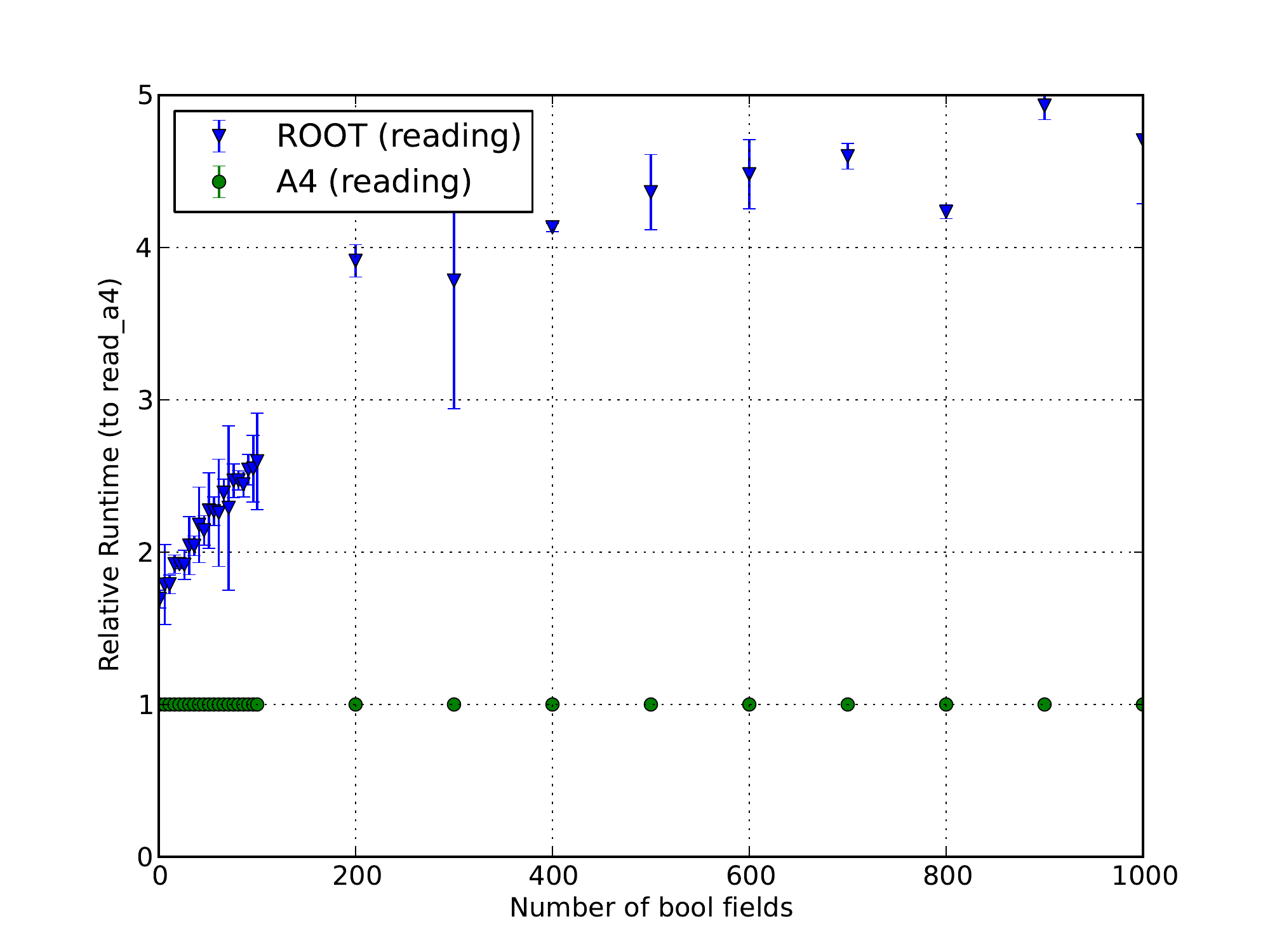}
\caption{Processing time in seconds for $100000$ events versus $n_{flat}$, for floats, doubles, integers and booleans from top left to bottom right. The top row shows absolute runtime, the lower row runtime relative to {\scshape a4}. Compression is enabled.}
\end{center}\end{figure}
\begin{figure}[ht]\begin{center}
\includegraphics[width=0.40\textwidth]{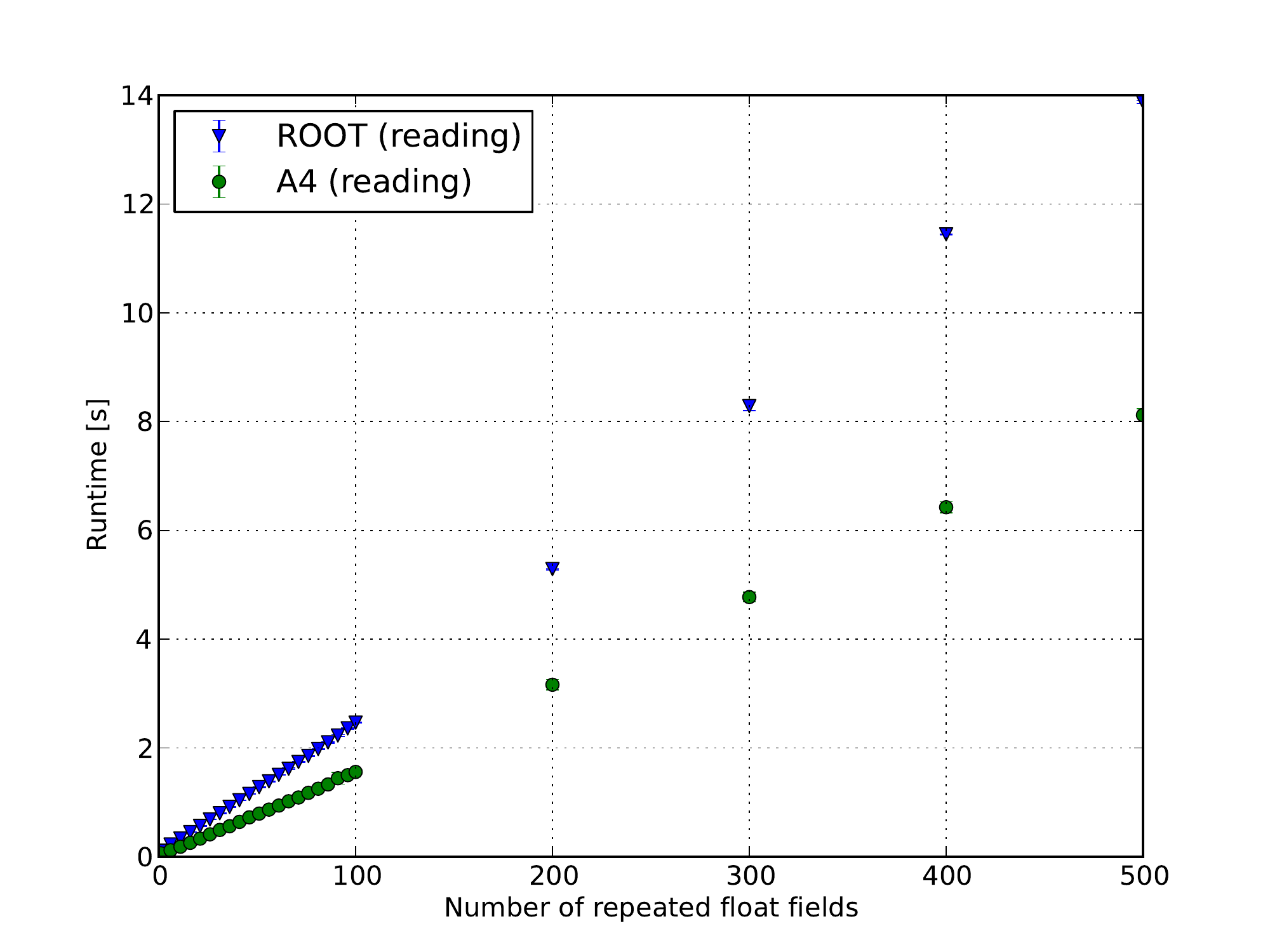}
\includegraphics[width=0.40\textwidth]{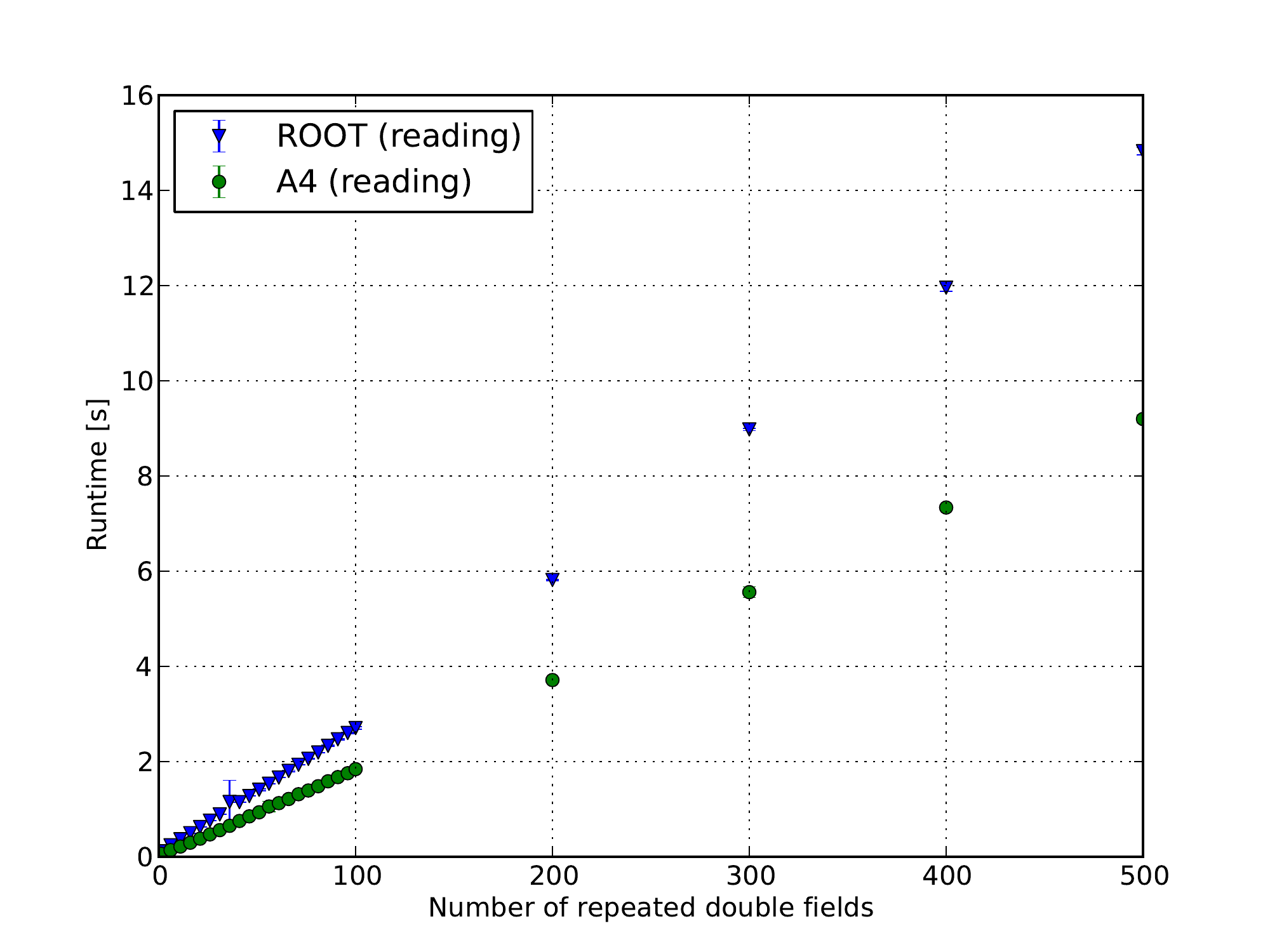}
\includegraphics[width=0.40\textwidth]{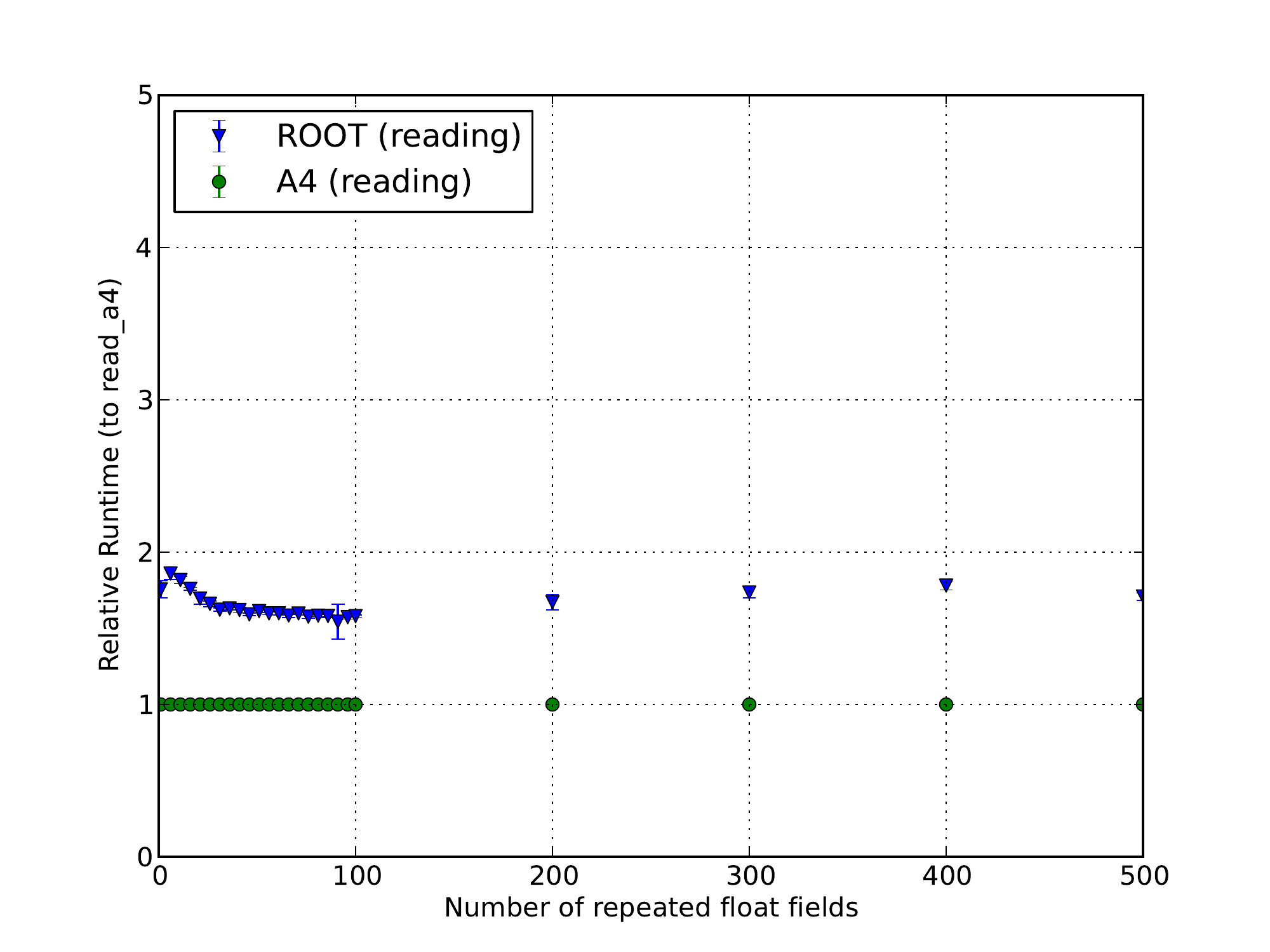}
\includegraphics[width=0.40\textwidth]{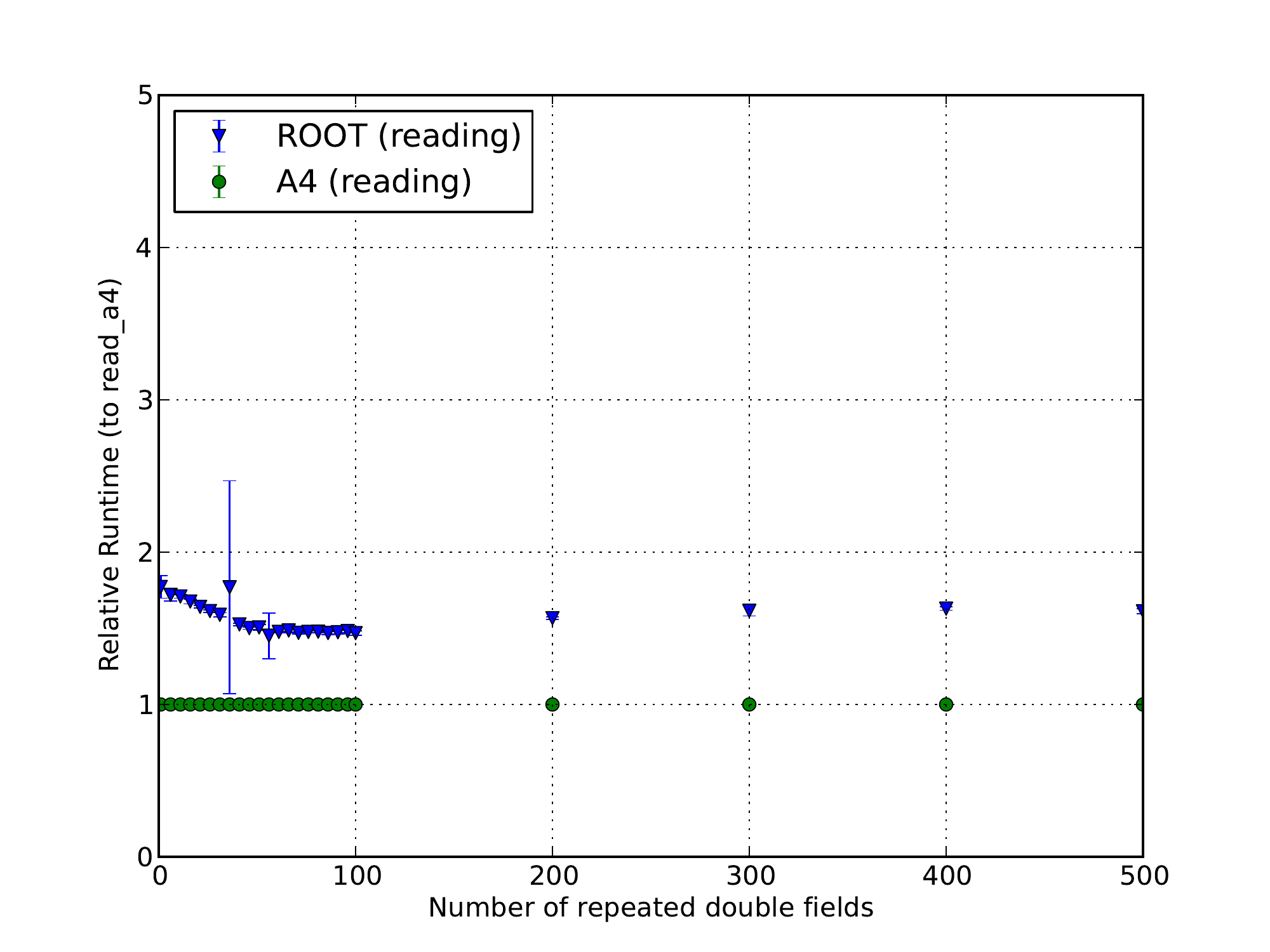}
\begin{tabular*}{\textwidth}{c}\hline\end{tabular*}
\includegraphics[width=0.40\textwidth]{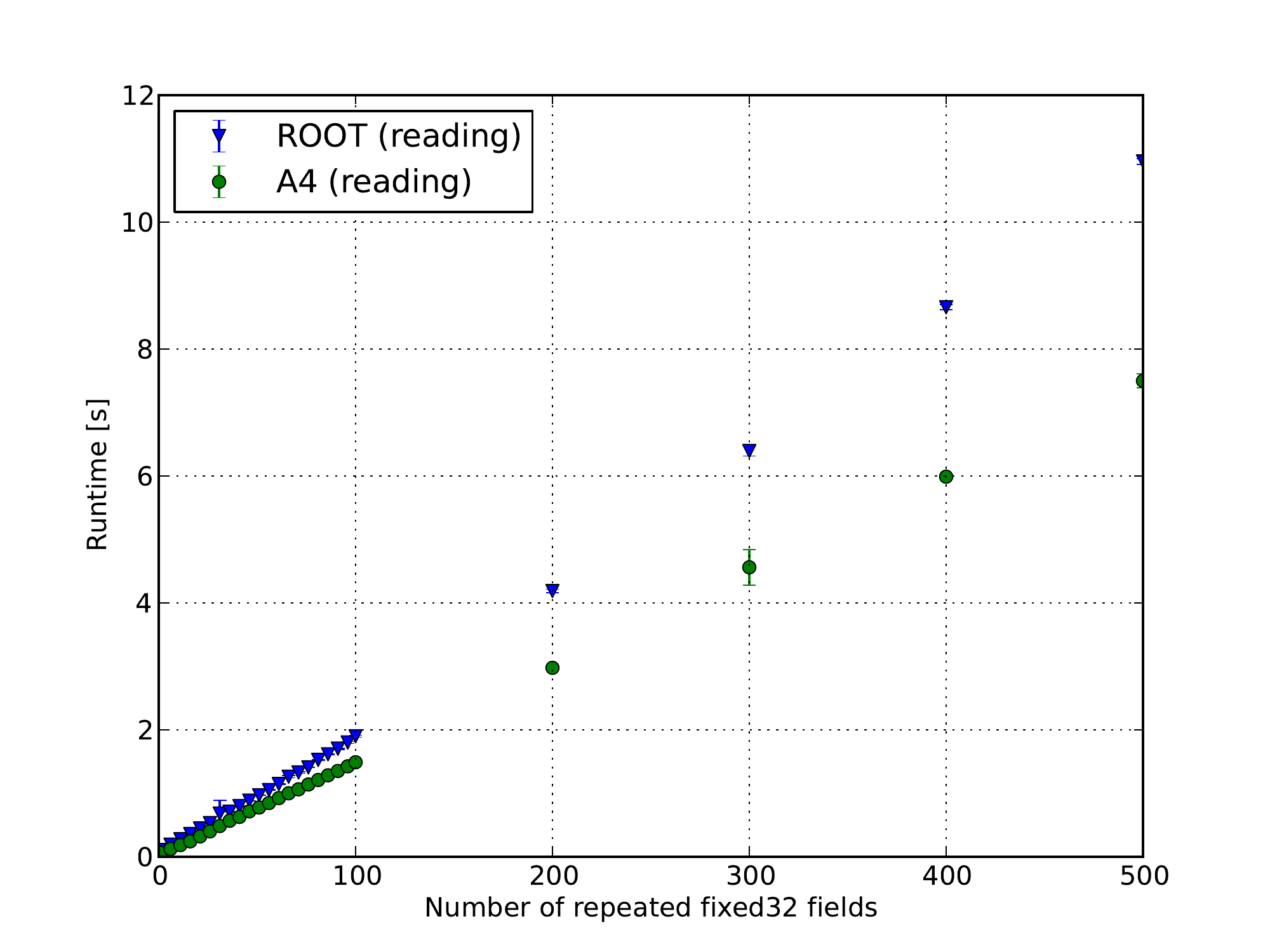}
\includegraphics[width=0.40\textwidth]{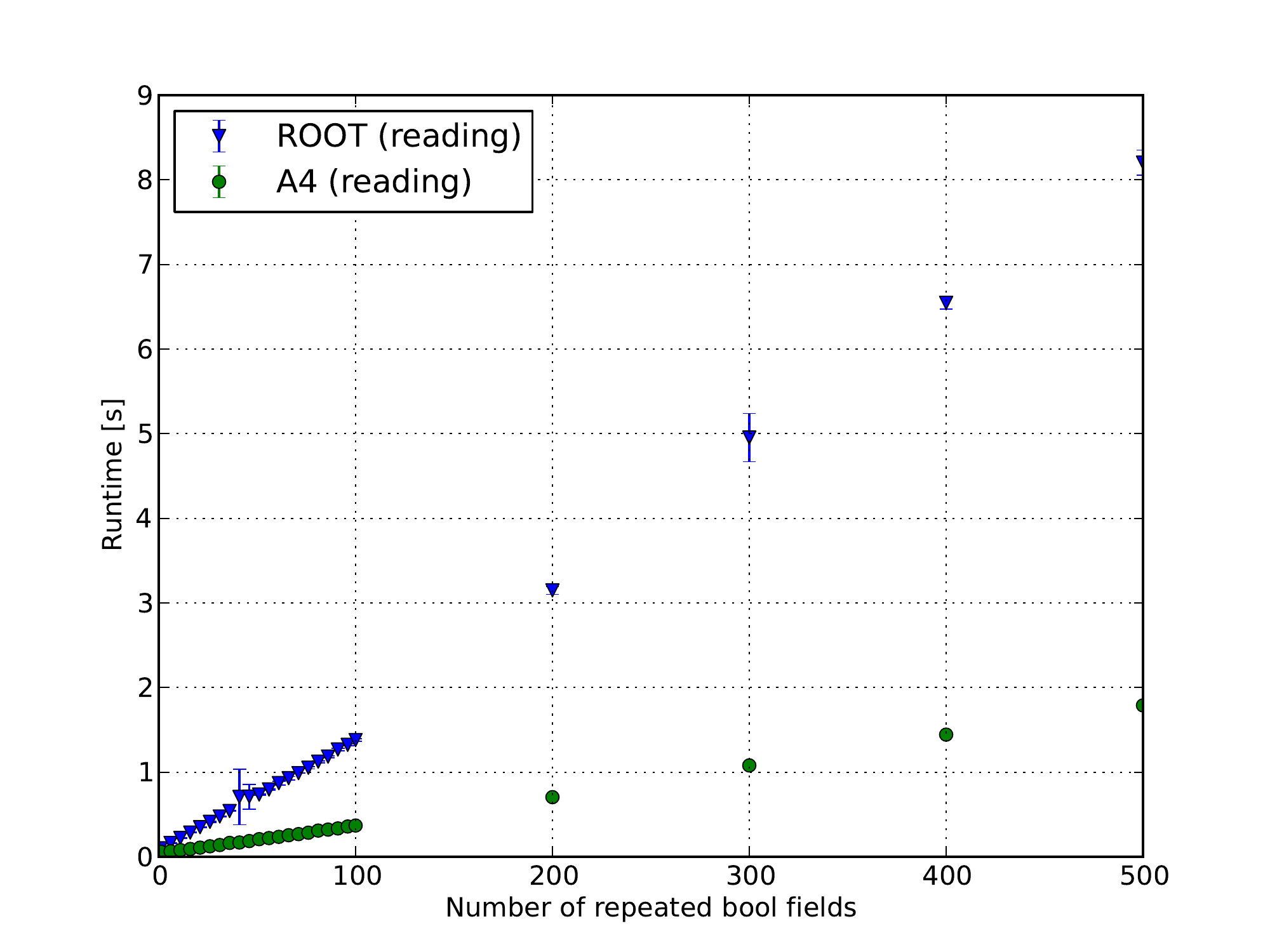}\\
\includegraphics[width=0.40\textwidth]{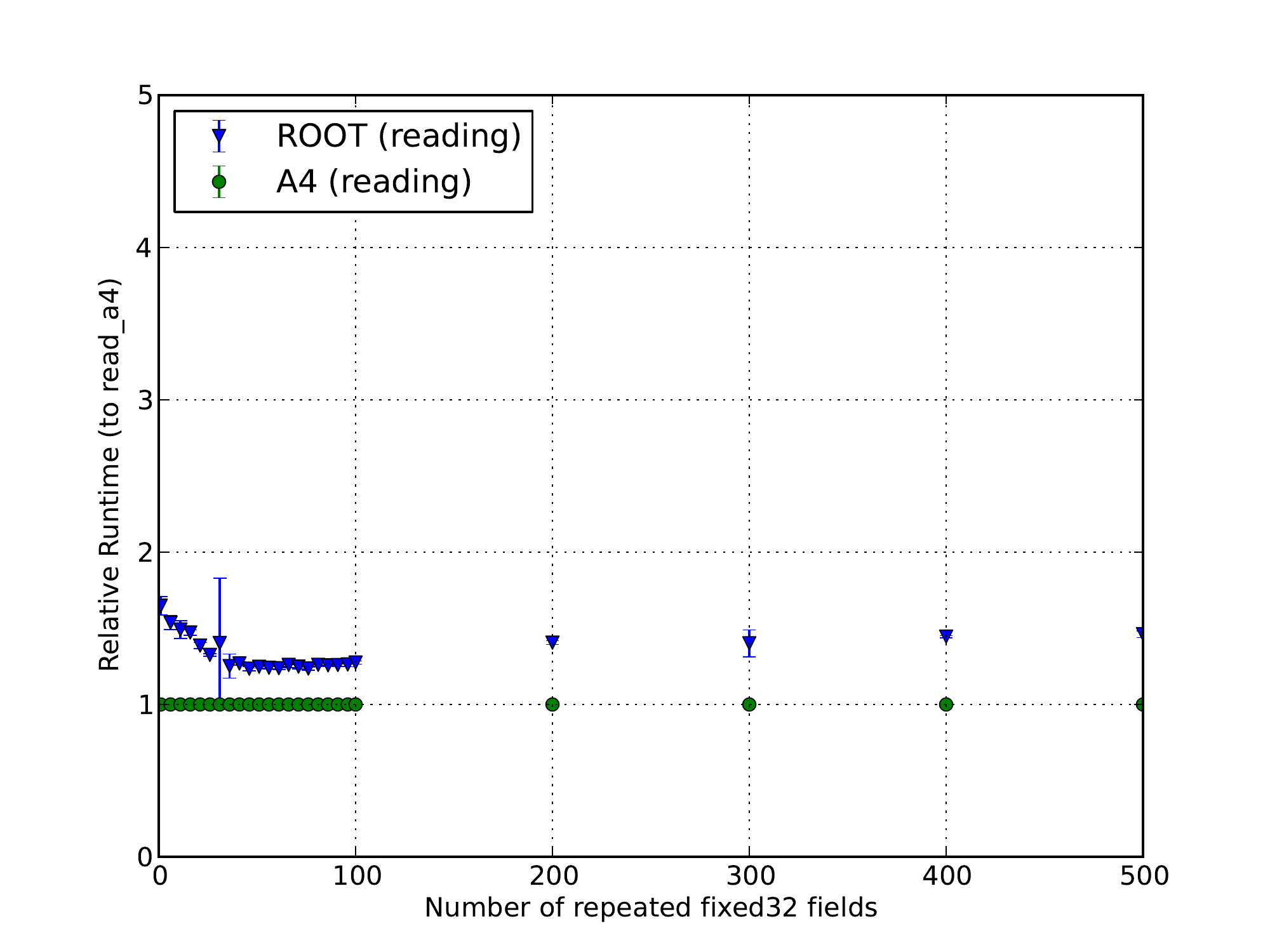}
\includegraphics[width=0.40\textwidth]{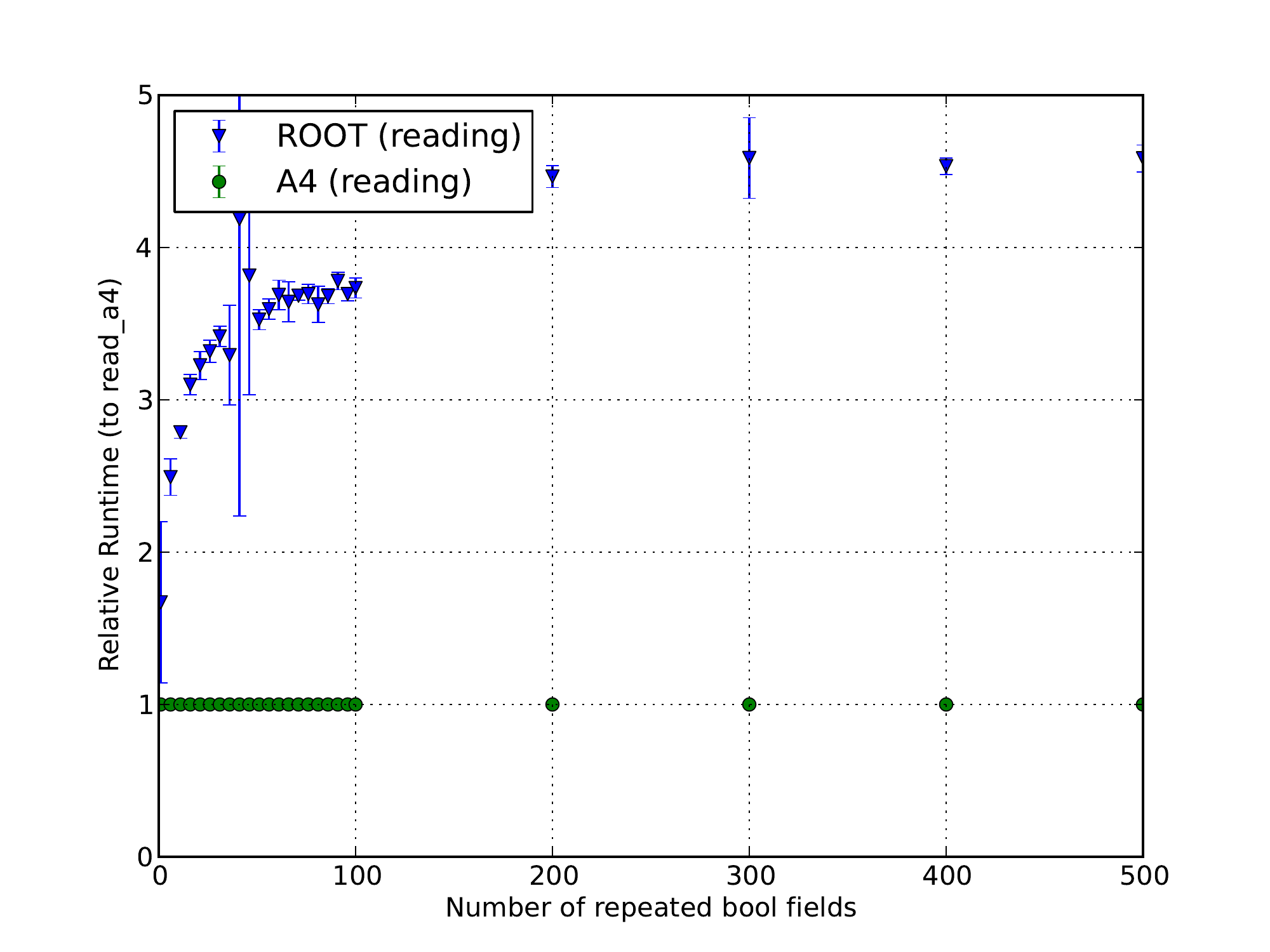}
\caption{Processing time ins seconds for $100000$ events versus $n_{rep}$, for floats, doubles, integers and booleans from top left to bottom right. The top row shows absolute runtime, the lower row runtime relative to {\scshape a4}. Compression is enabled.}
\end{center}\end{figure}
\begin{figure}[ht]\begin{center}
\includegraphics[width=0.40\textwidth]{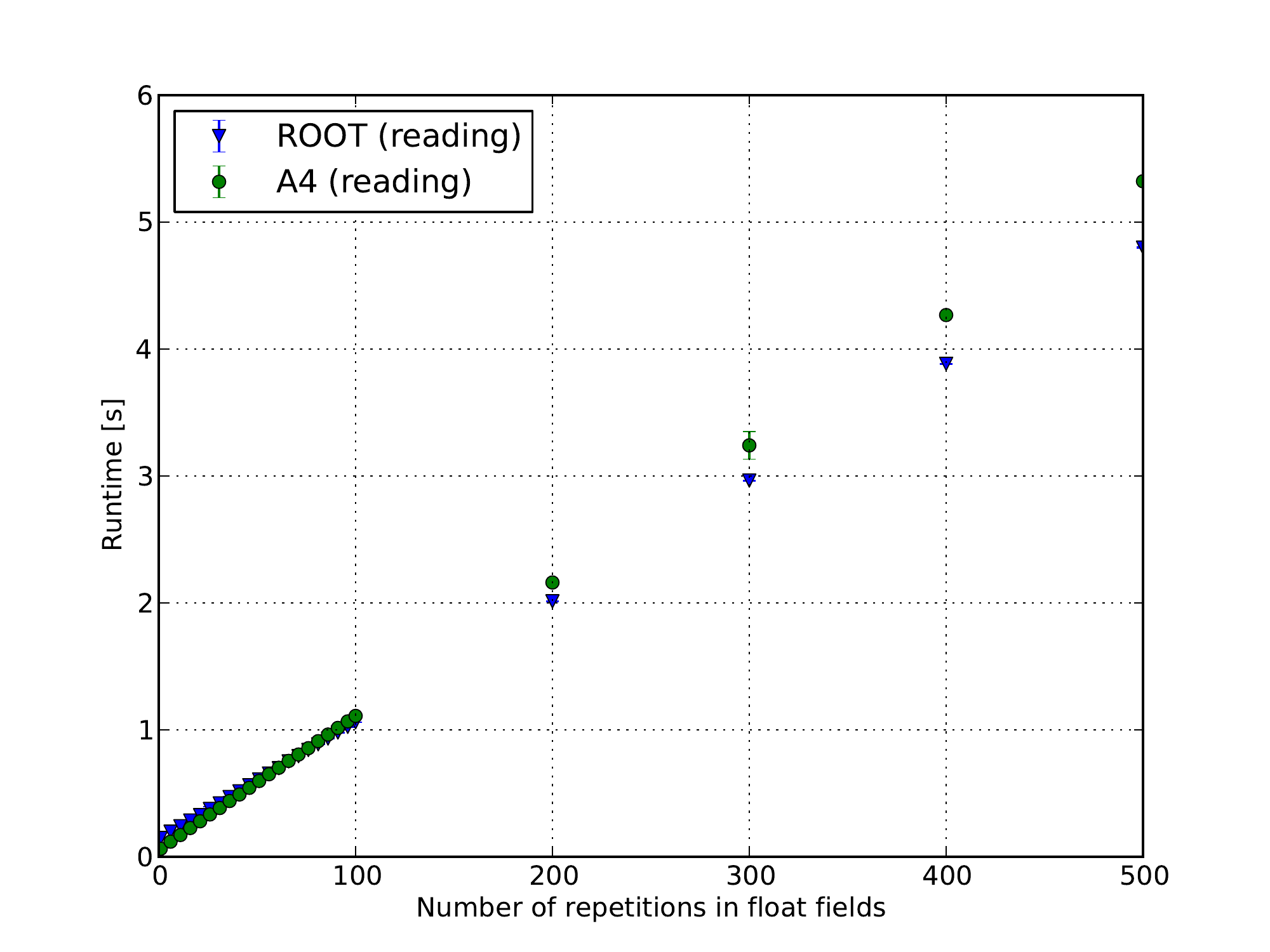}
\includegraphics[width=0.40\textwidth]{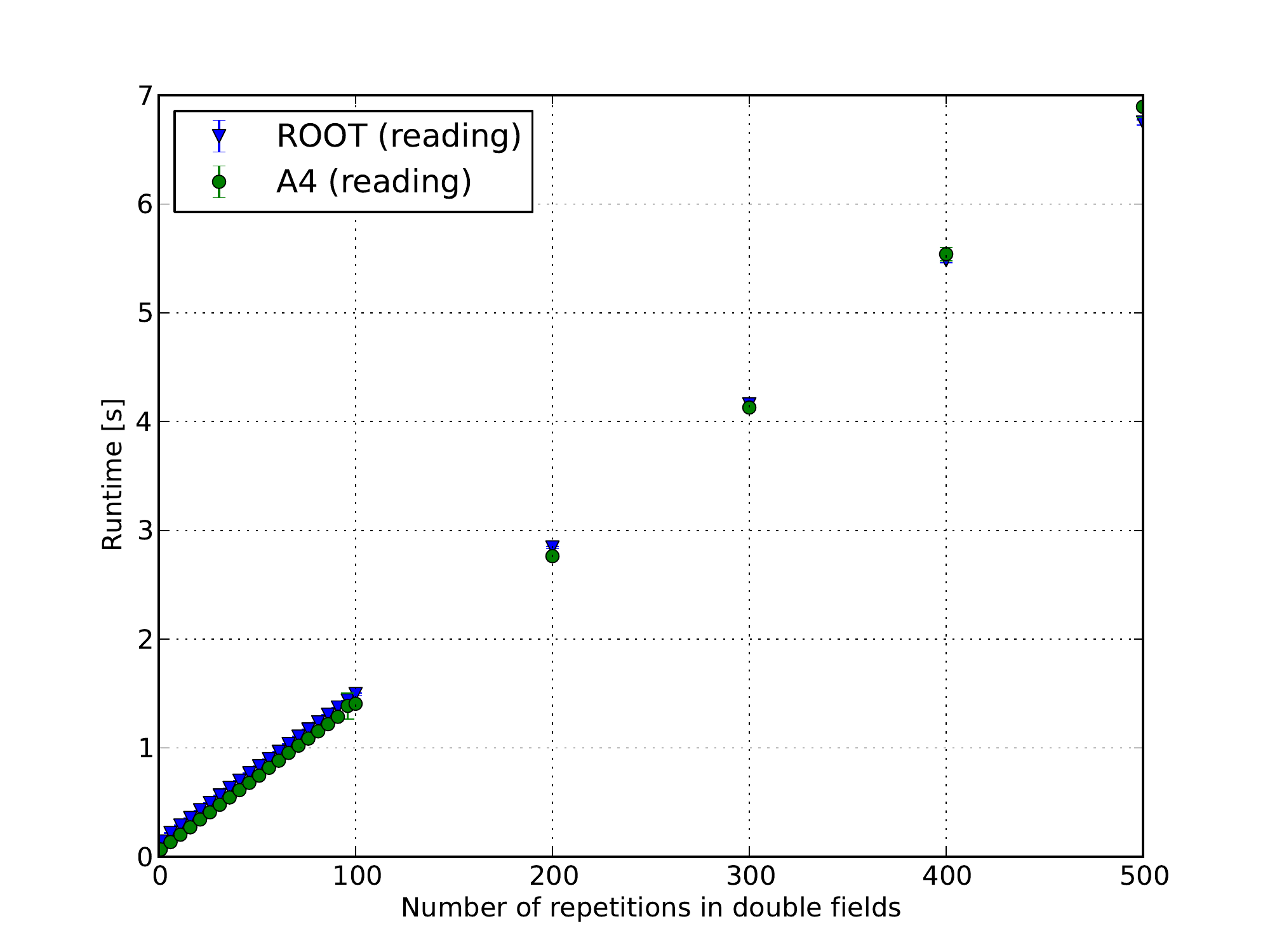}
\includegraphics[width=0.40\textwidth]{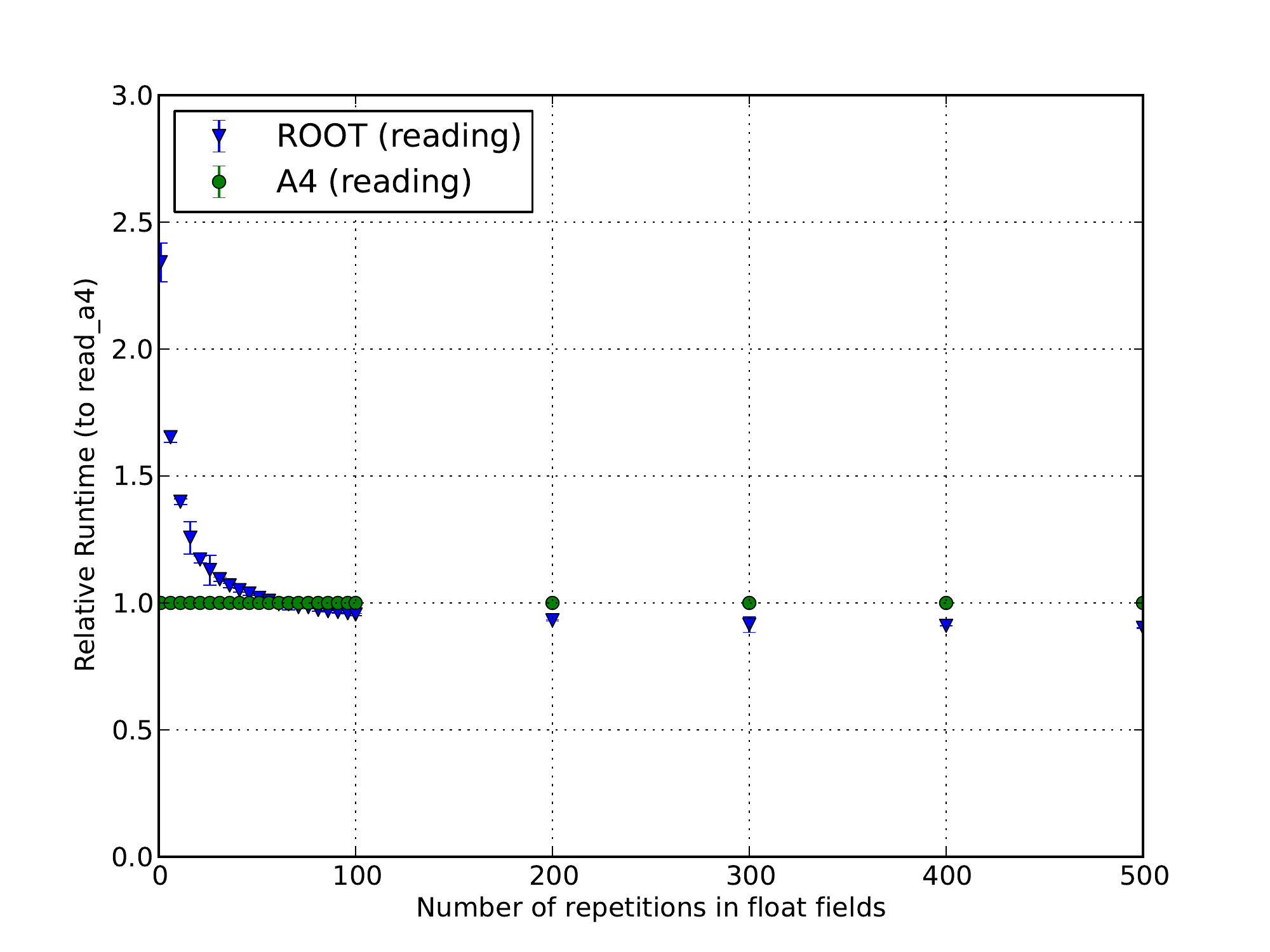}
\includegraphics[width=0.40\textwidth]{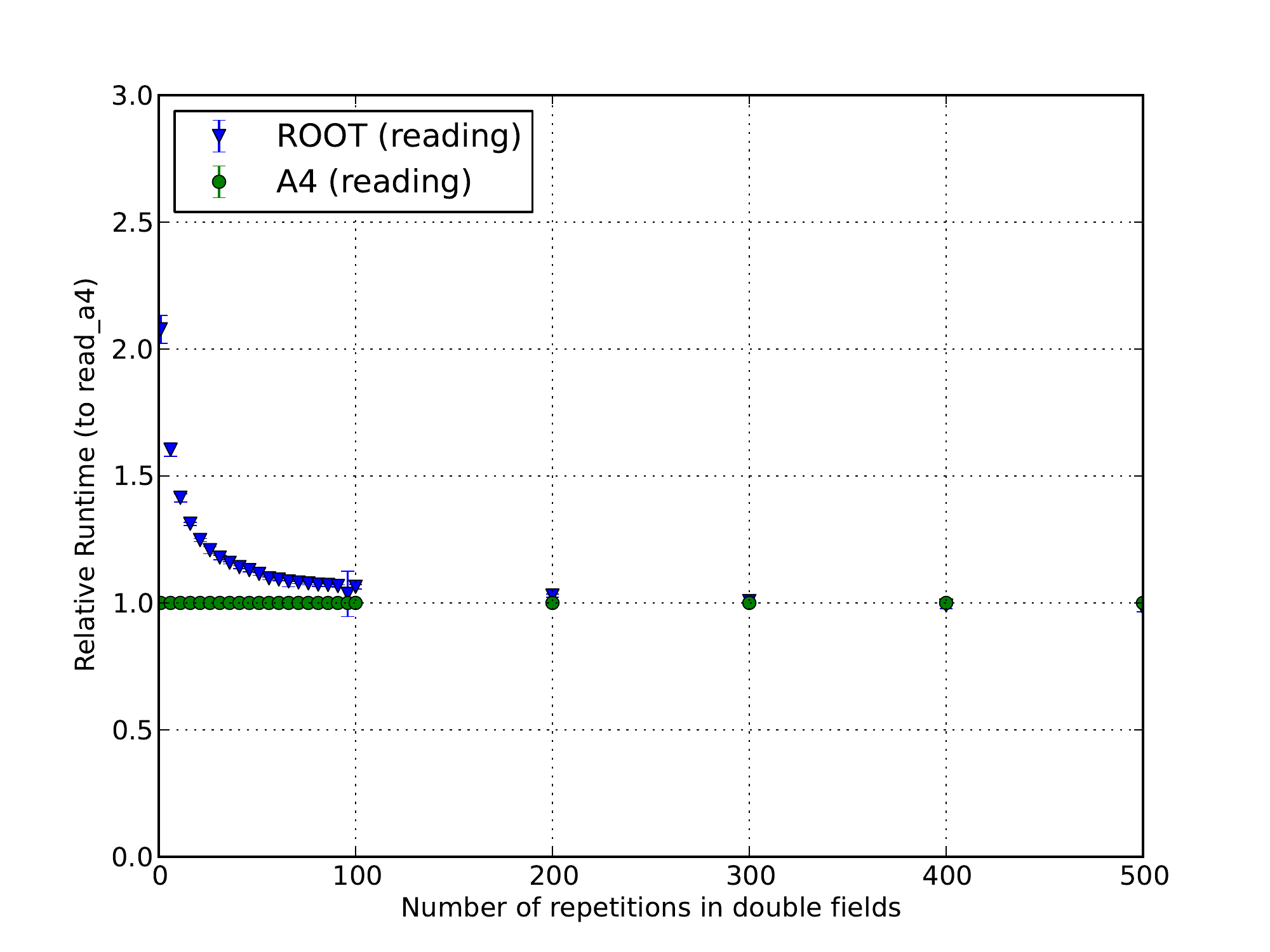}
\begin{tabular*}{\textwidth}{c}\hline\end{tabular*}
\includegraphics[width=0.40\textwidth]{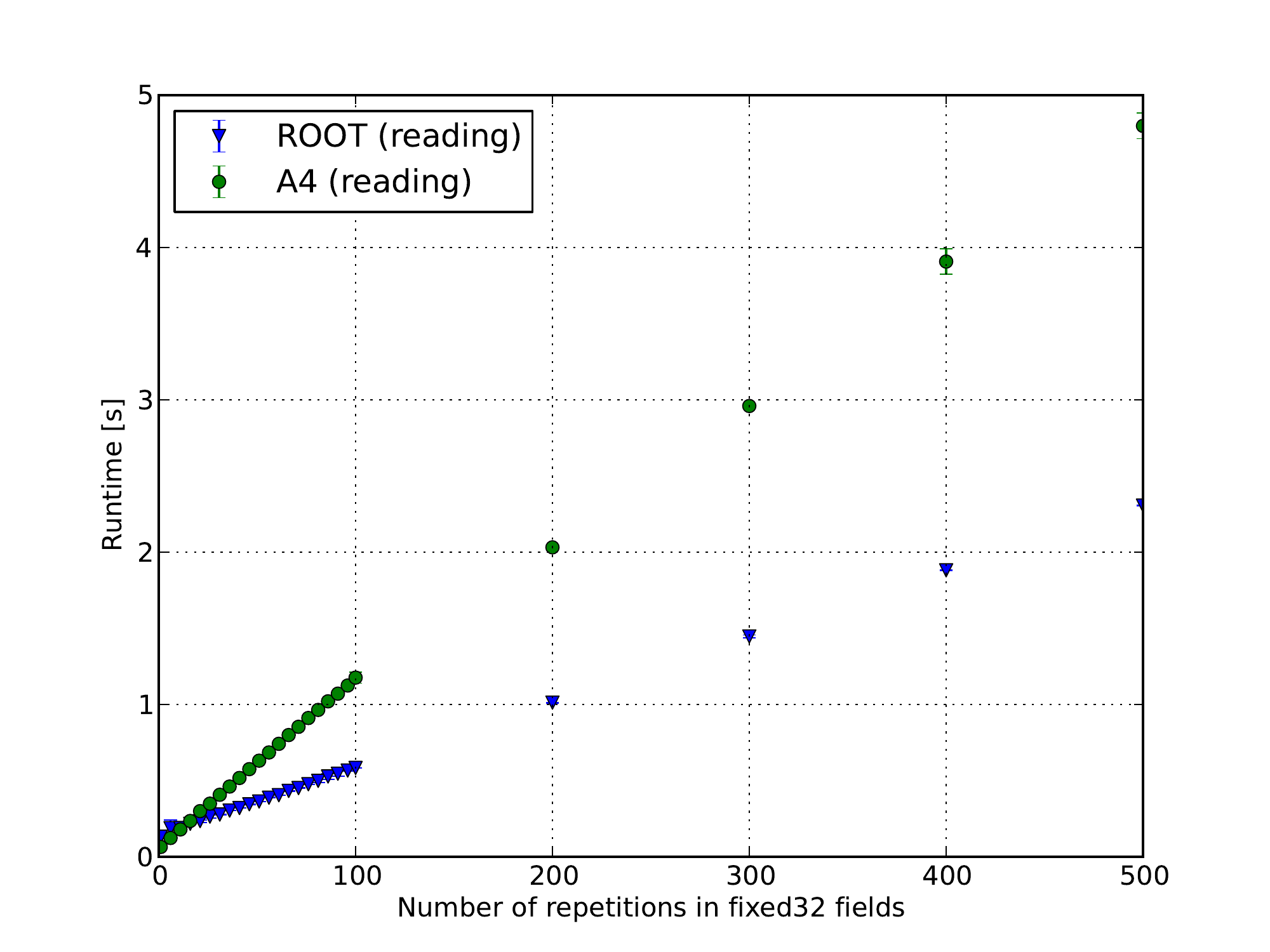}
\includegraphics[width=0.40\textwidth]{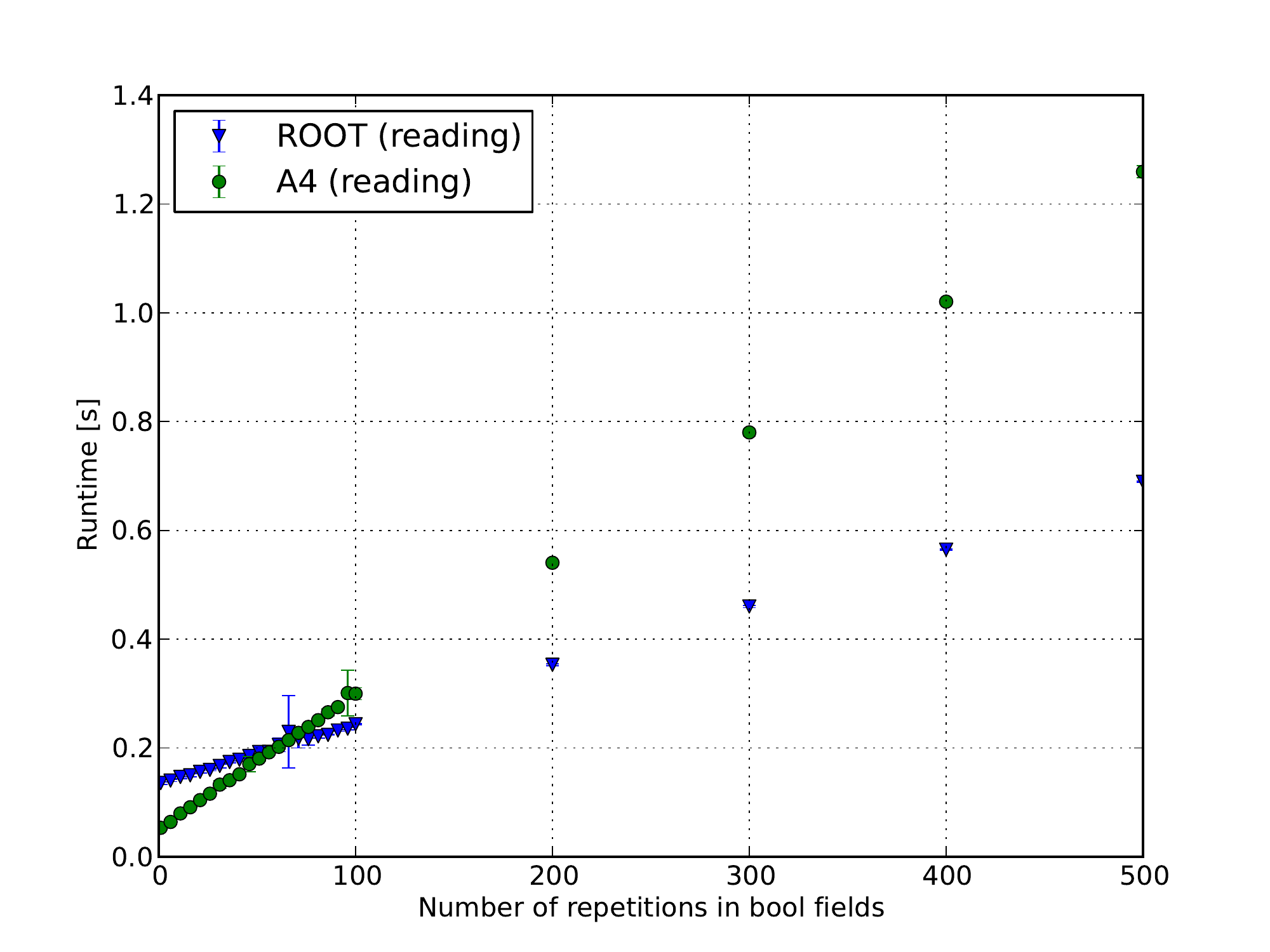}\\
\includegraphics[width=0.40\textwidth]{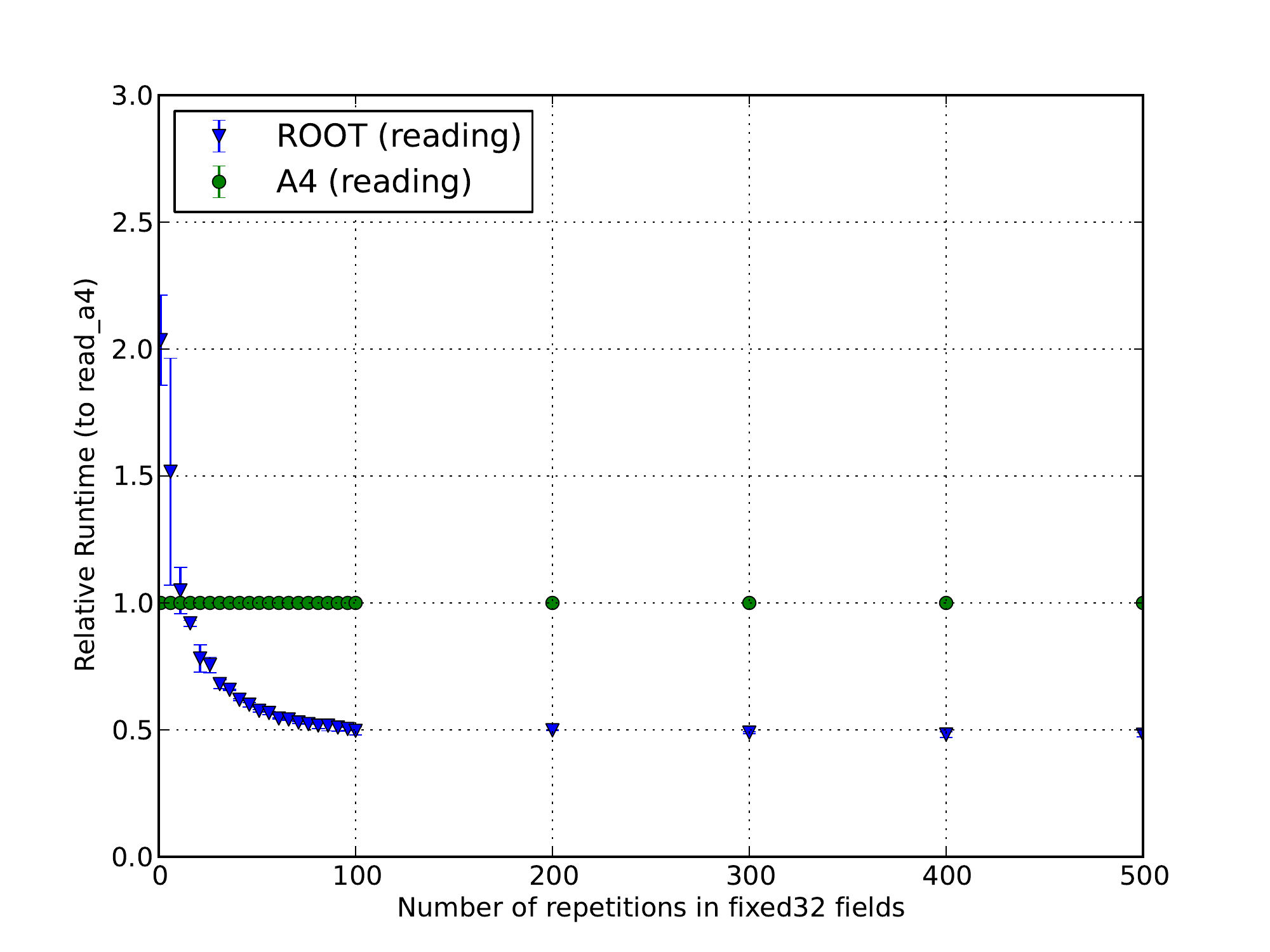}
\includegraphics[width=0.40\textwidth]{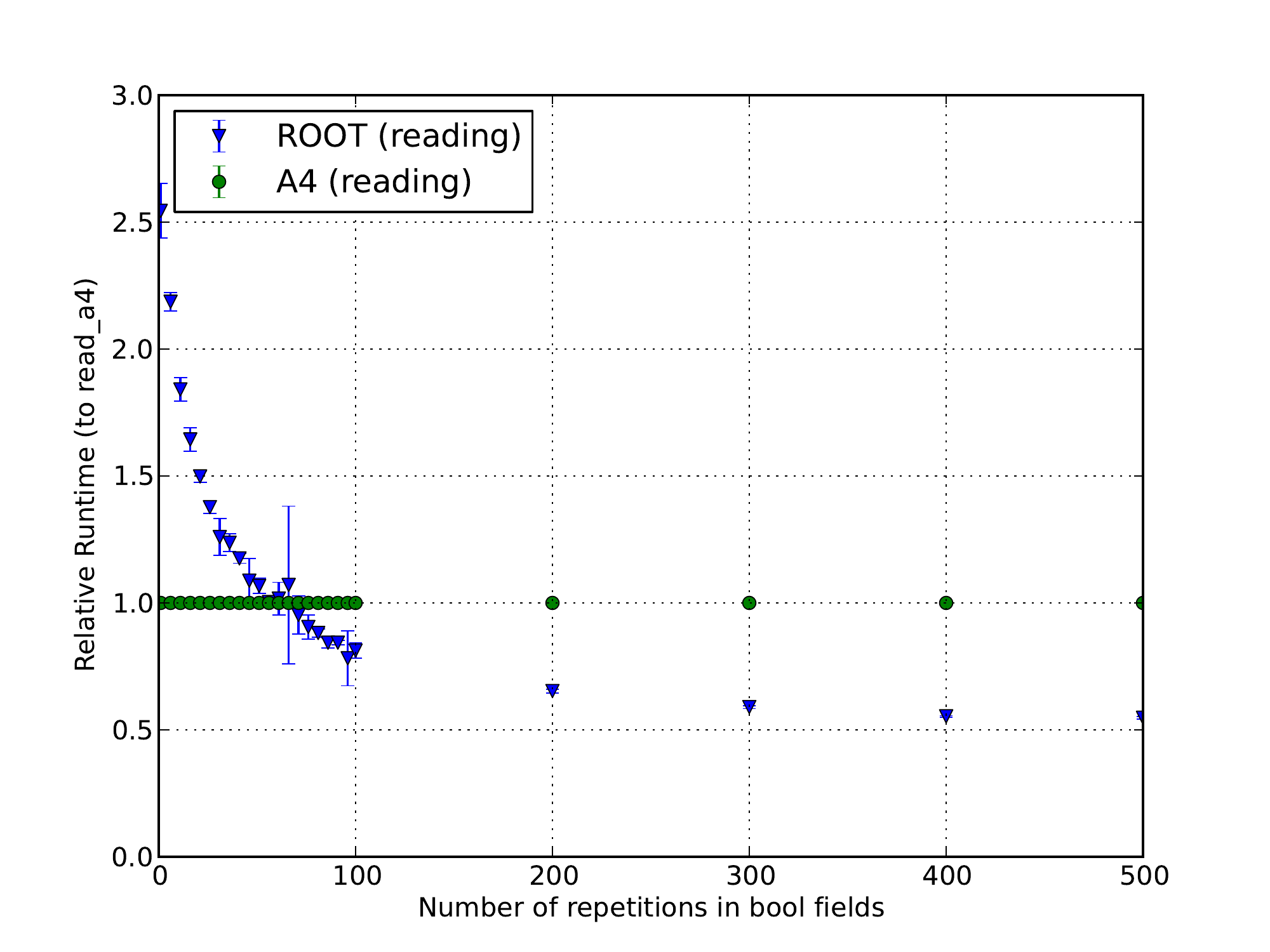}
\caption{Processing time in seconds for $100000$ events versus $n_{float}$ for $n_{rep} = 4$, for floats, doubles, integers and booleans from top left to bottom right. The top row shows absolute runtime, the lower row runtime relative to {\scshape a4}. Compression is enabled.}
\end{center}\end{figure}
\begin{figure}[ht]\begin{center}
\includegraphics[width=0.40\textwidth]{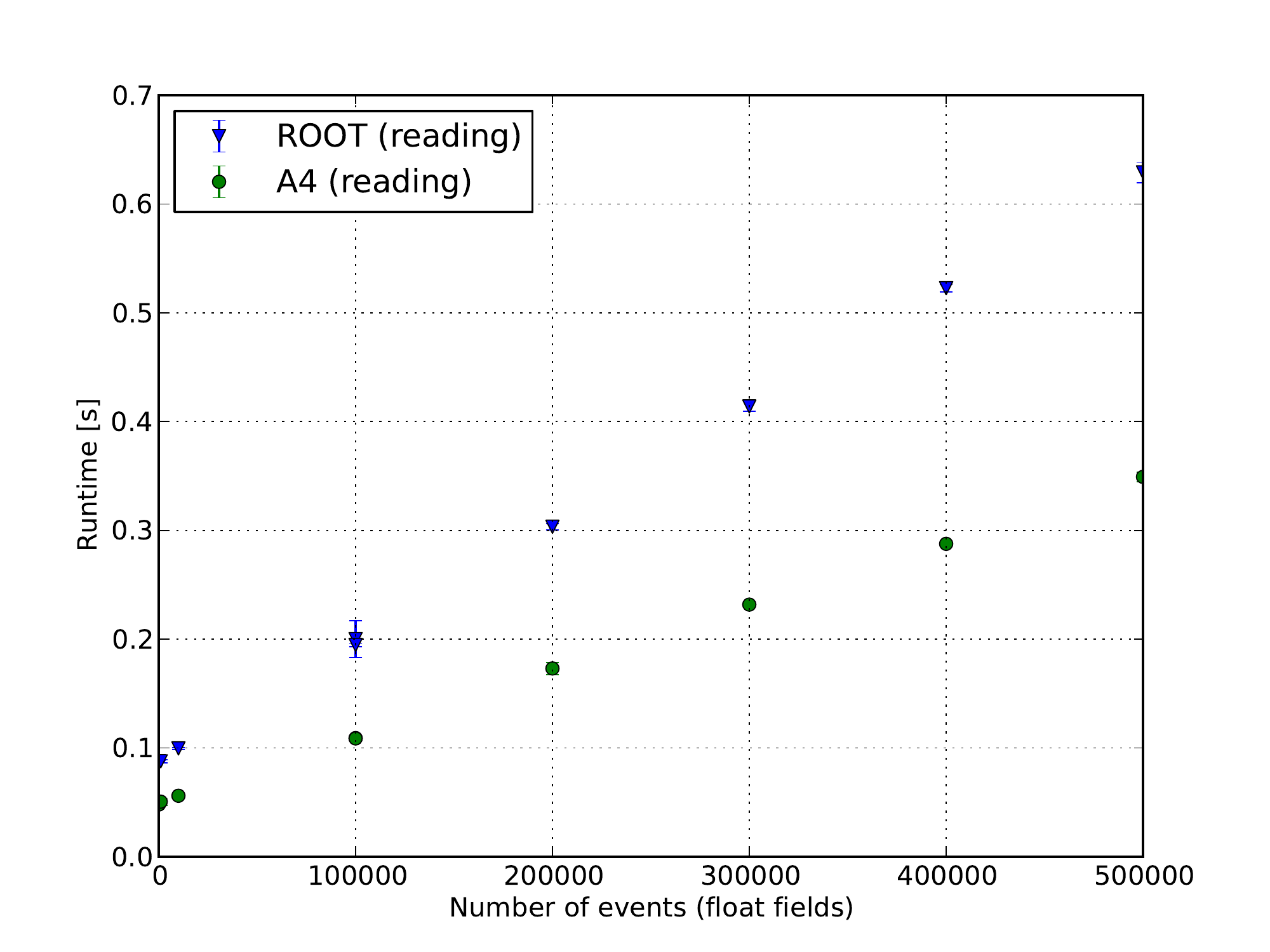}
\includegraphics[width=0.40\textwidth]{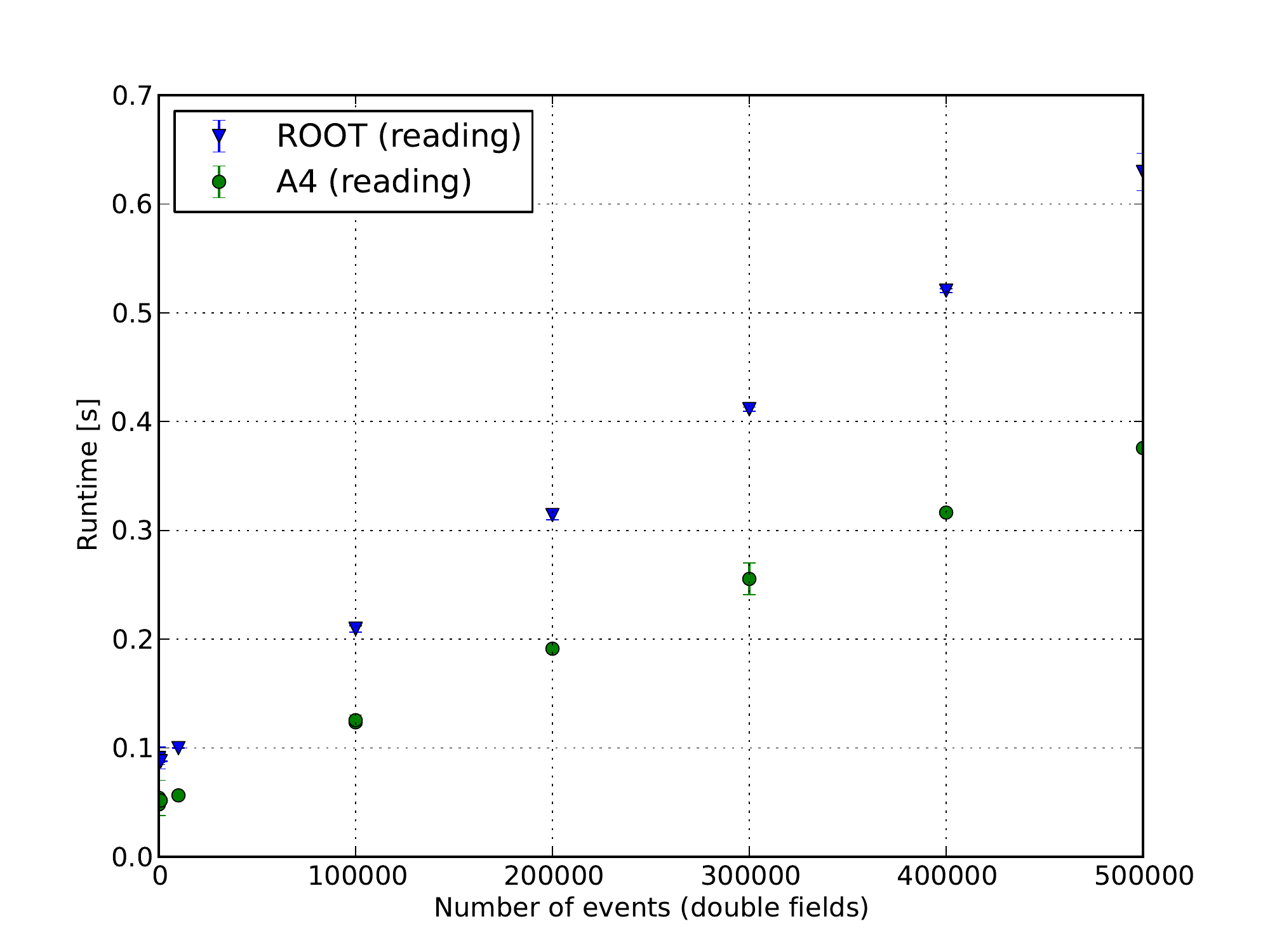}
\includegraphics[width=0.40\textwidth]{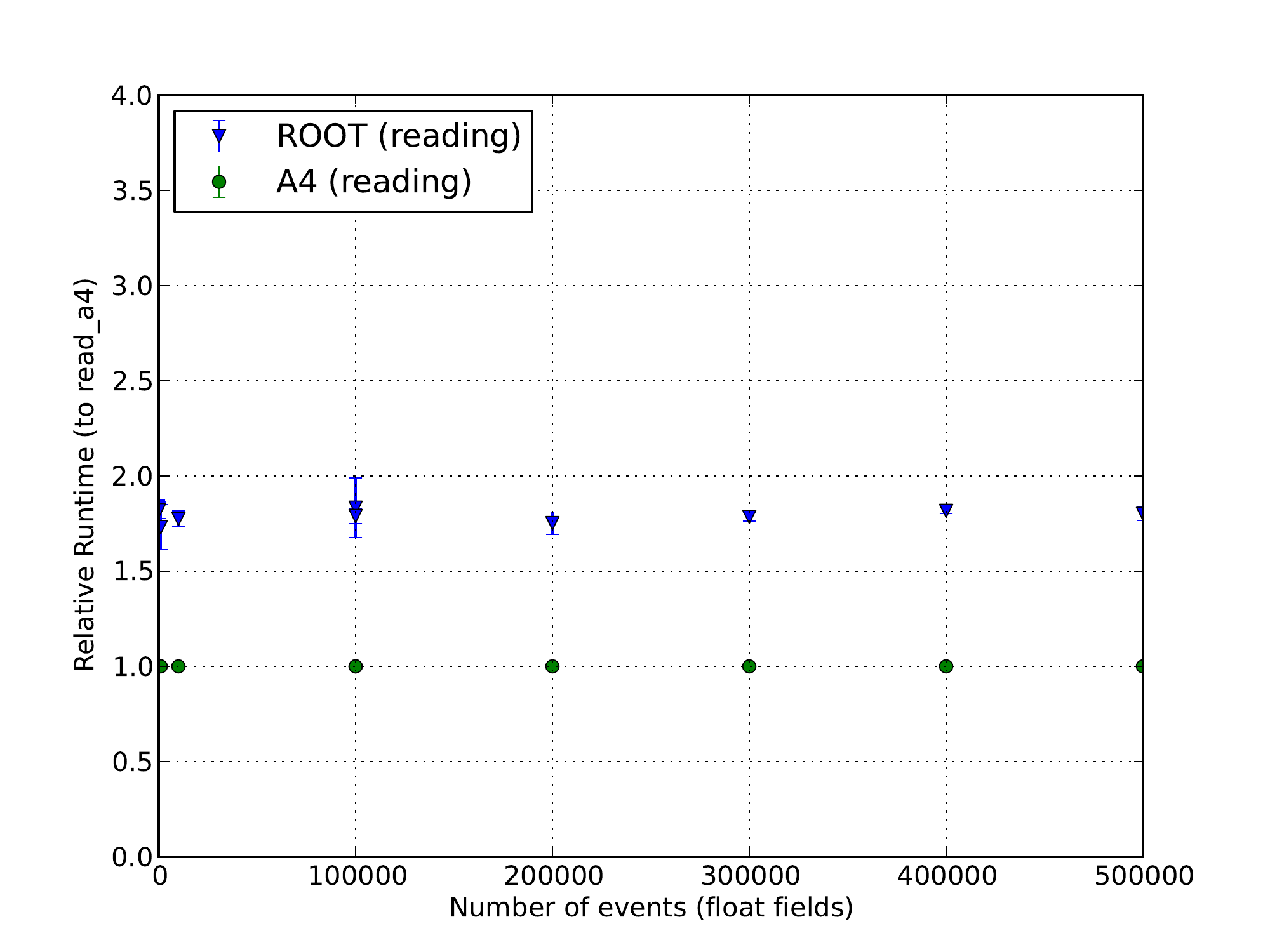}
\includegraphics[width=0.40\textwidth]{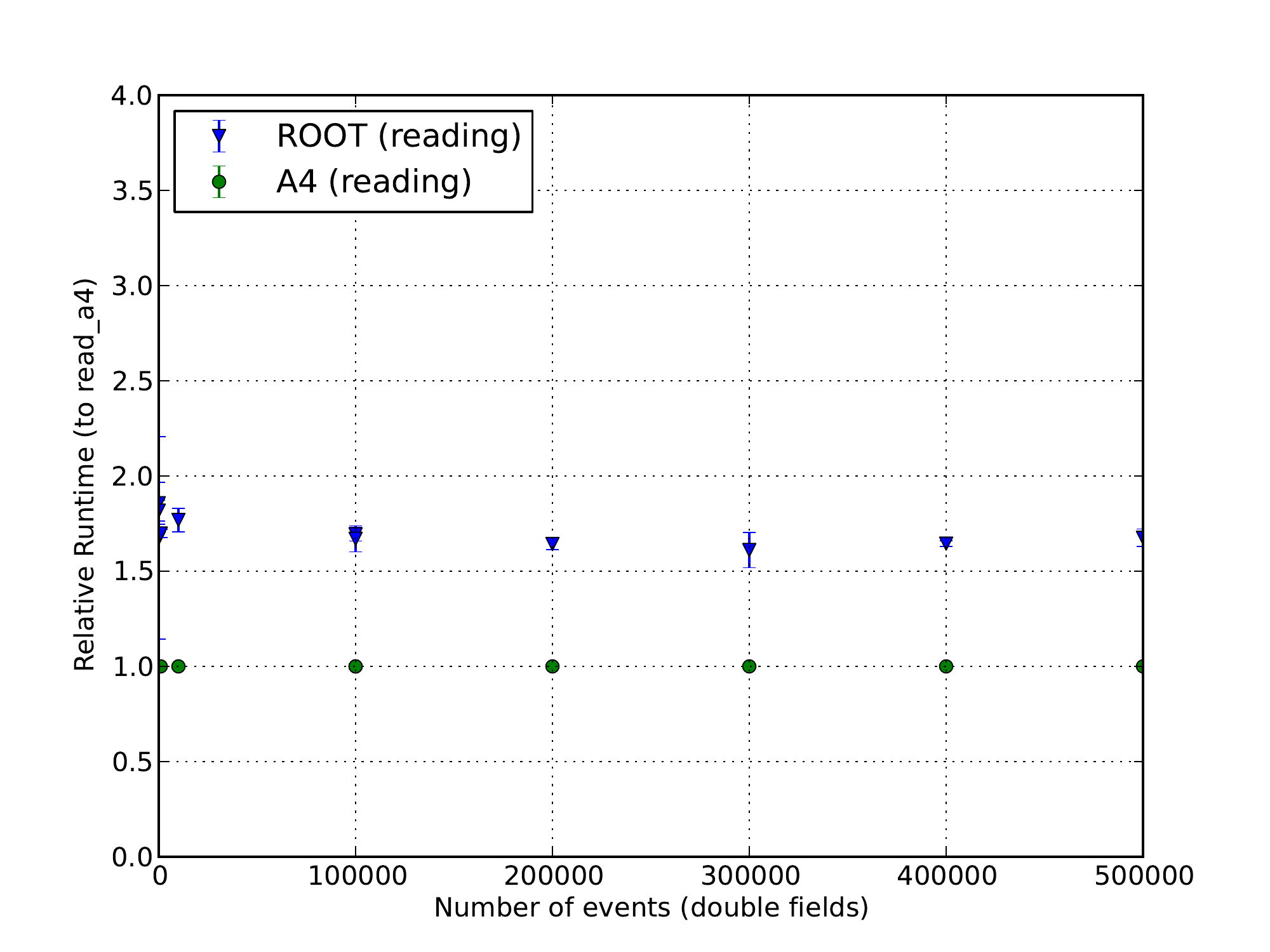}
\begin{tabular*}{\textwidth}{c}\hline\end{tabular*}
\includegraphics[width=0.40\textwidth]{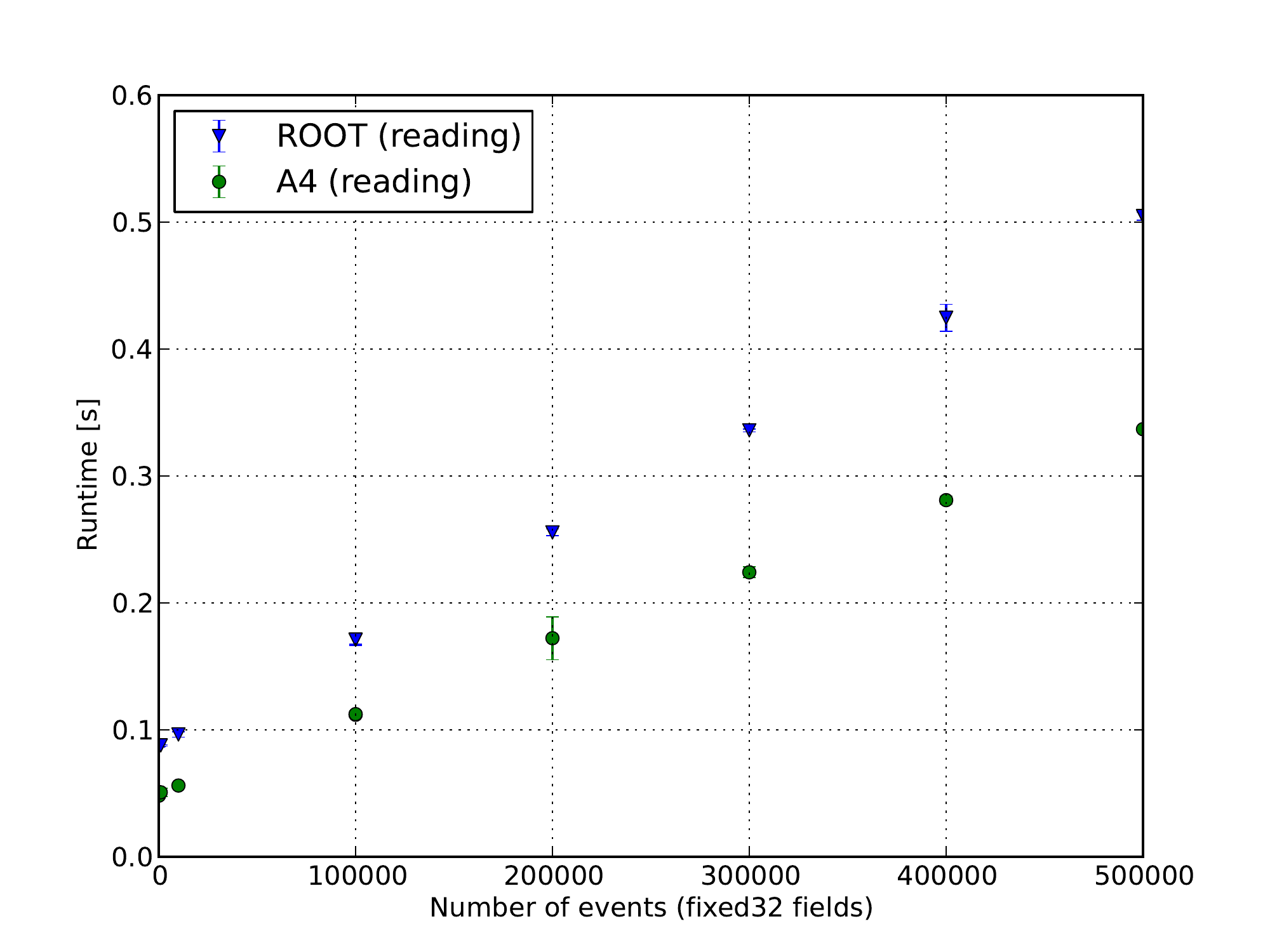}
\includegraphics[width=0.40\textwidth]{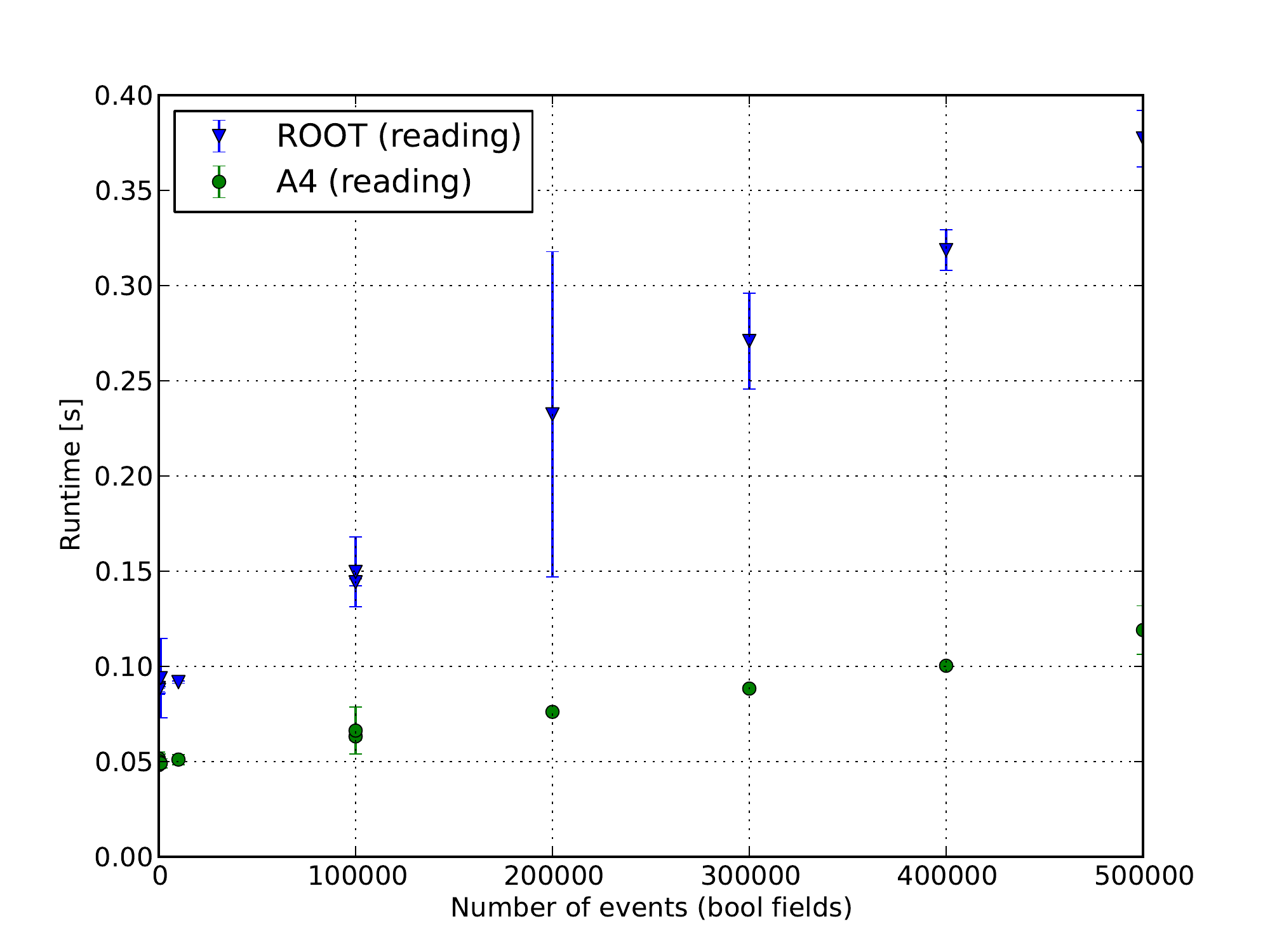}\\
\includegraphics[width=0.40\textwidth]{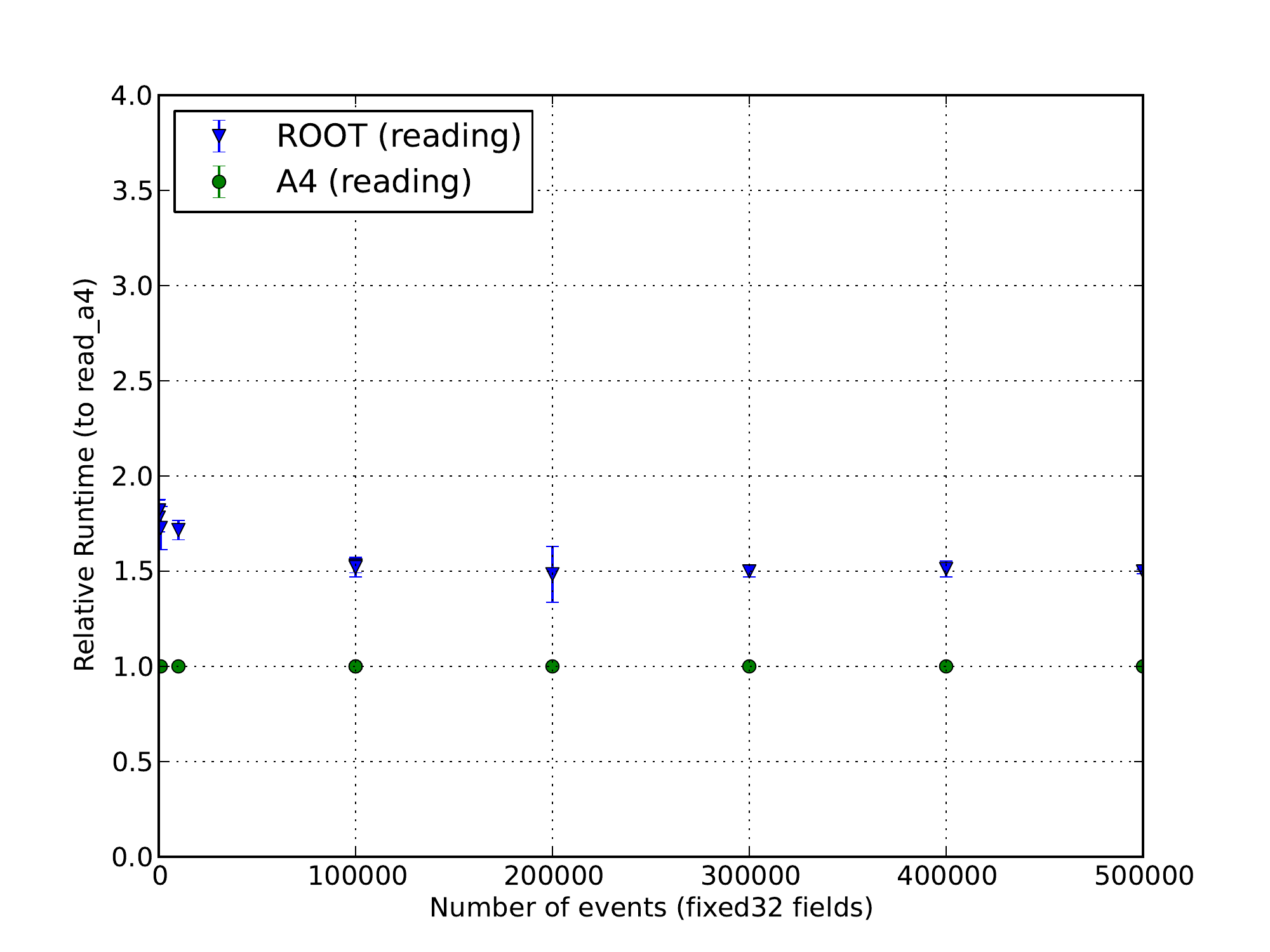}
\includegraphics[width=0.40\textwidth]{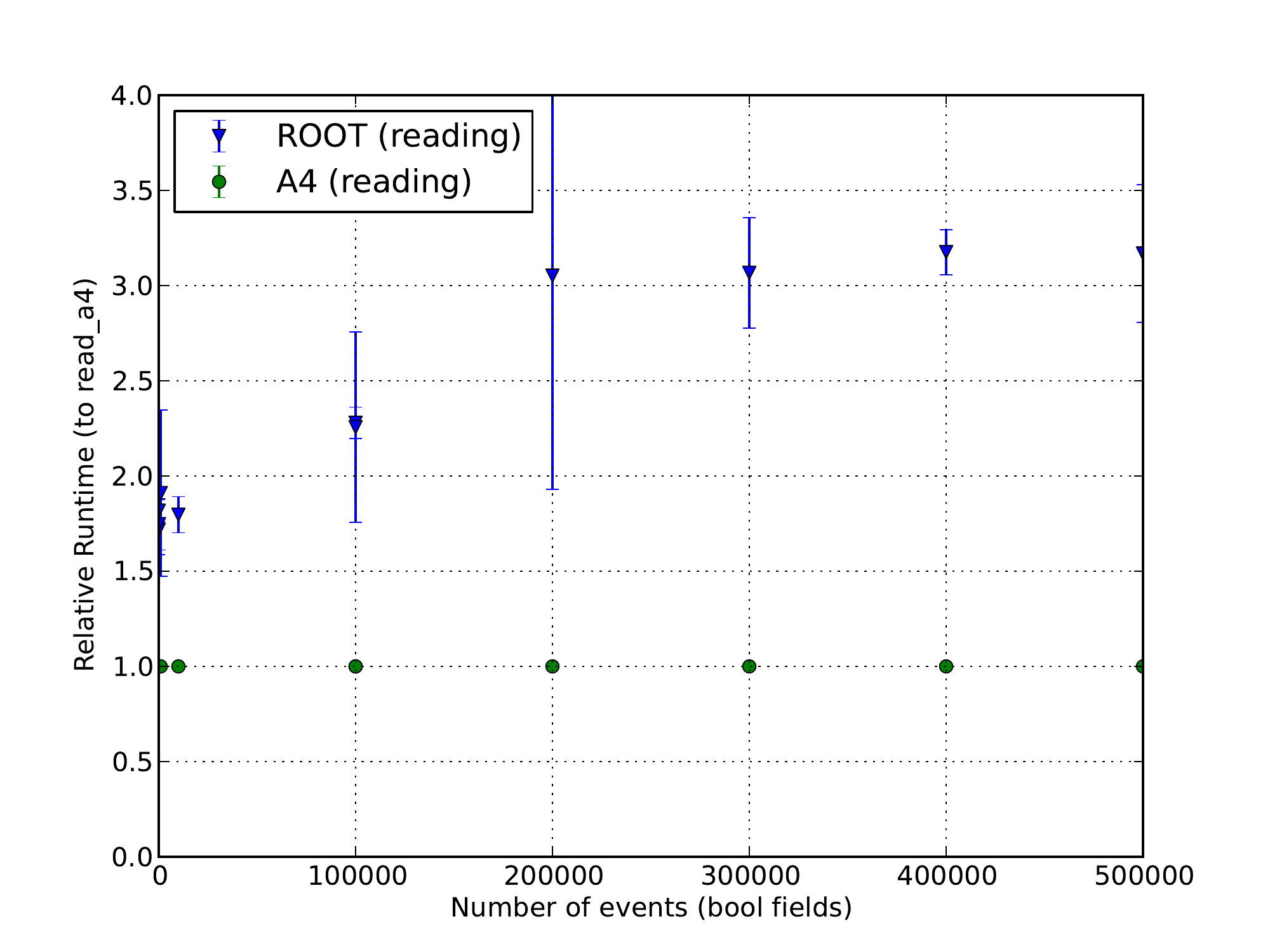}
\caption{Processing time in seconds vs number of events for $n_{flat} = 4, n_{rep} = 4$ and $n_{nfill} = 4$, for floats, doubles, integers and booleans from top left to bottom right. The top row shows absolute runtime, the lower row runtime relative to {\scshape a4}. Compression is enabled.}
\end{center}\end{figure}

\end{document}